\documentstyle[graphics] {article}
\begin{document}

\title{
Statics and Dynamics of Strongly  Charged Soft Matter}

\author{H. Boroudjerdi, Y.-W. Kim,  A. Naji, 
R.R. Netz\footnote{E-mail : netz@ph.tum.de}, X. Schlagberger, A. Serr\\
  Physics Department, TU Munich, 85748 Garching, Germany}

\date{\today}
\maketitle

\begin{abstract}
Soft matter materials, such as polymers, membranes, proteins, are
often electrically charged. This makes them water soluble, which is
of great importance in technological application and a prerequisite 
for biological function. We discuss a few static and dynamic systems
that are dominated by charge effects. One class comprises complexation
between oppositely charged objects, for example the adsorption of charged 
ions or charged polymers on oppositely  charged substrates of different
geometry. Here the main questions are whether adsorption occurs and
what the effective charge of the resulting complex is. We explicitly 
discuss the adsorption
behavior of polyelectrolytes on
substrates of planar, cylindrical and spherical geometry with
specific reference to DNA adsorption on supported charged lipid layers,
DNA adsorption on oppositely charged cylindrical dendro-polymers, and
DNA binding on globular histone proteins, respectively. 
In all these systems salt plays an important role, and some of the important features
can  already be obtained on the linear Debye-H\"uckel level.
The second class comprises effective interactions between similarly 
charged objects. Here the main theme is to understand
the experimental finding
that similarly and highly charged bodies attract each other
in the presence of multi-valent counterions. 
This is demonstrated using field-theoretic arguments as well as
Monte-Carlo simulations for the case of two homogeneously charged bodies. 
Realistic  surfaces, on the other hand, 
are corrugated and also exhibit modulated charge distributions,
which is important for static properties such as the counterion-density distribution,
but has even more pronounced consequences for dynamic properties such as
the counterion mobility. 
More pronounced dynamic effects are obtained with highly condensed charged
systems in strong electric fields. 
 Likewise, an electrostatically collapsed highly 
charged polymer is unfolded and oriented in strong electric fields. 
All charged systems occur in water, and water by itself is not a very well
understood material. At the end of this review, we give a very brief and 
incomplete account of the behavior of water at planar surfaces. 
The coupling between water structure and charge effects is largely unexplored,
and a few directions for future research are sketched.
On an even more nanoscopic level,
we demonstrate using ab-initio methods that specific interactions
between oppositely charged groups (which occur when their electron
orbitals start to overlap) are important 
and cause ion-specific effects that have recently moved into the focus of 
interest.
\end{abstract}

\tableofcontents

\section{Introduction}

Processes and structures involving 
electrostatic interactions are abundant 
in  soft matter and play an important role 
in colloidal, polymeric, and biological 
systems\cite{Holm,attardrev,Loewrev,Bellrev,Levrev,Shklovrev,Linserev,Alirev}. 
This is because charges tend to make objects soluble in water.
Even the ubiquitous van-der-Waals or dispersion interactions are in fact due to locally 
fluctuating electric fields (or, equivalently, spontaneous polarization charges)\cite{Mahanty}.
Soft materials are easily deformed or rearranged by potentials comparable to  thermal energy;
examples include polymers, self-assembled membranes or micelles and complexes
formed by the binding of oppositely charged macromolecular components.
It becomes clear  that interactions caused
and mediated by permanent and induced charges constitute prominent 
factors determining the behavior and properties of soft matter at the
mesoscopic scale,
since they are strong enough to control and modify soft matter structures.
We list  three examples to demonstrate the diversity of phenomena 
we have in mind:

\begin{itemize}
\item Colloids\footnote{The term {\em colloid} refers to an object that 
is larger than $1 nm$ and smaller than a few microns and thus encompasses 
proteins, polymers, clusters, micelles, viruses and so on.}  that are 
dispersed in aqueous solvents experience mutual attractions 
due to van-der-Waals forces\cite{Mahanty,Verwey} and additional
solvent-structure-induced forces\cite{Netz0}.
They thus tend to aggregate and form large agglomerates\cite{Weitz}.
Large aggregates typically sediment,
thereby destroying the dispersion. In colloidal science, 
this process is called coagulation or flocculation, depending on the 
strength and range of the inter-colloidal forces involved.
In many industrial applications (for example dispersion paints, 
food emulsions such as mayonnaise or milk), stability of a dispersion
is a desirable property, in other applications (such as sewage or 
waste-water treatment) it is not\cite{Napper,Horn}. One way to stabilize a 
colloidal dispersion against coagulation is to impart permanent charges 
to the colloids: Similarly charged particles typically repel each other  such that 
van-der-Waals attraction (which is always stronger than electrostatic 
repulsion at small distances) cannot induce 
aggregation\cite{Verwey}.
Every rule has an exception, and in this particular case
it is an interesting exception: It has been found over
the years that strongly charged colloids in certain cases
attract strongly, which caused considerable confusion at first and
is now quite well understood due to intense research over the last
years (more of this in Section 3)\cite{Bellrev,Levrev,Shklovrev,Linserev,Alirev}.
A second method of stabilizing
a colloidal dispersion is to graft polymers to the surface 
of the colloids. If the polymers are under good-solvent conditions,
they will swell and inhibit close contacts between two 
colloids. For this task, charged polymers 
are ideal, since they swell a lot in water\cite{PEbrushrev}.
Many structures obtained with
charged colloids bear resemblance with atomic structures, but occur on
length and time scales that are much easier to observe experimentally.
To some extent, colloidal systems have been used as models
for ordering  phenomena  on the atomistic scale.

\item Polymer science and technology have revolutionized the design, 
fabrication, and processing of modern materials and form an integral part 
of every-day life\cite{Degennesbook,Edwardsbook,Grosbergbook}. 
Classical polymer synthesis is based on hydro-carbon 
chemistry and thus leads to polymers which are typically insoluble in 
water. In the quest for cheap, environmentally friendly, and non-toxic 
materials, attention has shifted to  charged polymers, so-called
{\em polyelectrolytes}, since they are typically 
water-soluble\cite{Dautzen,Foerster,Barrat}. 
The mechanism behind this water-solubility is connected with the 
translational entropy of mobile ions that are trapped in the polyelelctrolyte solution\cite{Netzrep}.
For some polyelectrolytes, the resulting 
affinity for water is so high that they are righteously called 
{\em super-adsorbing  polymers}: They can bind amounts of water in multiple
excess of their own weight\cite{Linsesuper}. This property is put to good use in many
practical applications such as diapers.

\item Human DNA, the storage medium of all genetic information, is a
semiflexible biopolymer with a total length of roughly $2 m$, 
bearing a total negative 
charge of about $10^{10} e$ (where $e$ denotes an elementary charge), 
which is contained inside the cell nucleus with a diameter of less than $10 \mu 
m$. In addition to the task of confining such a large, strongly charged 
object in a very small compartment, the DNA is incessantly replicated, 
repaired, and transcribed, which seems to pose an unsurmountable 
DNA-packaging problem. Nature has solved this by an ingenious 
multi-hierarchical  structure. On the lowest level, a short
section of the DNA molecule, consisting of 146 base pairs (corresponding
to a length of roughly $50 nm$) is 
wrapped twice around a positively charged protein (the so-called 
histone). By this, the DNA is both compactified and partially neutralized.
 In experiments\cite{Yager}, it has been shown that
a tightly wrapped state is only stable for intermediate, physiological salt 
concentrations. Since salt modulates the 
electrostatic interactions, it is suggested that
electrostatics are responsible for this interesting behavior.
Indeed, as is explained in Section 6, only at intermediate
salt concentration is an optimal balance between 
electrostatic DNA--DNA repulsion
(favoring a straight DNA conformation) and the DNA--histone attraction
achieved. 
Similar complexes between charged spherical objects 
and oppositely charged polymers are also studied experimentally
in the context of micelle-polymer\cite{Dubin,McQuigg} and
colloid-polymer\cite{Haronska,Zhang,Gittins1,Gittins2} interactions.
\end{itemize}

In these examples, electrostatic interactions dominate, they are
responsible for the salient features and the characteristic properties 
 and therefore have to be included in any theoretical description.
This is the type of system we aim at in this review, and this is also 
the operational definition of a strongly charged system: a system
where it makes sense to neglect other interactions than 
Coulombic in a first approximation (a more quantitative
definition will be introduced in Section 3).
Of course, the boundary to materials where other interactions
come into play as well is diffuse: water structures at neutral
and charged interfaces exhibit surprising properties and 
can often not be neglected, as is discussed in Section 10. 
Likewise, almost all phenomena involving
charges in aqueous solution show a characteristic  ion-specificity\cite{Hofintro},
namely a poorly understood dependence on the specific 
ion type present in the bulk, which is somehow related
to the quantum-chemical properties of  solvated ions (see Section11).

Our viewpoint is that it makes sense to use the whole scenario
of simplified models theoretical physicists love and are used to,
namely to treat charged macroions as smooth, featureless 
and homogeneously charged bodies, ions as point-like
or (on a higher level) as charged spheres, and to replace water 
by a continuum medium.
This was very successful in the past  (as is reviewed in Sections 3 and 5-8)
and there are many lessons still to be learned  on this level. 
At the same time, many of the presently pressing experimental questions
can only be answered if one leaves this level and treats
water as a discrete solvent with the capability to 
rearrange at surfaces and close to charged particles and
ions as complex objects that form weak bonds with other charges
or water molecules. It is as yet not clear whether fundamental insight
can be gained on this more microscopic level or whether one will be lost in the realm of
particularities (Sections 10 and 11 give testimony of the problems one 
encounters when dealing with charges in  the microscopic world).
The hope would be that a coarse-grained formulation in terms of effective parameters
will still be possible which would  nevertheless encompass ion-specific and solvation effects.

\section{Charges: Why and how}

Almost any material acquires a surface charge when dipped into water.
{\bf \em Permanent} charges on single molecules, surfaces, or interfaces in aqueous 
media arise via two routes: Firstly, the substance can contain dissociable surface 
groups, which under suitable pH conditions may donate protons (in which 
case one speaks of {\em acidic} groups), thereby imparting  negative
charges to the surface, or accept protons 
(these are called {\em basic} groups) and thus produce 
positive charges on the surface (the pH is a logarithmic measure of the bulk
proton concentration, as will be discussed at length in Section 9). 
What is the mechanism for this dissociation? Why should 
molecules fall apart spontaneously to produce charged parts and why
do these oppositely charged pieces not bind together again? 
As an example, consider the ionisation of hydrogen, which requires the energy of 
$E_{ion}=13.6 eV$ or (in units of the thermal energy at room temperatures) 
$E_{ion} \approx 500 k_BT$. Clearly, this ionization process cannot be thermally activated
at room temperatures. The situation is very different for chemical groups which have
acidic character: Here the energy needed to remove a proton from the 
molecule in an aqueous environment  is much smaller; 
to give a few examples,
it is roughly $14 k_BT$ for the carboxyl group  in the reaction
\begin{equation}
RCOOH +H_2O \rightarrow RCOO^-+H_3O^+
\end{equation}
 and 
$9 k_BT$ for the sulfonic group in the reaction 
\begin{equation}
RSO_3H +H_2O \rightarrow RSO_3^-+H_3O^+.
\end{equation}
The sulfonic group
is therefore said to be a stronger acid than the carboxylic group. 
The dielectric properties of the  surrounding
water are very important in these reactions, as without water  (i.e. in the gas phase) 
these reactions cost much more energy (see Section 11).
Still, energy has to be paid in order to crack the acids, but again water properties
come in: Since the concentration of water molecules in the condensed liquid state
(about $55 mol/l$) is much higher than of the other components, according to the
law of mass action the equilibrium is shifted to the right side and charged groups 
do indeed occur frequently. 
The equilibrium between association and dissociation can be fine-tuned by temperature and
the concentration of $H_3O^+$ ions in the solution (i.e. $pH$).
The second mechanism for the permanent 
charging of surfaces involves small charged molecules, such as salt ions, 
which physically or 
chemically adsorb to a surface, thereby leading to an effective surface 
charge. In practice, one typically encounters a mixture of these two
mechanisms, such that the effective charge of a surface is controlled by 
the distribution of acidic and basic surface groups, solution pH, and 
bulk concentration of charged solutes. 
{\em Induced} charges arise via the polarization of atoms, molecules, and  
macroscopic bodies\cite{Boettcher}. 
For molecules that possess a permanent
dipole moment (such as water), the macroscopic polarization 
contains a large contribution from the orientation of such 
molecular dipole moments.
The interaction between spontaneous polarization charges gives rise to van-der-Waals forces,
which act between all bodies and particles, regardless of whether they are
charged, contain permanent dipole moments or not\cite{Verwey,Mahanty}.

 The reduced electrostatic interaction 
between two spherically symmetric  charges in vacuum (throughout this review, all 
energies are given in units of the thermal energy $k_B T$)
 can be written as $U(r)= Q_1 Q_2 v(r)$ where
\begin{equation} \label{intro1}
v (r) = \frac{e^2}{4 \pi \varepsilon_0 k_B T r}
\end{equation}
is the Coulomb interaction between two elementary charges,
$Q_1$ and $Q_2$ are the reduced charges in units of the elementary 
charge $e$, and $\varepsilon_0$ is the vacuum dielectric 
constant\footnote{Note that the Syst\'{e}me International
(SI) is used, so that the factor $4 \pi$ appears in the Coulomb
law but not in the Poisson equation.}. 
The interaction only depends on the distance $r$ between the 
charges. Electrostatic interactions are additive, therefore the total
electrostatic energy of a given distribution of charges results from
adding up all pairwise interactions between charges according to 
Eq.(\ref{intro1}). {\em In principle}, the equilibrium behavior 
of an ensemble of charged  particles (e.g. a salt solution) follows from the 
partition function, i.e., the weighted sum over all different microscopic 
configurations, which ---via the Boltzmann factor---
depends on the electrostatic energy of each configuration.  
{\em In practice}, however, this route is complicated for several reasons:

 i) The Coulomb interaction, Eq.(\ref{intro1}), is very long-ranged, such 
that (even, and as turns out, especially
 for low densities) many particles are coupled due to their
simultaneous electrostatic interactions\footnote{The potential Eq.(\ref{intro1})
reaches unity at a distance of roughly $r \approx 56nm$, which in
the nanoscopic world is considered large.}.  Electrostatic problems are 
therefore typically {\em many-body problems}. As is well known, even the 
problem of only three bodies interacting via gravitational potentials (which 
are analogous to Eq.(\ref{intro1}) except that they are always attractive) 
defies closed-form solutions. To make the problem even worse,
even if we consider only two charged
particles, the problem effectively becomes a many-body problem,
for the following two reasons:

ii) In almost all cases, charged objects are dissolved in water. As all
molecules and atoms, water is polarizable and thus reacts to the presence
of a charge with polarization charges. In addition, and this is a by far
more important effect, water molecules carry a permanent dipole moment, 
and are thus partially oriented in the vicinity of charged objects.
The polarization
effect of the solvent can to a good approximation\footnote{ 
Deviations from this continuum linear approximation 
take the form of a momentum-dependent dielectric 
function $\tilde{\varepsilon}(k)$ and non-linear correction terms.
They are important for the solvation of ions.} 
be taken into account
by introducing a relative dielectric  constant 
$\varepsilon$\cite{Boettcher,Mossotti,Debye,Onsager2}. Note that for water,
$\varepsilon \approx 80$, so that electrostatic interactions
are much weaker in water than in air (or some other low-dielectric 
solvent). The Coulomb potential  now reads
\begin{equation} \label{intro2}
v (r) = \frac{e^2}{4 \pi \varepsilon_0 \varepsilon k_B T r} = \frac{\ell_B}{r}
\end{equation}
and the Bjerrum length $\ell_B = 1/(4 \pi  \varepsilon \varepsilon _0 k_B T)$,
which is a measure of the distance where the
interaction is of thermal strength, has the value $\ell_B \approx 0.7 nm$.

iii) In all biological and most industrial applications, water contains
mobile salt ions. Salt ions of opposite charge are 
drawn to charged objects and form loosely bound counter-ion clouds
and thus effectively reduce their charges; this process is called 
{\em screening}. The effect of charge screening is dramatically different
from the presence of a polarizable environment. As has been shown by
Debye and H\"uckel some 80 years  ago\cite{DH}, screening modifies the
electrostatic interaction such that it falls off exponentially with 
distance. 

The following points are important for the discussion in the subsequent  sections:
For each surface charge an
oppositely charged counterion is released into the aqueous solution. These counterions
form clouds that are loosely bound to the surface charges. 
The interactions between charged
bodies and their electric properties itself (such as their electrophoretic mobilities
in an electric driving field) are predominantly determined by the properties
of these counterion clouds, and an understanding of the properties of charged
bodies requires an understanding of the counterion clouds first. 
Highly and opppositely charged surfaces or particles with 
permanent charges  typically have interaction potentials that are
much stronger than thermal energy, one often obtains quasi-bound complexes
which have to be dealt with in a very different way than the rather 
diffuse and highly fluctuating counterion distributions. Typically,
charged soft matter (e.g. polymers, fluid membranes) is deformable and shows
thermally excited shape fluctuations, and one is dealing with the 
intricate interplay of shape and counterion fluctuations.
Electric fields are used in electrophoresis experiments
to analyze and purify charged soft matter. The electric field sets
charged ions and particles in motion and thus leads to dissipation of energy,
one is facing a non-equilibrium situation.
It also changes the equilibrium distribution functions, and can lead to
non-equilibrium phase transitions, as will be shown towards the end of this review.
Finally, oppositely charged chemical groups are often in intimate contact to each
other, for example in situations when oppositely charged bodies are bound to each 
other. The boundary between chemical binding and salt bridging is diffuse,
and quantum-mechanical effects which are caused by the overlap of electron
orbitals give sizeable and very specific contributions to the effective 
interaction between charged groups. For a detailed understanding of the statistics
and dynamics of charged soft matter, those quantum-mechanical effects 
in principle have to be taken into account.

\section{Interactions between charged objects}

\subsection{Attraction between similarly charged plates: a puzzle?}

\begin{figure}[t]
\begin{center}
\resizebox{12cm}{!}{
\includegraphics{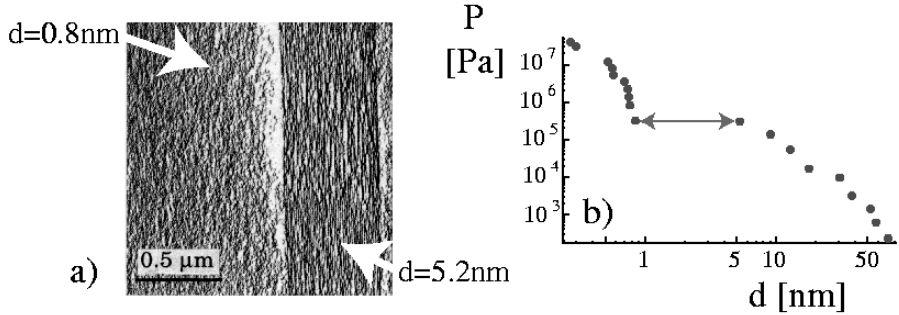}
}
\end{center}
\caption{a) Cryo-electron-micograph of a membrane stack
consisting of equal amounts of water and DDAB surfactant, 
frozen in from the equilibrated structure at room temperature,
exhibiting macroscopic phase separation between two lamellar phases
of different water content and thus different spacing between the bilayers
(adapted from Ref.\cite{Zemb3}).
b) Osmotic pressure as a function of the water-layer thickness $d$.
A pronounced plateau is apparent(adapted from Ref.\cite{Dubois}).}
\label{fig1}
\end{figure} 

Experimentally, the interaction between charged planar objects can be very
elegantly studied using a stack of charged, self-assembled 
membranes\cite{Zemb1}-\cite{Zemb6}. 
Such membranes spontaneously form in aqueous solution of charged amphiphilic
molecules (lipids or surfactants) and consist of bilayers which are
separated by water slabs of thickness $d$ (it is the same
structure that forms an integral part of biological cell 
walls)\cite{Lip}.
Since the membranes are highly charged (they typically contain one surface
charge per $0.6 nm^2$ and thus belong to the most highly charged surfaces
known), one would expect strong repulsion between them, or, which is
equivalent, a strongly positive and monotonically decaying osmotic
pressure in such a stack. In contrast, experiments using the cationic 
surfactant DDAB show that a mysterious attraction exists between the charged
lamellae\cite{Zemb3,Dubois}. 
This is seen in Fig.1a, where an electron-micrograph of a sample
containing 50 \% water and 50 \% DDAB, rapidly frozen from the equilibrated
structure at room temperature (and thus representative of the room-temperature
situation) is shown. One can discern
a two-phase coexistence between 
two macroscopic lamellar phases with different water-layer thicknesses 
$d$. In the corresponding pressure/surfactant concentration
isotherm (obtained at room temperature) in Fig.1b the osmotic pressure
shows a pronounced plateau as a function of the water-layer thickness,
equivalent to macroscopic coexistence of two lamellar phases with different water
content.
Such phase coexistences are best known from non-ideal gases and result
from an attraction between the gas molecules (compare the van-der-Waals equation
of state). In the present case, it means that an attractive force acts between
the highly charged membranes, strong enough to overcome the electrostatic
repulsion between the charges on the membrane (note that the dispersion 
attraction is too weak by orders of magnitude to account for this attraction).
This is quite surprising, and cannot be explained within classical theories
(based on a mean-field description for the counterion distribution). 
Clearly, the real membrane system is quite complex and contains a number 
of effects that we will not consider (such as shape fluctuations, chemical structure
of the surfactant heads, etc.).
But we will demonstrate in the following that a simple argument for the
counterion induced interaction between charged surfaces suffices to explain
the observed miscibility gap in an almost quantitative fashion. This will
lead us to a theoretical description of strongly coupled charged systems
which complements the classical mean-field theory. 
In all the above-cited experiments on charged lamellar 
phases monovalent counterions were employed.
We should add that a similar attraction is also seen with less strongly
charged bilayer systems when the mono-valent counterions
are replaced by divalent counterions\cite{Khan,Khan2}.

\subsection{Counterions at a single  charged plate}

The experimentally observed attraction between similarly charged surfaces
 requires a deeper understanding
of counterion layers at highly charged surfaces, we therefore start
our discussion with a single, planar charged plate with counterions
only (i.e. no additional salt ions).
The Hamiltonian for a system of $N$ counterions of valence $q$,
located at positions ${\bf r}_i$,  close to  a 
single oppositely charged planar wall of charge density $\sigma_s$ is 
(in units of $k_BT$) given by
\begin{equation}
  \label{Hone}
  {\mathcal H} = 
  \sum_{j=1}^{N-1} \sum_{k=j+1}^{N} 
  \frac{q^2 \ell_B }{|{\bf r}_j - {\bf r}_k|} + 
  2 \pi q \ell_B \sigma_s \sum^N_{j=1} {z}_j,
\end{equation}
where $\ell_B \equiv e^2/ 4 \pi  \varepsilon \varepsilon _0 k_B T$
is the Bjerrum length ($e$ is the elementary charge, $\varepsilon$ is 
the relative dielectric constant). 
In water, one typically has $\ell_B \approx 0.7 nm$.
For the sake of simplicity, the dielectric constant is assumed 
to be homogeneous throughout the system, the plate
is smooth, impenetrable to ions  and homogeneously charged, and the counterions
are assumed to be point-like. Still, the system is nontrivial 
and allows to understand the special features of strongly charged systems
in a very lucid manner.
The first term in Eq.(\ref{Hone}) contains the 
Coulombic repulsion between all ions, the second term accounts
for the electrostatic attraction to the wall (which is assumed to be of 
infinite lateral extent and located in the $z=0$ plane). 
The relevant length scale in the system is the Gouy-Chapman length, $\mu$,
which is defined as  the distance from the charged wall at which the potential
energy of one isolated counterion equals the thermal energy $k_B T$.
As will turn out later, it is a measure of the typical height of the counterion 
layer\footnote{In fact, within mean-field theory, it is the distance
up to which half of the counterions are confined.}.
From Equation (\ref{Hone}) it can be read of to be
\begin{equation}
  \mu = \frac{1}{2 \pi q \ell_B \sigma_s}.
\end{equation}
If one expresses all lengths in units of the Gouy-Chapman
length and rescales them according to 
\begin{equation}
\tilde{r}=r / \mu,
\end{equation}
the Hamiltonian Equation (\ref{Hone}) can be rewritten as
\begin{equation}
  \label{H-one-mu}
  {\mathcal H} = \sum_{j=1}^{N-1} \sum_{k=j+1}^{N} 
  \frac{\Xi}{|\tilde{\bf r}_j - \tilde{\bf r}_k|} + 
  \sum^N_{j=1} \tilde{z}_j.
\end{equation}
Now the Hamiltonian only depends on a single parameter, the coupling 
parameter
\begin{equation}
\Xi=2 \pi q^3 \ell_B^2 \sigma_s \sim \frac{ q^3 \sigma_s}{T^2},
\end{equation}
which includes the effects of varying temperature $T$ (via the Bjerrum length
$\ell_B$), surface charge density $\sigma_s$, and counterion valence $q$.
The counterion valence enters the coupling parameter as a cube, showing 
that this is an experimental parameter which decisively controls the 
resultant behavior of the double layer (compare the experiments
with charged lamellar systems where the counterion valency has been
increased\cite{Khan,Khan2}).
Small values of $\Xi$ define the weak-coupling regime (where, as we will 
demonstrate later on, the mean-field Poisson-Boltzmann (PB)
theory becomes valid), large values define the
strong-coupling (SC) regime, where surface-adsorbed 
ions are strongly correlated\cite{Rouzina,joanny,Shklovskii}.
This strong-coupling regime
constitutes a sound physical 
limit with behavior very different from the PB
 limit, as can be shown rigorously using
 field-theoretic methods\cite{moreiraepl}-\cite{andreepl}.

\begin{figure}[t]
\begin{center}
\resizebox{12cm}{!}{\includegraphics{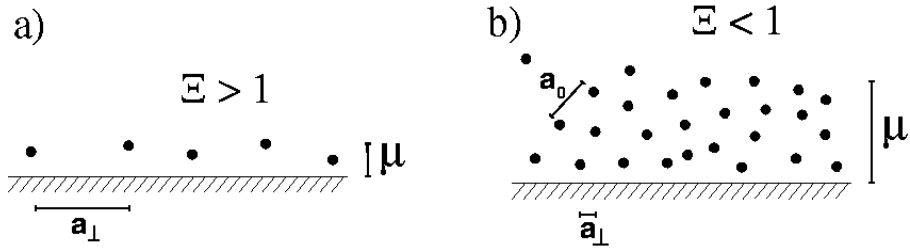}}
\end{center}
\caption{ a) For 
large coupling parameter $\Xi >1$ the
lateral distance between ions $a_\perp$ is larger
than their average separation from the wall, 
proportional to the Gouy-Chapman length $\mu$.
In rescaled units, this lateral distance reads
$\tilde{a}_\perp = a_\perp / \mu = \protect \sqrt{8 \Xi}$.
The layer is essentially flat, two-dimensional and strongly correlated.
b) For $\Xi <1$, the lateral ion separation $a_\perp$ 
is smaller than the layer height $\mu$. 
Within the counter-ion layer, the inter-ionic distance $a_0$ 
scales as $\tilde{a}_0  = a_0 / \mu \sim  \Xi^{1/3}$ and 
the ion-ion correlations are rather weak.}
\label{schema1}\end{figure}

\begin{figure}[t]
\begin{center}
\resizebox{12cm}{!}{
\includegraphics{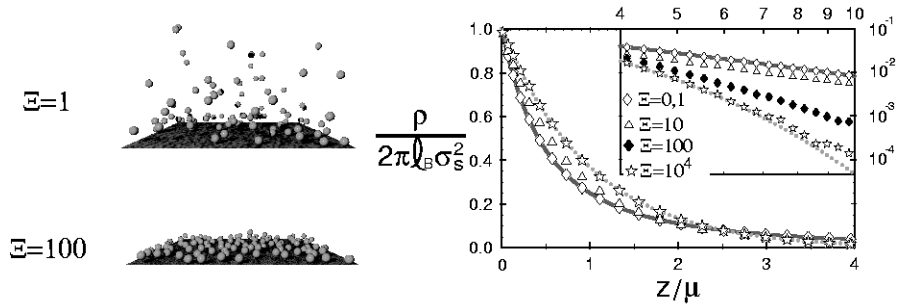}
}
\end{center}
\caption{a) Snapshots of counterion distributions at a charged
surface for two different values of the coupling parameter, showing a rather
diffuse distribution for small $\Xi$ and a flat quasi-two dimensional
layer for large $\Xi$.
b) Numerically determined counterion density profiles (data points)
as a function of the distance from the surface for different values
of the coupling parameter $\Xi$ in comparison with the asymptotic 
predictions in the mean-field (solid curve) and strong-coupling (broken line)
limits (adapted from Ref.\cite{moreiraepl}).}
\label{fig2}
\end{figure}

The mean lateral area per counter-ion is determined by the
surface charge density and defines a length scale (which we associate
with the lateral distance between ions), $a_\perp$, via the relation
\begin{equation}\label{aperp1}
\pi (a_\perp/2)^2 = q/\sigma_s.
\end{equation}
In rescaled units, this lateral distance reads
\begin{equation} \label{aperp}
\tilde{a}_\perp = a_\perp / \mu = \sqrt{8 \Xi}.
\end{equation}
Since the height of the bound counterion cloud  is unity in reduced
units, it follows from 
equation (\ref{aperp}) that for 
coupling parameters larger than unity, $\Xi > 1$, the
lateral distance between ions is larger than their separation from the
wall and thus the layer is essentially flat and 
two-dimensional, as is shown schematically
in Fig. \ref{schema1}a\cite{Rouzina,Shklovskii}.
For $\Xi <1$, on the other hand, the lateral ion separation $a_\perp$ 
is smaller than the layer height $\mu$, which means that within 
the counter-ion layer
the ion-ion correlations should be rather 3D fluid-like,
as depicted  schematically in Fig. \ref{schema1}b. 
The two different limits are visualized  in Figure \ref{fig2}, where we show snapshots 
of counterion distributions obtained in Monte-Carlo simulations
for two different values of the coupling parameter,
$\Xi= 1, 100$.
For small $\Xi$, 
the ion distribution is indeed rather diffuse and disordered 
and mean-field theory should work, since each ion moves
in a weakly varying potential due to the 
diffuse cloud of neighboring ions. For large $\Xi$, on the other hand, 
ion-ion distances are large 
compared to the distance from  the wall;
the ions form a flat layer on the charged wall.
For large $\Xi$, 
the repulsion between condensed ions at a typical
distance $a_\perp$, proportional to 
$\ell_B q^2 / a_\perp$,
 is large compared with thermal energy, as can be
seen from the fact that 
\begin{equation}
\frac{\ell_B q^2}{a_\perp} \sim \sqrt{\Xi} \sim \frac{a _{\perp}}{\mu}.
\end{equation}
The layer is thus flat and also strongly coupled\cite{Shklovskii}\footnote{Along
the same lines, for $\Xi < 1$, in the three-dimensional diffuse counterion cloud,
depicted schematically in Fig.\ref{schema1}b,
the  typical inter-ionic distance is 
$\tilde{a}_0  \ a_0 / \mu \sim  \Xi^{1/3}$
and the interaction at such distance scales as 
$\ell_B q^2 / a_0 \sim \Xi^{2/3}$\cite{Netz4}. In this case the counterion 
cloud is weakly coupled and thus only weakly correlated.}.
As will be shown in Section 3.5, the counterion layer forms a crystal around
$\Xi \approx 31000$\cite{Baus}, meaning that there is a wide range of coupling
parameters, $1 < \Xi < 31000$,
 where the counterion layer is highly correlated but still liquid.
Nevertheless, mean-field theory,
which can pictorially be viewed as an approximation where one laterally
smears out the counterion charge distribution,
 is expected to break 
down, at least for the system with $\Xi=100$; this is so 
because each ion moves, though confined by its immediate neighbors
in the lateral directions, almost independently from the other ions
along the vertical direction (which constitutes the soft mode). 
We stress that this continuous crossover from a three-dimensional,
disordered counterion distribution for small $\Xi$, to a two-dimensional
correlated counterion distribution for large values $\Xi$ (which
will be discussed in more detail later on) is a pure consequence 
of scaling analysis; as the only input, it requires the rescaled
counterion layer height to be of order unity, which is
true irrespective of the precise
value of $\Xi$ as will be demonstrated next. 

Using Monte-Carlo simulation techniques, we have obtained
counterion density profiles by averaging over statistically sampled
counterion configurations for different values of $\Xi$.
Since the surface charge density is homogeneous, the counterion
density profile $\rho(z)$ only depends on the distance from the wall, $z$.
The counterions exactly neutralize the surface charges, the 
integral over the counterion density profile is therefore given by 
(in unrescaled units) $\int_0^\infty {\rm d} z \; \rho(z) = \sigma_s /q$.
Using the rescaled distance coordinate $\tilde{z}=z/\mu$,
the integral gives  $\int_0^\infty {\rm d} \tilde{z} \; \rho(\tilde{z}) = 
2 \pi \ell_B \sigma_s^2$, which suggests to define the rescaled density
profile as
\begin{equation}
  \label{rescale}
\tilde{\rho}(\tilde{z}) =
  \frac{\rho (\tilde{z})}{2 \pi \ell_B \sigma_s^2}  
\end{equation}
which, via the condition of electroneutrality, is normalized to unity,
\begin{equation} \label{condit1}
\int_0^\infty {\rm d} \tilde{z} \; \tilde{\rho}(\tilde{z}) = 1.
\end{equation}
In Figure \ref{fig2}b
we show rescaled counterion density profiles obtained using
Monte Carlo simulations for various values of the coupling parameter
$\Xi=0.1$, $10$, $100$  and $10^4$.
One notes that all profiles saturate at a rescaled density of unity
at the charged wall. This is in 
accord with the contact-value theorem, which states that
the counterion density  at the wall is ---for the case of a single
homogeneously
charged wall--- exactly given by $\rho(0) = 2 \pi \ell_B \sigma_s^2$,
or, in rescaled units, 
\begin{equation} \label{condit2}
\tilde{\rho}(0) = 1
\end{equation}
(incidentally in agreement with the Poisson-Boltzmann 
prediction)\cite{Henderson2,Carnie2,Wenner}.
The contact value theorem Eq.(\ref{condit2})
 follows from the requirement of vanishing
net force acting on the wall, which means that the osmotic pressure,
in units of $k_BT$
given by the counterion density at the wall, $P_{os}= \rho(0)$, has
to cancel the electrostatic attractive force between wall and 
counterion layer, which is given by $P_{el}=-2 \pi \ell_B \sigma_s^2$,
i.e. $P_{os}+P_{el}=0$, from which Eq. (\ref{condit2}) directly follows.
Given the two constraints on the rescaled density profile, 
namely being normalized to unity and reaching
a contact density of unity at the wall, 
Equations (\ref{condit1}) and (\ref{condit2}), it is clear that the profiles 
in the units chosen by us have to be quite similar to each other even for 
vastly different coupling parameters, as indeed observed in
Figure \ref{fig2}b. Also, it is a rather trivial consequence
of both constraints that the typical decay length of the profiles
is always given by unity in rescaled units
(though, strictly speaking, the first moment $\langle \tilde{z}\rangle $
of the density distribution diverges logarithmically  within PB theory).
Still, the asymptotic predictions for vanishing coupling constant
($\Xi \rightarrow 0$, PB theory, solid line in Figure \ref{fig2}b) 
and diverging coupling 
constant ($\Xi \rightarrow \infty$, SC theory, broken line)
are as different as they can be from a functional point of view,
while still obeying the constraints mentioned above,
as we will now recapitulate.

At low coupling, the counterion density distribution is well 
described by the Poisson-Boltzmann (PB) theory, which predicts an
algebraically decaying profile\cite{gouy,chapman,andelman:review}
\begin{equation}
  \label{PBone}
\tilde{\rho}_{\rm PB}(\tilde{z}) =
\frac{1}{\left( 1+\tilde{z} \right)^2},
\end{equation}
while in the opposite limit of high coupling the strong coupling (SC) theory,
predicting an exponentially decaying profile\cite{moreiraepl,Netz4}
\begin{equation}
  \label{SCone}
\tilde{\rho}_{\rm SC}(\tilde{z}) 
= \exp \bigr(- \tilde{z} \bigr),
\end{equation}
becomes asymptotically exact.
An exponential density profile (although with a different pre-factor)
has also been obtained by Shklovskii\cite{Shklovskii} using a heuristic model
for a highly charged surface, where counterions bound to the wall are in chemical
equilibrium with free counterions. 
The intuitive explanation for the exponential density profile
Equation (\ref{SCone})
uses the fact that for large values of the coupling constant, the lateral 
distance between counterions is large and therefore
a counterion mostly interacts with the charged plate and 
experiences the bare linear wall potential, the second
term in Equation (\ref{H-one-mu}), with only small corrections
due to other ions. 
The single-ion distribution function follows by exponentiating
the linear wall potential, similar to the derivation of the barometric
height formula for the atmospheric density, and in agreement
with the result in Equation (\ref{SCone}).
It is important to note, though, that Equation (\ref{SCone})
 has been obtained as the leading term in a systematic field-theoretic 
derivation which also gives correction terms\cite{Netz4} which in turn
have been favorably compared with simulation results\cite{Andrelong}. 
As can be seen from Figure \ref{fig2}b, 
the PB density profile Equation (\ref{PBone}) 
is only realized for $\Xi <1$, while  the strong-coupling profile
Equation (\ref{SCone}) is indeed the asymptotic solution and
agrees with simulation results for $\Xi > 10^4$ over the distance
range considered in the simulations. 
In fact, there is a crossover between
the two asymptotic theories which is distance dependent\cite{Netz4,Andrelong}, 
as we will briefly discuss now.

In the strong coupling limit $\Xi > 1$ an expansion of all
observables in inverse powers of $\Xi$ can be set up
that has much in common with a virial expansion\cite{Netz4,Andrelong}. 
The density distribution can thus be written as
\begin{equation} \label{expand2}
\tilde{\rho}(\tilde{z}) = \tilde{\rho}_{SC}(\tilde{z})
+ \frac{1}{\Xi} \tilde{\rho}_{SC}^{(1)}(\tilde{z}) +{\cal O}(\Xi^{-2})
\end{equation}
with the leading correction to the asymptotic strong-coupling profile given by\cite{Netz4}
\begin{equation} \label{strong7}
\tilde{\rho}_{SC}^{(1)}(\tilde{z}) =
{\rm e}^{- \tilde{z}} \left( \frac{\tilde{z}^2}{2} - \tilde{z} \right).
\end{equation}
A systematic estimate of  the limits of accuracy of the asymptotic SC theory is
furnished by comparing the leading and next-leading contributions,
Eqs.(\ref{SCone}) and (\ref{strong7}), which enter the 
systematic SC-expansion of the counter-ion density Eq.(\ref{expand2}).
This limit of applicability turns out to be distance-dependent.
For large separations $\tilde{z} \gg 1$ the SC theory should be valid for 
\begin{equation} \label{SCvalid2}
\Xi > \tilde{z}^2.
\end{equation}
Using the relation between the lateral distance between 
counter-ions, $\tilde{a}_\perp$,
and the coupling parameter, Eq.(\ref{aperp}),
the latter threshold can be transformed into $\tilde{a}_\perp > 
\tilde{z}$ or $a_\perp > z$. This means that the SC approach should
be valid as long as one considers distances from the wall, $z$,  smaller
than the average lateral distance between counter-ions, $a_\perp$. 
This is in accord with the intuitive expectation since the bare 
wall potential prevails for these distances.

In the small-coupling regime, $\Xi<1$, a similar expansion can
be performed using the field-theoretic tool of a loop-expansion\cite{NetzO,andrebook,Netz4}.
We  obtain for the density profile the expansion in powers of the coupling parameter
\begin{equation} \label{dens9}
\tilde{\rho}(\tilde{\bf r})  = \tilde{\rho}_{PB}(\tilde{\bf r}) +
\Xi \tilde{\rho}_{PB}^{(1)} (\tilde{\bf r}) +{\cal O}(\Xi^2).
\end{equation}
This shows directly that the saddle-point (or mean-field) 
method, which yields the first (leading) term, is good 
when the coupling parameter $\Xi$ is small. For large values
of $\Xi$, higher-order terms become important.
For large separations from the wall,
the asymptotic behavior has been determined explicitly 
as\cite{NetzO}
\begin{equation} \label{hasym2}
\tilde{\rho}_{PB}^{(1)} (\tilde{z}) \simeq - \frac{\ln \tilde{z}}{\tilde{z}^3}.
\end{equation}
The correction in Eq.(\ref{hasym2}) decays faster than the leading term in
Eq.(\ref{PBone}). By comparing the two expressions, one obtains that 
for large  separations from the plate, $\tilde{z} \gg 1$,
the PB prediction for the  density, Eq.(\ref{PBone}),
should be valid for coupling parameters
\begin{equation} \label{PBvalid2}
\Xi < \frac{\tilde{z}}{ \ln(\tilde{z})}.
\end{equation}
This shows that it does not make sense to talk about the accuracy of the PB or SC approach
per se for a given coupling parameter $\Xi$. Rather, from Eq.(\ref{PBvalid2})
it is seen that the PB solution becomes
more accurate as one moves further away from the plate. 
Conversely,  from Eq.(\ref{SCvalid2})  the SC solution becomes
more accurate as one moves closer to the plate.
By comparing Eqs. (\ref{SCvalid2}) and (\ref{PBvalid2}) one realizes that for
large distances from the wall (or for large coupling strengths), 
a  gap appears  over the distance range
\begin{equation} \label{gap}
\sqrt{\Xi} < \tilde{z} < \Xi
\end{equation}
where neither of the asymptotic 
theories is applicable. This gap widens as the coupling strength increases and can 
be interpreted as a distance range where the density distribution is neither 
described by the SC result $\tilde{\rho} \simeq e^{-\tilde{z}}$,
see Eq.(\ref{SCone}), nor the PB result, Eq.(\ref{PBone}), which for large separations
reads $\tilde{\rho} \simeq \tilde{z}^{-2}$. That an intermediate scaling range
has to exist already follows from the fact that the asymptotic  density profiles cross only 
once at a rescaled distance from the plate of the order of unity. In order to connect 
the SC  and PB profiles continuously at much larger distances, 
one needs an intermediate distance range
where the density decays slower than the inverse square with distance.
Some ideas on how to understand and analytically describe this intermediate
regime have been brought forward in Refs.\cite{Shklovskii,Netz4}
In a  number of recently published papers counterion density profiles
were calculated for intermediate coupling parameter 
using various approximate theories and successfully
compared with numerical data\cite{yoram2,pellicer,santangelo2}.

In summary, the strong-coupling theory is a theory that becomes
asymptotically exact in the opposite limit when the mean-field 
or Poisson-Boltzmann theory is valid. The two theories therefore
describe the two extreme situations, as can be seen most clearly in
Figure \ref{fig2}.
Experimentally, a coupling parameter $\Xi= 100$, which is already
quite close to the strong-coupling limit,
is reached with divalent ions for a surface charged density
$\sigma_s \approx 3.9 nm^{-2}$, which is feasible with compressed
charged monolayers, and with trivalent counter ions for 
$\sigma_s \approx 1.2 nm^{-2}$, which is a typical value.
The strong-coupling limit is therefore experimentally accessible
and not only interesting from a fundamental point of view.

\subsection{Charged plate in the presence of salt}

\begin{figure}[t]
\begin{center}
\resizebox{6cm}{!}{\includegraphics{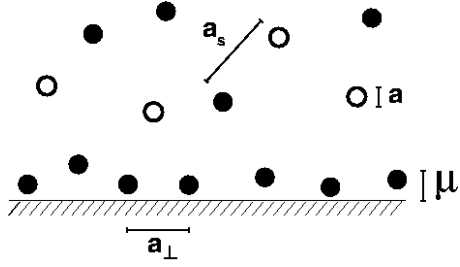}}
\end{center}
\caption{ Schematic view of the relevant
length scales for a charged wall in the presence of salt.
The lateral distance between counterions in a
neutralizing layer at the surface is denoted  by $a_\perp$.
 The Gouy-Chapman length $\mu$ is the height 
 of the counterion layer, and $a_s$ is the 
 distance between salt ions in the bulk. 
 Finally, the ion diameter is denoted by $a$. 
 In the picture we chose $a_\perp > \mu$, equivalent 
 to $ \Xi > 1$, 
 meaning that we have a strong-coupling situation. We also chose
  $a_s > a_\perp$, which together with 
 $\Xi > 1$ means that the screening length $\kappa^{-1}$
 satisfies $\kappa^{-1} > \mu $ and thus the counterion height
 is indeed given by the Gouy Chapman length.
}
\label{schema2}\end{figure}

The case of counterions at a wall is particularly simple, since the two length scales in
the problem, namely the Gouy-Chapman length, $\mu$,
 and the mean-lateral distance between charges,
$a_\perp$, can be combined into a single parameter according to $\Xi \sim (a_\perp / \mu)^2$.
Experimentally, one is always dealing with aqueous solutions at finite salt concentration 
(and if it was only for ions due to the auto-dissociation of water, which 
gives rise to an ionic  concentration of at least $10^{-7}$ mol/l and thus to a screening length
of the order of a micrometer), so we have to have a look at how our arguments in 
the preceding section are modified in the presence of salt. 
Salt adds an additional length scale, namely the mean distance between salt ions in the bulk,
see Figure \ref{schema2}, which we denote by $a_s$ and which is related to 
the salt concentration $c_s$ via $c_s \sim a_s^{-3}$. 
In principle, if the bulk contains oppositely charged ions, one also needs to give the ions
a finite diameter $a$ to prevent them from collapsing into each other; however, in order to 
concentrate on the essentials, we will largely neglect the finite ion diameter in this Section. 
Thus we confine ourselves to three length scales, $\mu$, $a_\perp$, and $a_s$, that
can be combined into two unitless parameters which fully define the problem. 
The actual physics, however, is quite rich, since from the three geometric length
scales we define in Fig. \ref{schema2}, 
one can derive  two additional  length scales which play an important role,
namely the screening length $\kappa^{-1}$ defined by $\kappa^2 = 8 \pi \ell_B q^2 c_s$,
and the length at which two ions interact with thermal energy, $q^2 \ell_B$. 

Within mean-field, i.e. the Poisson-Boltzmann theory\cite{gouy,chapman,andelman:review},
the electrostatic potential $\psi(z)$ at a charged wall decays as
$q e \psi(z) / k_BT = 2 \ln [(1-\gamma e^{-z \kappa})/(1+\gamma e^{-z \kappa})]$.
The counter and coion density distributions at a charged wall follow within mean-field as 
\begin{equation} \label{PBsalt}
\rho^+_{PB}=c_s^2/ \rho^-_{PB} =c_s e^{- q e \psi(z) / k_BT} = c_s
 \left( \frac{1+\gamma e^{-z \kappa } }{ 1-\gamma e^{-z \kappa}}\right)^2
 \end{equation}
where the constant $\gamma$ is determined by the equation
\begin{equation} \label{gameq}
2 \gamma /(1-\gamma^2) = 1/(\kappa \mu).
\end{equation}
It is seen that the screening length gives the scale over which 
the ionic charge distribution decays towards the bulk value 
as one moves far away from the charged wall; in other words,
the screening length is the correlation length of the salt solution\footnote{The potential
and the ion densities are also related by the Poisson equation according
to $\psi''(z) = - q e [\rho^+_{PB}(z) - \rho^-_{PB} (z)]/\varepsilon \varepsilon_0$.}.

In the following we will discuss various crossover boundaries for the 
system under investigation, which will eventually  be summed up in a scaling 
diagram.

i) In the Debye-H\"uckel (DH) limit defined by 
\begin{equation} \label{rel1}
\kappa^{-1} < \mu   
\end{equation}
the screening length is smaller than the Gouy Chapman length;
the charged surface perturbs the ionic densities only slightly, the 
mean-field equations can be linearized and 
the linear superposition principle for densities and potentials is valid.
Eq.(\ref{gameq}) is solved by $\gamma \simeq 1/(2 \mu \kappa)$ and
the potential is $q e \psi(z) / k_BT  \simeq 2 e^{-z \kappa} / (\mu \kappa)$
and the ion densities follow as  
$\rho^{\pm}_{PB}  = c_s (1 \pm 2 e^{-\kappa z} / (\mu \kappa))$.
When inequality Eq.(\ref{rel1}) is not satisified, i.e. when the DH approximation
is not valid, the algebraic density profile Eq.(\ref{PBone}) is realized
for the counterions
at distances smaller than the screening length.

ii) If the interaction between salt ions at their mean separation $a_s$
is larger than thermal energy, we have a strongly coupled salt solution
and mean-field theory breaks down, even in the bulk and in the absence
of a charged surface\footnote{This defines the realm of large plasma parameters
and where an electrolyte solution exhibits a critical condensation 
transition\cite{Baus,Levincrit,Netz-1}. Experimentally, such a transition is
reached with organic solvents.}.
This condition reads $\ell_B q^2 / a_s > 1$ 
and can be reexpressed as 
\begin{equation}    \label{rel2}
\Xi > ( \kappa \mu  )^{-1}.
\end{equation}
In practice, an effective mean-field theory can be defined where the 
screening length is renormalized from its bare value\cite{DHrenorm}. 
Such a  modified DH theory with renormalized screening length we denote
by DH$^*$.
Since the intermediate distance range, where the counterion density
profile is neither described by SC nor PB,  is given by
$\sqrt{\Xi}< \tilde{z} < \Xi$,  Eq.(\ref{gap}),
it follows that when Eq.(\ref{rel2}) holds,
the counterion density profile at large  distances  $ \tilde{z} > \Xi$
can be described by a linear DH* theory since the non-linear
PB regime is preempted by the intermediate regime where neither
SC nor PB works.

iii)  When the screening length becomes smaller than $a_\perp$,
we expect the intermediate distance range, which 
is expected for the range ${\tilde a}_\perp \sim \sqrt{\Xi}< \tilde{z} < \Xi$,  to disappear.
The condition $\kappa^{-1} < a_\perp$ is equivalent to 
\begin{equation}    \label{rel3}
\Xi > ( \kappa \mu  )^{-2}.
\end{equation}

\begin{figure}[t]
\begin{center}
\resizebox{6cm}{!}{\includegraphics{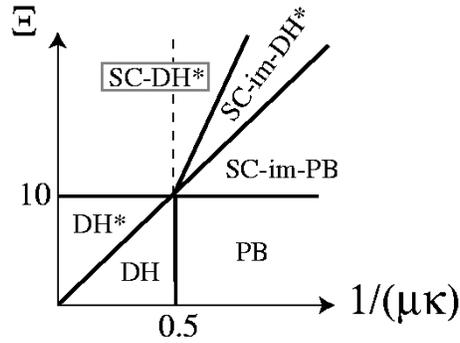}}
\end{center}
\caption{ 
Scaling diagram for the behavior of a salt solution at
a charged surface as a function of coupling parameter
$\Xi$ and ratio of screening length $\kappa^{-1}$ and
Gouy-Chapman length $\mu$. Axes are logarithmic,
meaning that power laws are straight lines.
Effects of bulk correlations 
between salt ions and correlations between counterions at
the surface are included.
The various phases and scaling boundaries are
explained in the text. The crossover between
PB and DH is located at  $1/ \mu \kappa  \approx 1/2$,
since there the electrostatic potential at the surface is roughly 
unity in terms of the thermal energy,
i.e., $q e \psi /k_BT \approx 1$; 
 for monovalent ions this corresponds to $\psi = 25 mV$.
}
\label{schema3}\end{figure}

All three scaling boundarie Eqs.(\ref{rel1}- \ref{rel3}) 
are represented in Fig. \ref{schema3}, where we chose as axes 
the coupling parameter $\Xi$ and the ratio of screening length and 
Gouy-Chapman length, $\kappa^{-1} / \mu$.  The horizontal
line in addition denotes the boundary between weak coupling 
and strong coupling regimes, which roughly occurs at
$\Xi  \approx 10$, see Fig. \ref{fig2}b. In the scaling 
regime 'PB' the ordinary Poisson-Boltzmann theory
is valid and the ion densities are correctly described by Eq.(\ref{PBsalt}).
In the Debye-H\"uckel regime denoted by 'DH',
the linearized version of PB is sufficient.
In the phase 'DH*' the salt is strongly coupled, and ion pairs proliferate. 
This can be taken care of by a renormalized screening length.
Now we move to the phases for strong coupling constant $\Xi > 1$,
where things are more interesting but also less certain.
In the phase 'SC-im-PB'  the ion density profile exhibits three different
scaling ranges: for $\tilde{z} < \sqrt{\Xi}$ the strong-coupling profile
is realized,  $\sqrt{\Xi} < \tilde{z} < \Xi $ defines  the intermediate range
(where predictions based on a Gaussian theory have been advanced
in Ref.\cite{Netz4}), and for $\Xi < \tilde{z}$ the Poisson Boltzmann profile
is valid (note that the PB profile itself is subdivided into a nonlinear
range $\Xi < \tilde{z} < 1/(\kappa \mu)$ and a linear DH range
$ 1/(\kappa \mu) <  \tilde{z} $). In the 'SC-im-DH*' phase
the non-linear PB range has disappeared, and finally, in the 
'SC-DH*'  phase the intermediate range has been swallowed up by the
DH* scaling range. The SC-DH* phase is curious, since the 
counterion density profile is expected to show a crossover between
two  exponential decays governed by two different decay lengths,
namely the Gouy-Chapman length (for small distances) and the
screening length (for large distances). It is itself subdivided  by a broken line
into  two subregimes. The right regime is more interesting, since here the 
charged wall induces counterion concentrations much higher than the 
bulk concentration and thus a quite visible effect (as will be shown
shortly in simulation data). 
The crossover
between the two exponential decays, however,  will be hard to observe in practice.

\begin{figure}[t]
\begin{center}
\resizebox{12cm}{!}{\includegraphics{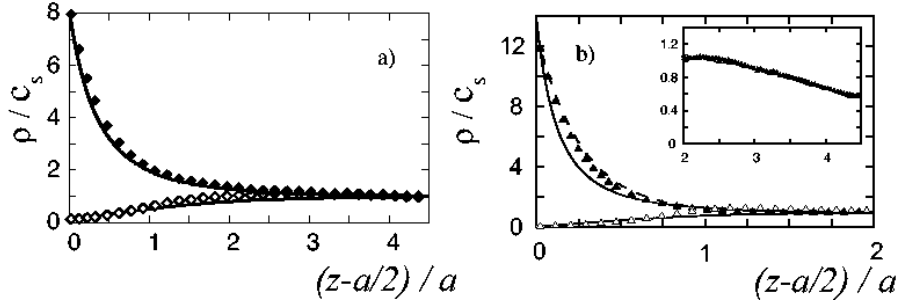}}
\end{center}
\caption{ 
Counterion (filled symbols) and coion (open symbols) density
profiles obtained within simulations as a function of the distance from
the charged wall divided by the ion diameter, $(z-a/2)/a$.
a) PB regime with weak coupling $\Xi =2.3$ where PB profiles, Eqs.(\ref{PBsalt}),
 solid lines, are accurate.
The  Coulomb interaction between two ions at contact is $\zeta = q^2 \ell_B / a = 1.75$
and the rescaled screening length is $\kappa^{-1}/ \mu = 1.26$.  The Gouy-Chapman 
length in units of the ion diameter  is $\mu/a = 0.758$.
b) Strong coupling regime 'SC-im-DH*'  with $\Xi =18.9$ where PB (solid lines) breaks down.
Coulomb strength is $q^2 \ell_B / a = 5$ and the rescaled screening length is 
 $\kappa^{-1}/ \mu = 1.71$ and $\mu/a = 0.265$. The
 broken line is the salt-modified strong-coupling profile for the counterions, Eq.(\ref{SCsalt}).
 The inset shows a gradual density depression at the uncharged upper system boundary,
 which is caused by correlation effects (see discussion in text).
 }
\label{schema4}\end{figure}

In Fig.\ref{schema4} we show counterion and coion density profiles at a charged
wall as obtained in Brownian dynamics simulations\cite{Woon}. 
The profile in Fig.\ref{schema4}a
is obtained for a coupling parameter $\Xi=2.3$ and a rescaled screening length
$\kappa^{-1}/ \mu = 1.26$.  According to our scaling arguments advanced above,
this system belongs to the Poisson-Boltzmann regime, and indeed the PB profiles
Eq.(\ref{PBsalt}), solid lines, match the simulation results very nicely.
The data in  Fig.\ref{schema4}b
are obtained for  $\Xi=18.9$ and $\kappa^{-1}/ \mu = 1.71$.  Since the crossover
in $\Xi$ occurs for $\Xi \approx 10$, the system belongs to the SC
regime and indeed the PB prediction (solid lines) performs poorly. 
In order to compare the data with  the strong-coupling profile, which was derived 
in the counterion-only-case, we have to use additional information.
First of all, the counterion profile saturates at a finite value far away from 
the surface which is given by the bulk salt concentration. Secondly, the 
ion density at the wall still obeys the contact-value theorem, which is slightly
modified in the presence of salt: The net pressure acting on the wall
is not zero, as with counterions only, compare Eq. (\ref{condit2}), 
but equals the bulk osmotic pressure $P_{bulk}$.
In the limit of a weakly coupled salt solution (i.e. for a small bulk-plasma parameter
or for $\Xi < 1/\mu \kappa$), the bulk osmotic pressure
is that of an ideal gas, $P_{bulk}  = 2 c_s$. 
Neglecting also correlation effects at the surface, which are similar to the
 Onsager-Samarras effect\cite{Samarras}, 
the pressure acting
on the surface equals the sum of the surface osmotic pressure,
$P_{os}=\rho^+(0)+\rho^-(0)$, proportional to the surface ion densities, and the
electrostatic double layer attraction
attraction $P_{el}= - 2 \pi \ell_B \sigma_s^2$, compare our discussion after Eq. (\ref{condit2}).
Equating surface and bulk pressures,  $P_{bulk}=P_{os}+ P_{el}$,
we obtain $2 c_s =\rho^+(0)+ \rho^-(0)  - 2 \pi \ell_B \sigma_s^2$.
Using that for a highly charged surface the
coion surface density $\rho^-(0)$ vanishes, we obtain
$\rho^+(0) \approx 2 c_s + 2 \pi \ell_B \sigma_s^2$. 
The simplest functional satisfying the surface and the bulk constraints,
and which decays according to the SC prediction Eq.(\ref{SCone}), is
\begin{equation} \label{SCsalt}
\frac{ \rho_{SC}^+(\tilde{z})}{c_s} = 1 + (1 + 4 \mu^{-2} \kappa^{-2} ) e^{-\tilde{z}}
 \end{equation}
 which is shown in Fig.\ref{schema4}b as a broken line and
 describes the data quite well. The coion distribution is quite featureless
close to the wall and equally well described by the PB or by a modified SC expression.

A pronounced density depression of both coions and counterions
is seen in the inset of Fig.\ref{schema4}b
at the uncharged boundary surface located at $z/a=5$. This is analogous  to 
the aforementioned Onsager-Samarras effect according to which the ions
in an electrolyte solution
are repelled from a low-dielectric substrate\cite{Samarras,Levintension}. 
In the present case the 
dielectric constant is uniform, but still the ions are repelled from
the bounding surface since the effective polarizability of the salt solution
is higher than that of the half-space devoid of ions\cite{Netz1}. 

After having discussed the counterion distribution at a single charged
wall, it is now time to go on to the experimentally relevant case
of two charged walls.

\subsection{Counterions between  two charged plates}

A great deal of work has been devoted in the past twenty years to 
understanding the interaction between two double layers. Specifically,
it has been known for some time that two similarly and strongly charged
plates can attract each other in the presence of multivalent counterions
or even with monovalent counterions when the surface charge density
is extremely high. 
This has been seen in Monte Carlo 
simulations\cite{guldbrand-al:84,bratko-al:86}, observed 
experimentally with the surface force apparatus\cite{kekicheff-al:93} 
and also deduced from the phase diagrams of charged lamellar
systems\cite{Khan,Khan2,Dubois}, as has been discussed in Section 3.1.
A similar attraction is theoretically predicted for highly charged
cylinders\cite{attcyl1}-\cite{attcyl13},
flexible polymers\cite{Molnar} and spheres as well\cite{attsph1}-\cite{attsph11}
and is thus by no means confined to the planar geometry.
Experimentally, a long-ranged
attraction has also been seen for charged spherical colloids
confined by walls\cite{att1,att2,att3,att4},
although it has been shown in the mean time that 
for some setups the effect can be  caused by hydrodynamic artifacts.
For other setups the long-ranged interaction persists.
It was very recently argued that  optical artifacts caused by the 
imaging process can lead to minute distortions in the particle distances as obtained by 
digital video microscopy. Those distortions in turn result in an apparent minimum in 
the interaction energy\cite{att5}.
The general occurrence of 
like-charge attraction is quite relevant concerning the stability
of colloidal solutions, since it means that the stabilization 
of colloids with charges  can fail if the surfaces are too highly charged. 
Such behavior strongly contradicts 
the Poisson-Boltzmann theory, which predicts that the electrostatic 
interaction between similarly charged surfaces is always repulsive\cite{andelman:review}. 
Most theoretical approaches (beyond PB)
tried to include the correlations between counterions,
which were thought to be the reason for the discrepancy between
the mean-field and the experimental/simulation results and which are
neglected on the mean-field level\cite{kjellander:96,attardrev}. 
The first theoretical approach that demonstrated the existence of attraction between
equally charged plates (with electrostatic origin) is due to 
Kjellander and Mar{\v{c}}elja\cite{kjellander-marcelja:84},
who used a sophisticated integral-equation
theory (with HNC closure) and obtained results that compared very well with 
simulations\cite{guldbrand-al:84,kjellander-marcelja:84,kjellander-al:92}. 
Also perturbative expansions
around the PB solution\cite{attard-mitchell-ninham:88,podgornik:90,NetzO} and 
density-functional 
theory\cite{stevens-robbins:90,diehl-al:99} were used, and predicted as well
the existence of an attractive interaction. For plates far away from each other, 
i.e., at distances such that the two double layers weakly overlap, 
the attractive force was obtained within the approximation of 
two-dimensional counterion layers  by including
in-plane Gaussian 
fluctuations\cite{Attard3,pincus-safran:98,lukatsky-safran:99,lau02} and, 
more recently, plasmon 
fluctuations at zero temperature\cite{lau-levine-pincus:00}
and at non-zero temperatures\cite{levine}.
Fluctuation-induced interactions between macroscopic objects
constitute a quite general phenomenon, which is present whenever 
objects couple to a fluctuating background field\cite{golestanian},
giving rise to a wide range of interesting phenomena including
colloidal flocculation in binary mixtures\cite{Netz0}.

The rescaled pressure $\tilde{P}$ between two 
plates in the presence of counterions only
is given by the contact value theorem
\begin{equation}
  \label{cvt}
\tilde{P} = 
\frac{P}{2 \pi \ell_B \sigma_s^2} = 
\frac{\rho(0)}{2 \pi \ell_B \sigma_s^2} - 1,
\end{equation}
which relates the pressure in units of $k_BT$, $P$,  acting on one wall
to the counterion density at that wall, $\rho(0)$ (which in a simulation
can be extracted via a suitable extrapolation scheme). 
As has been discussed before,
the first term on the right-hand side 
is the osmotic pressure due to counterion
confinement, the second term is the double layer attraction
between the counter-ions and the charged plates.
This theorem can be formulated in different ways and is 
exact\cite{Henderson2,Carnie2,Wenner}. Clearly, the pressure depends
on the rescaled distance $\tilde{d}=d/\mu$ between the two walls.

The mean-field (Poisson-Boltzmann) prediction for the pressure follows from
equation (\ref{cvt}) as
\begin{equation} \label{PBpress}
\tilde{P}_{\rm PB}(\tilde{d}) = \Lambda
\end{equation}
where $\Lambda$ is determined  by the transcendental equation\cite{Engstrom}
\begin{equation}
  \label{trans}
  \frac{1}{\Lambda^{1/2}} = \tan \biggl( \frac{\tilde{d}}{2} \Lambda^{1/2} \biggr)
\end{equation}
which is solved by 
\begin{equation}
\label{P0}
\tilde{P}_{\rm PB}(\tilde{d}) =
     \left\{
\begin{array}{llll}
     & 2 \tilde{d}^{-1} -1/3  & {\rm for} &  \tilde{d} \ll 1  \\
     & \pi^2 \tilde{d}^{-2} & {\rm for} &  \tilde{d} \gg 1  .\\
\end{array} \right.
\end{equation}
As is well-known, the PB pressure is always repulsive\cite{andelman:review}.

\begin{figure}[t]
\begin{center}
\resizebox{12cm}{!}{
\includegraphics{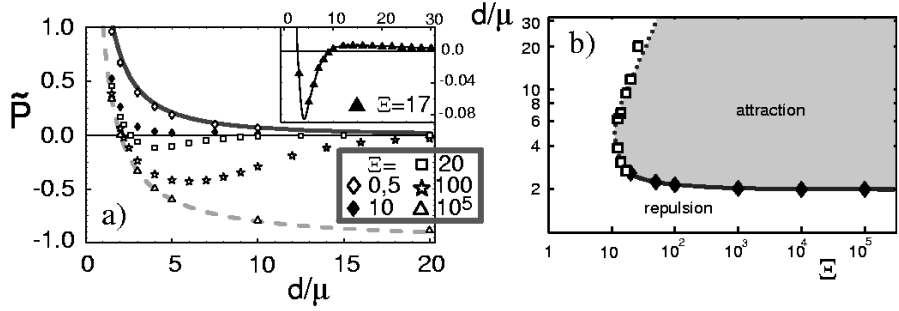}
}
\end{center}
\caption{a) Simulation results for the rescaled pressure 
$\tilde{P} \equiv P / 2 \pi \ell_B \sigma_s^2$ 
as a function of the plate separation
$\tilde{d} = d/\mu$ for different values of the coupling parameter $\Xi$.
The solid line denotes the PB prediction Equation~(\ref{PBpress}) and the broken
line the SC prediction Equation~(\ref{SCpress}).
The error bars are smaller than the symbols. Note that for large enough distances,
all MC data exhibit positive pressures.
b) The global behavior of the inter-wall pressure
as obtained from the simulations, showing the region where the
pressure between two charged walls is negative (attractive)
and where the pressure is positive (repulsive).
The dividing line denotes a line of vanishing pressure.
The filled diamonds (and full line) 
denote the thermodynamically stable distance between the plates.
The open
squares correspond to a metastable local
minimum (lower branch) and a maximum (upper branch) in the
free energy.
For couplings $\Xi> \Xi_n \approx 12$, there is a range in $\tilde{d}$ where the 
pressure is negative (attraction). At $\Xi = \Xi_u \approx 17$ a first-order 
unbinding transition occurs (as follows from the Maxwell construction).
}
\label{fig3}
\end{figure}

Within the strong-coupling theory,
the leading result for the pressure is\cite{moreiraprl,Netz4}
\begin{equation}
  \label{SCpress}
    \tilde{P}_{\rm SC}(\tilde{d})= \frac{2}{\tilde{d}}-1
\end{equation}
and will be derived using simple arguments below.
While the PB theory predicts that the pressure is always positive (only repulsion), 
the SC theory
gives attraction between the plates for $\tilde{d} > 2$ (negative pressure)
and thus predicts a bound state (free energy minimum) at a distance
$\tilde{d}^* =2$. In analogy to the strong-coupling result 
for the counterion density
profile at a single charged wall, and
as is explained in detail in Ref.\cite{Netz4},
the leading term of the SC expansion for the pressure, equation 
(\ref{SCpress}), is the first virial
term and thus follows from the partition function of
a single  counterion sandwiched between two charged plates. 

Since the lateral distance between two counterions is of the order of 
$\tilde{a}_\perp \simeq \sqrt{\Xi}$ in rescaled coordinates, see Eq.(\ref{aperp}),
and since we expect the SC theory to be a good approximation as long as
the lateral distance between counterions is larger than the plate distance,
i.e. for $\tilde{a}_\perp > \tilde{d}$,
the SC result should be valid for 
\begin{equation} \label{SCvalid}
\Xi \gg \tilde{d}^2
\end{equation}
(this argument can be substantiated by a Ginzburg argument based on comparing
different orders in the SC perturbation expansion\cite{Netz4}).
The SC theory at the same time predicts a bound state
at a rescaled plate separation $\tilde{d}^*=2$.
This prediction for the bound state is thus within the domain of validity 
of the SC theory for coupling constants $\Xi > 4$.
One could therefore argue
that the mechanism of the attraction between similarly charged bodies
is contained in the SC theory. To gain intuitive insight into this
mechanism, we reconsider the   partition function of
a single  counterion sandwiched between two charged plates
which we now explicitly evaluate.
Denoting the distance between the counterion and the
plates (of area $A$) as $x$ and $d-x$,
respectively, we obtain for the electrostatic interaction
between the ion and the plates
(note that all energies and forces are given in units of $k_BT$)
for $d \ll \sqrt{A}$  the results
$W_1=2 \pi \ell_B q \sigma_s x $ and
$W_2=2 \pi \ell_B q \sigma_s (d- x) $, respectively,
as follows from the potential of an infinite charged wall
and omitting constant terms.
The sum of the two interactions is
$W_{1+2}=W_1+W_2 = 2 \pi \ell_B q \sigma_s d $, which shows that
i) no pressure is acting on the counter-ion since the forces
exerted by the two plates exactly cancel and ii) that
the counter-ion  mediates an effective attraction between the two
plates. The interaction between the two plates is
proportional to the total charge on one plate, $A \sigma_s$, and
for $d \ll \sqrt{A}$ given by
$W_{12} = -2 \pi A \ell_B \sigma_s^2 d$.
Since the system
is electro-neutral, $q = 2 A \sigma_s$, the total energy is
$W= W_{12}+W_1+W_2 = 2 \pi A \ell_B \sigma_s^2 d$, leading
to an electrostatic pressure $P_{el} = -\partial (W/A)/\partial d
=-2\pi \ell_B \sigma_s^2$ per unit area.
{\em The two plates attract each other!}
The osmotic pressure due to counter-ion confinement
is $P_{os} = 1/ A d = 2\sigma_s/q d$.
The total pressure is given by the sum $P_{\rm SC} =
P_{el}+P_{os}$ and reads in rescaled units
$\tilde{P}_{\rm SC} = 2/\tilde{d}-1$ and thus
agrees exactly with the result in Equation (\ref{SCpress}).
The equilibrium plate separation is characterized by zero total
pressure, $P_{\rm SC}=P_{el}+P_{os}=0$,
leading to an equilibrium plate separation
$d^* = 1/\pi \ell_B q \sigma_s$, or, in rescaled units,
$\tilde{d}^* =2$.

We collect the simulation results, as well as the 
asymptotic strong-coupling and Poisson-Boltzmann predictions 
in Figure~\ref{fig3}a,
where the pressure as a function of the distance between the charged 
walls is plotted for different values of the coupling. 
For a small coupling $\Xi=0.5$, PB (solid line), Equation (\ref{PBpress}),
describes very well the MC results, while at very high coupling
($\Xi=10^5$) the SC theory (broken line), Equation (\ref{SCpress}),
gives the correct prediction. Intermediate values of the coupling
lead to values of the pressure that are bounded by the two asymptotic
predictions, similarly to our findings for the single charged wall in 
the preceding section.

\begin{figure}[t]
\begin{center}
\resizebox{12cm}{!}{
\includegraphics{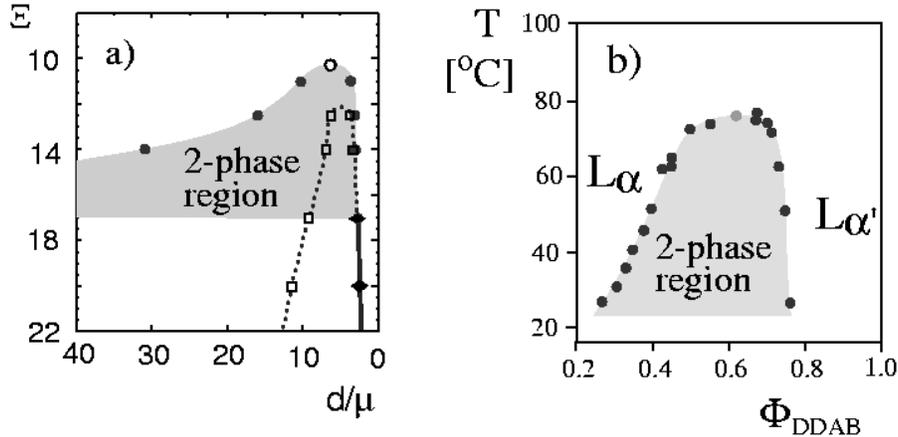}
}
\end{center}
\caption{a) The same theoretical phase diagram as in
Figure~\ref{fig3}b, but enlarged around the 
region where the first-order unbinding transition occurs. The circles
denote the binodal, determined according to the Maxwell
construction, the open squares and filled diamonds are the points where the pressure
is zero, corresponding to extrema of  the free energy. At a coupling
$\Xi=\Xi_u \approx 17$ a discontinuous unbinding transition occurs as one
comes from higher $\Xi$ (correspondingly, one branch of the binodal moves to 
infinity as one comes from lower $\Xi$).
Notice that a critical
point is present at $\Xi=\Xi_c \approx 10.25$ (denoted 
by an open circle).  The pressure is strictly  positive for $\Xi < \Xi_n \approx 12$.
The full and broken lines are guides to the eye.
b) Experimentally determined binodal for a two component mixture
of the cationic surfactant DDAB and water (reproduced after \cite{Zemb3}).
}
\label{fig4}
\end{figure} 

We summarize the behavior of the pressure in  the 
phase diagram Figure~\ref{fig3}b, which 
shows the region of negative (attractive) pressure, separated
from the region of positive (repulsive) pressure by a line on
which the pressure is zero.  This line can correspond to 
a thermodynamically stable, metastable, or instable state,
as will be discussed in detail now.
For couplings larger than $\Xi=\Xi_n \simeq 12$, 
there is a range of $\tilde{d}$ within which the pressure
is negative and the two plates attract each other. The boundary
between attraction and repulsion in Figure~\ref{fig3}b 
is given by the points where the pressure is zero: 
the filled diamonds 
(connected by a solid line which serves as a guide to the eyes)
correspond to thermodynamically stable bound states 
(absolute minima of the free energy at finite $\tilde{d}$), 
while the open squares 
(connected by a broken line) are local, metastable
 minima (lower branch) and maxima (upper branch) of the free energy.
At a coupling $\Xi=\Xi_u \simeq 17$ a first-order unbinding transition occurs, 
where the free energy has two minima of equal depth, 
one at finite separation $\tilde{d} \simeq 3$ and the other at  
infinite separation $\tilde{d}=\infty$. 
Below this value of the coupling the absolute minimum of the free energy is 
at infinite plate separation, 
i.e., the thermodynamically stable state of the system is the 
unbound state, above this value, the thermodynamically stable state exhibits
a finite value of the separation and is denoted  by the solid line.
We note that we determine the free energies from our data
by integration of the pressure curve from infinite distance
to a finite distance value.
In the limit of large values of $\Xi$, the lower zero-pressure branch  saturates at 
$\tilde{d} \approx 2$, in agreement with the asymptotic result of the
SC theory.

The upper branch of the zero-pressure line
can  be estimated by field-theoretic methods:
Within a loop expansion, the pressure is expanded in 
powers of the coupling parameter $\Xi$ according to
\begin{equation} \label{Ploop}
\tilde{P} = \tilde{P}_{PB} + \Xi \tilde{P}_{PB}^{(1)} + {\cal O}(\Xi^2).
\end{equation}
The zero-loop prediction for the pressure follows from PB theory
and is given in Eq.(\ref{P0}).
The  one-loop correction to the pressure
has been calculated by Attard et al.\cite{attard-mitchell-ninham:88} and 
by Podgornik\cite{podgornik:90} and is in reduced
units given by
\begin{equation}  \label{P1}
\tilde{P}_{PB}^{(1)}  (\tilde{d}) =
     \left\{
\begin{array}{llll}
     & - 2 \tilde{d}^{-1}  & {\rm for} &  \tilde{d} \ll 1  \\
     & - \pi^2 \tilde{d}^{-3} \ln \tilde{d} & {\rm for} &  \tilde{d} \gg 1  .\\
\end{array} \right. 
\end{equation}
The correction to the asymptotic PB result is attractive.
By equating the two orders for large distances $\tilde{d}$
one obtains an estimate for the zero pressure
line as
\begin{equation}  \label{PBvalidity}
\Xi \simeq \tilde{d} / \ln(\tilde{d})
\end{equation}
which agrees almost quantitatively with the numerical results in Figure~\ref{fig3}b.
However, one has to meet this result with all due suspicion
and it receives credibility only due to the good agreement with the numerics,
since the onset of attraction at the same time
signals the break-down of the loop-expansion. 

\begin{figure}[t]
\begin{center}
\resizebox{12cm}{!}{
\includegraphics{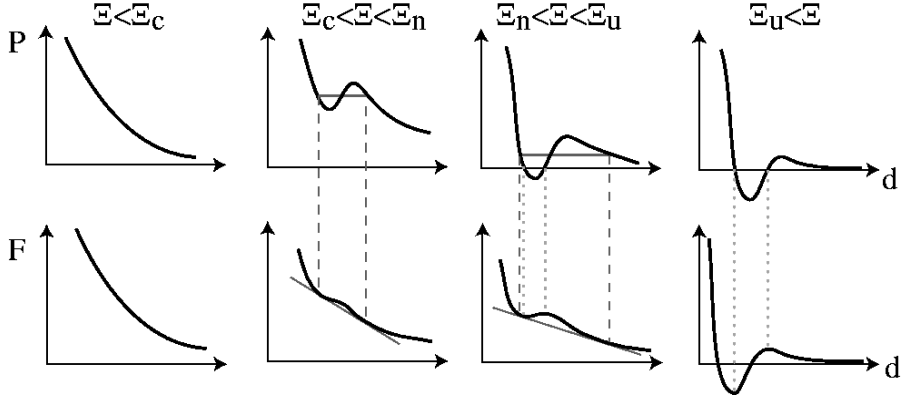}
}
\end{center}
\caption{Schematic scenario for the evolution of
the interplate pressure $P$ and the free energy $F$ as the 
coupling constant increases. The straight lines visualize
the Maxwell construction for the pressure profile (top row)
and the equivalent common-tangent construction for the
free-energy profile (bottom row). The broken lines denote
the thermodynamically coexisting state (i.e. points on the binodal)
while the dotted lines denote states with vanishing pressure.
The mechanism for the unbinding transition at a coupling constant
$\Xi_u \approx 17$ is the repulsion (positive pressure) 
which is always observed at large distances.
}
\label{fig5}
\end{figure}

Experimentally, the solid line in Figure~\ref{fig3}b
describes the distance between charged plates in the thermodynamic
ensemble when the external pressure is zero (this corresponds to the case where
a lamellar phase is in equilibrium with excess water).
If the plate-distance is controlled by some pressure acting
on the system (which is relevant to the experimental
situation where the total water content
of a lamellar phase is fixed), the system exhibits a
critical point and a
binodal where two lamellar states with different spacings
coexist thermodynamically.
This is shown in Figure~\ref{fig4}a, where in addition to
the boundary between negative and positive pressures (shown as a broken and solid
line) we also show
the binodal, which has been numerically determined
for a finite set of coupling constants (circles) and which
corresponds to the boundary of the shaded region for values
of coupling constant $\Xi < \Xi_u \approx 17$.
The binodal corresponds 
to coexisting states, which are located  through a Maxwell construction.
This is demonstrated in Figure~\ref{fig5}, where we schematically show
the free energy and the corresponding inter-plate pressure  for four
different representative values of the coupling constant $\Xi$. 
The binodal exhibits a critical point (denoted by an open circle)
 at a coupling constant
$\Xi_c \approx 10.25$ and at a plate separation $\tilde{d} \approx 6$.
For smaller coupling constants, $\Xi< \Xi_c$ the pressure is strictly positive
and decays monotonically.
In the coupling constant range $\Xi_c< \Xi < \Xi_n \approx 12$ 
the thermodynamically coexisting states can be  located using 
the Maxwell construction for the pressure profile 
(i.e. by enforcing the areas above and
below the horizontal  line in Fig.~\ref{fig5} to be the same) or for 
the free energy profile by the equivalent common-tangent construction
(see Fig.~\ref{fig5}; note that in this coupling range the 
free energy decays monotonically and the pressure is thus strictly
positive).
In the coupling constant range  $\Xi_n< \Xi < \Xi_u \approx 17$
the pressure is negative for a 
range of distances limited by the open squares in Figure~\ref{fig4}a.
It is important to note that the pressure becomes positive for 
large distances, which reflects the fact that the mean-field theory
becomes valid at large distances between the plates\cite{Netz4}.
As the coupling constant increases, the binodal branch at large distances
moves out to infinity.
For the pressure
data for $\Xi =\Xi_c =17$, which are shown in the inset in Figure~\ref{fig3}a,
the Maxwell construction leads to a coexisting state at infinite separation,
which demarks the unbinding transition. From our arguments given above, 
the unbinding transition is a quite generic feature, caused by
the fact that PB becomes valid and thus 
the pressure becomes repulsive at large separations. 
The ratio of the unbinding and the critical coupling is 
$\Xi_u/\Xi_c \approx 1.7$, leading to a temperature ratio 
of roughly $T_c / T_u \approx 1.3$.

In Figure~\ref{fig4}b we reproduce the binodal of the cationic surfactant system
DDAB (which also contains only counterions since salt has been carefully
removed from the system)\cite{Dubois}. The general shape of the experimental
binodal qualitatively agrees with the theoretical one in Figure~\ref{fig4}a.
It is interesting to note
that for this experimental system,
the critical point roughly occurs at a temperature of $T_c = 75^o C$
or $348 K$, which points (using the above estimate
$T_c / T_u \approx 1.3$)
to an unbinding transition of $T_u = 268K$ or
$-5^oC$, a little bit below freezing. The binodal in the experimental
phase diagram somewhat follows  the predicted unbinding behavior,
since the binodal branch of the dilute lamellar phase indeed moves 
progressively to the left as the temperature is decreased\cite{Zemb3,Dubois}.
The critical surface charge density for monovalent counterions and 
at room temperature follows from our estimate $\Xi_c \approx 10$
to be equivalent to  one surface charge per area 30 \AA$^2$.
The membrane charge density in the above
mentioned  experiments is between 60 \AA$^2$ and 70 \AA$^2$
and therefore differs by a factor of two.
Our comment about the ratio of the critical and unbinding 
temperatures therefore has to be taken as a rough
estimate. The deviations might be caused
by effects associated with dielectric boundaries and 
inhomogeneous surface charge distributions (which are both not included in
our simple model) and which are likely to shift the critical point to larger
values of the area per surface group. The distance between the charged
surfaces at the critical point is given by $\tilde{d}_c \approx 6$,
which for monovalent counterions is equivalent to roughly $0.6 nm$. 
We note that the finite size of the ions is not really important 
for average-size ions, since the spacing $d$ used in our simulations 
corresponds to the vertical height available for the ionic centers.
In other words, $d$ denotes the difference between the distance
between the plate surfaces and the ionic diameters. Adding an
ionic diameter of roughly $0.3 nm$ to the theoretically predicted
plate distance at criticality, one arrives at a plate separation of
roughly $0.9 nm$ which is indeed very close to what is seen experimentally.

\subsection{Wigner crystallization}

\begin{figure}[t]
\begin{center}
\resizebox{12cm}{!}{
\includegraphics{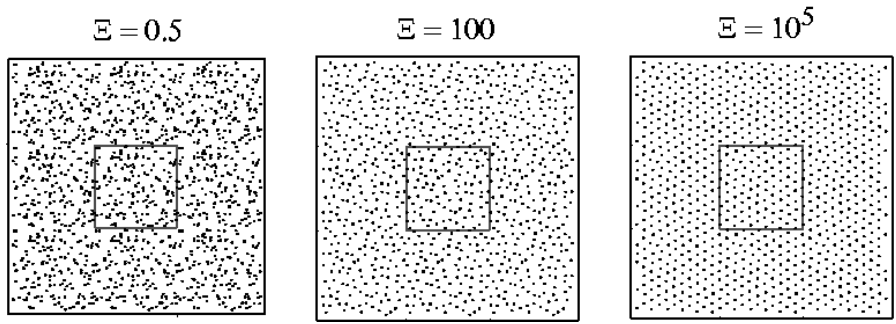}
}
\end{center}
\caption{
Top-view snapshots of counterions
 between two plates at separation $\tilde{d} =2$
for coupling parameters $\Xi=0.5$ and $N=150$ particles,
showing only weak lateral correlations, $\Xi=100$ and $N=100$, 
showing short-ranged correlations,  
and $\Xi=10^5$ and $N=100$, exhibiting crystalline order,
indicative of Wigner crystallization. The central square is the actual
simulation box, the outer square visualizes the first shell of
periodic neighbors.
}
\label{fig6}
\end{figure}

Recently, there has been an active
discussion about the significance
of Wigner crystallization for the behavior of strongly charged matter
such as the attraction between  similarly charged plates\cite{Rouzina,Shklovskii}.
A two-dimensional one-component plasma is known to crystallize for a value
of the plasma parameter $\Gamma \approx 125$\cite{Baus}. From the
definition of the two-dimensional plasma parameter\cite{Baus},
$\Gamma = \ell_B q^2 / (q/\pi \sigma_s)^{1/2}$,
we obtain the relation $\Xi = 2 \Gamma^2 $. This leads to 
a crystallization threshold (in units of our  coupling parameter) of 
$\Xi \approx 31000$.
For the system with two charged plates
the crystallization is in the limit $\tilde{d} \rightarrow 0$
predicted to occur at $\Xi \simeq 15600$.
In Figure \ref{fig6} we show top-view snapshots 
for ions sandwiched between two plates,
obtained within the Monte-Carlo
simulations for $\Xi=0.5$, $\Xi=100$ and $\Xi=10^5$ at fixed
inter-plate distance  $\tilde{d} =2$.
In agreement with the estimated Wigner crystallization threshold,
$\Xi \simeq 15600$, the snapshots for $\Xi=0.5$ and $\Xi=100$
show liquid behavior, while the snapshot for  $\Xi=10^5$
exhibits crystalline order. Since the experimentally relevant
attraction occurs for values $\Xi < 100$, it seems that Wigner
crystallization is not connected or responsible for the
attraction between similarly charged plates\cite{Linserev}.
On the other hand, treating the strongly correlated
liquid layer of counter-ions  like a Wigner crystal
is in many cases a reasonable approximation\cite{Shklovskii}.

\begin{figure}[t]
\begin{center}
\resizebox{12cm}{!}{\includegraphics{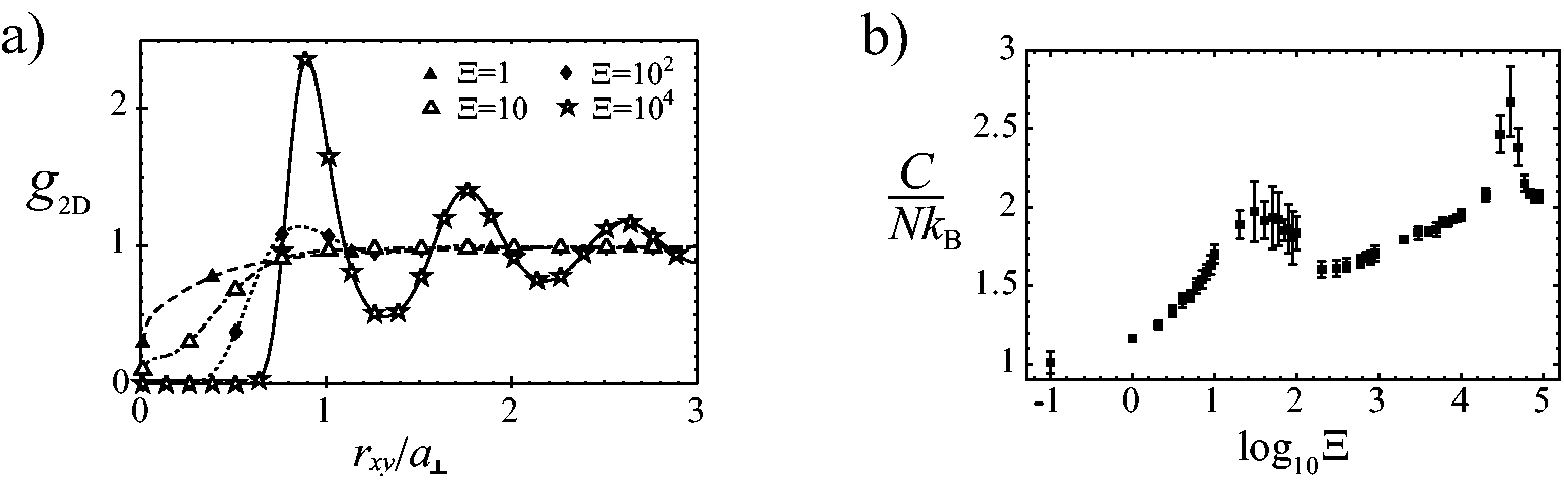}}
\end{center}
\caption{
a) The two-dimensional pair distribution function of 
counterions at a single charged planar wall plotted as a function of the
lateral distance $r_{xy}$ 
between counterions, obtained by averaging over the height z. 
Coupling parameters are $\Xi = 1$ (filled triangles),  $\Xi = 10$ (open triangles),   
$\Xi = 100$ (filled diamonds) and $\Xi = 10^4$ (open stars). 
The lateral distance is rescaled by the scale $a_\perp = 2 \sqrt{q/(\pi \sigma_s)}$.
b) Excess heat capacity  as a function of the coupling parameter. 
The number of counterions is N = 100 in a periodic square simulation box.
The broad hump at intermediate coupling $10 < \Xi  < 100$ reflects the structural
change in the counterionic layer due to increasing correlations between counterions. 
At a coupling strength  $\Xi \approx 31000 $
the counterion layer crystallizes, as indicated by a 
pronounced peak in the heat capacity. Reproduced after \cite{Alirev}.}
\label{ali}
\end{figure} 

To gain more quantitative information on the correlations in the counterion layer, we 
present results for the lateral two-point correlation function $g_{2D}$ at a single charged plate.
Physically,  $g_{2D}$ gives the normalized  probability of finding two counterions at a
certain lateral distance from each other.
The Monte-Carlo results for this quantity are shown in  Fig.\ref{ali}.
For small coupling parameter, $\Xi   = 1$, filled triangles, 
there is only a very short-range depletion zone at small separations between counterions. 
A pronounced correlation hole is created for coupling parameters 
$10 < \Xi   < 100$, where the distribution function vanishes over a finite range at 
small inter-particle separations. For larger coupling strengths, the correlation
hole becomes more pronounced and is followed by an oscillatory behavior 
in the pair distribution function, $\Xi = 10^4$, open stars. 
This indicates a liquid-like order in the counterionic structure in agreement with 
qualitative considerations in the preceding Sections. 
Note that the distance coordinate in Fig. \ref{ali}a is rescaled by 
 $a_\perp = 2 \sqrt{ q/(\pi \sigma_s)}$ as defined in Eq.(\ref{aperp1}).
The location of the
first peak of $g_{2D}$  for  $\Xi  = 10^4$  appears at a distance of 
$r_{xy}/a_\perp \approx 0.9$. 
In a perfect hexagonal crystal, the peak is expected to occur at
$r_{xy}/ a_\perp = \sqrt{\pi /(2 \sqrt{3})} \approx 0.95$, 
and in a perfect square crystal at
 $r_{xy}/ a_\perp =  \sqrt{\pi}/2  \approx 0.89$.
The crystallization in fact occurs at even larger
coupling parameters, which can best be derived  from the behavior 
of the heat capacity as a function of the coupling parameter $\Xi$.
In Fig. \ref{ali}b, the simulated excess heat capacity of the counterion-wall system 
(obtained by omitting the trivial kinetic energy contribution $3k_B/2$) 
is shown for various coupling parameters. The crystallization of counterions at
the wall is reflected by a pronounced peak at large coupling parameters about 
$\Xi  \approx 31000$, in good agreement with our estimate based on the 2D one-component
plasma.
The characteristic properties of the crystallization transition in the counterion-wall system 
are yet to be specified, which requires a detailed finite-size
scaling analysis in the vicinity of the transition point. 
Another interesting behavior is observed in Fig. \ref{ali}b at the
range of coupling parameters  $10 < \Xi  < 100$, where the heat capacity exhibits a broad hump. 
This hump probably does not represent a phase transition\cite{Dimaphase}, 
but  is most likely associated with the formation
of the correlation hole around counterions and
the structural changes in the counterionic layer from three-dimensional
at low couplings to  quasi-2D at large couplings. 
In the region between the hump and the crystallization 
peak, for $200 < \Xi  < 10^4$, the heat
capacity is found to increase almost logarithmically with  $\Xi$. 
The reason for this behavior is at present not clear.

The results in this section demonstrate that the Wigner crystallization 
transition, which has been studied extensively for a
two-dimensional system of charged particles, also exists
for a $2 \frac{1}{2}$-dimensional system where the counterions
are confined to one half space but attracted to a charged surface.
This is a non-trivial result, and for the system of counterions sandwiched
between two plates one expects interesting phase transitions between
different crystal structures as the plate distance is varied and becomes
of the order of the lateral distance between ions.

\subsection{The zero-temperature limit} 

\begin{figure}[t]
\begin{center}
\resizebox{8cm}{!}{\includegraphics{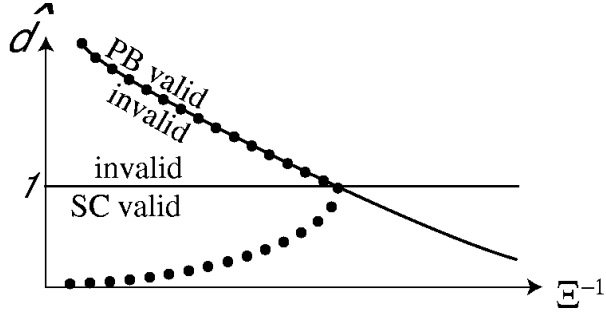}}
\end{center}
\caption{ Schematic phase diagram for counterions between two plates
as a function of the inverse coupling parameter 
$\Xi^{-1} \sim T^2 /  \sigma_s q^3 $ 
and the temperature independent rescaled plate distance
$\hat{d} =d/a_\perp$ where $a_\perp$ is the lateral distance
between counterions. Shown are the rescaled conditions for
validity of SC theory, Eq.(\ref{SCvalid}), $\hat{d} <1$, 
validity of PB theory, Eq.(\ref{PBvalidity}), $\hat{d} > \Xi^{1/2}$,
 (which coincides with the upper branch of the zero-pressure curve, dotted line)
 and the lower branch of the zero-pressure
line, Eq.(\ref{SCpress}), $\hat{d} \sim \Xi^{-1/2}$, dotted line.
The zero temperature-limit corresponds to the limit
$\Xi^{-1}  \rightarrow 0$.}
\label{zeroT}
\end{figure}

A word is in order  on the connection of our strong-coupling theory to
zero-temperature arguments for the pressure between charged surfaces
which involve two mutually interacting Wigner lattices\cite{Rouzina}
and which were extended by including  plasmon 
fluctuations at zero temperature\cite{lau-levine-pincus:00}
and at non-zero temperatures\cite{levine}.
Is the SC theory in fact a zero-temperature limit? No, it is not, as can be
seen from the asymptotically limiting pressure in  Eq.(\ref{SCpress}):
the first term is the confinement entropy of counterions, which 
clearly only exists at finite temperatures. Is the zero-temperature
contained in the SC theory and can it be derived from it? 
Only partially:
At zero temperature, the coupling constant tends to infinity, but 
on the other hand the Gouy-Chapman length (which sets
the spatial scale) tends to zero, and thus  all
rescaled lengths blow up. Coming back to the pressure in
Eq.(\ref{SCpress}), this means that the first, entropic term
disappears and only the second, energetic term remains. 
This is in exact accord with the predictions of the 
zero-temperature Wigner-lattice theory for small plate separation\cite{Rouzina}.
For plate separations larger than the lateral ion separation,
the Wigner-lattice theory predicts an exponential decay
of the attraction, which however is not contained within SC theory
since this is precisely the distance where SC starts to break down 
and an infinite resummation of all terms in the perturbation series would be needed.
To make things more transparent, let us construct from the
two parameters used for the two-plate system so far,
$\Xi$ and $\tilde{d}$, which both depend on temperature,
a parameter that does not depend on
temperature: it is given by $\hat{d} = \tilde{d}/\Xi^{1/2} \sim d / a_\perp$
and thus is a purely geometric parameter describing the ratio
of the distance between the  plates to the lateral distance between ions.
Sending $\Xi \rightarrow \infty$ at fixed $\hat{d}$ is the zero-temperature
limit and corresponds to finding the ground state of a counterion arrangement 
at  a fixed aspect ratio of the counterion-plate  unit cell. The condition for
validity of SC theory, Eq.(\ref{SCvalid}), translates into $\hat{d} < 1 $, 
while from Eq.(\ref{PBvalidity}) the PB theory  follows to be accurate for
$\hat{d} > \Xi^{1/2}$ (which coincides with the upper branch 
of the zero-pressure curve). The lower branch of the zero-pressure
line, Eq.(\ref{SCpress}), is given by $\hat{d} \sim \Xi^{-1/2}$.  These 
scaling predictions are assembled  in Fig. \ref{zeroT}, where the zero-pressure
lines are drawn as dotted lines and the limits of validity as solid lines.
For large $\Xi^{-1}$ a regime appears where both regimes of validity 
overlap, as was discussed in Ref.\cite{Netz4}, for small $\Xi^{-1}$ 
a large gap appears where non of the asymptotic PB and SC  theories
is valid. The zero-temperature limit is obtained for 
$\Xi^{-1} \rightarrow 0 $ in this diagram and thus complements
the PB and SC theories in that limit.

\section{Charged structured surfaces}

In the previous section we looked at the somewhat artificial model
where the charged surface is smooth and homogeneously charged, and 
where the counterions are pointlike and thus only interact
via Coulomb interactions. In reality, even an atomically flat
surface exhibits some degree of corrugation, and counterions have
a finite extent and thus experience some type of excluded-volume interaction.

\begin{figure}
\begin{center}
\resizebox{14. cm}{!}{\includegraphics{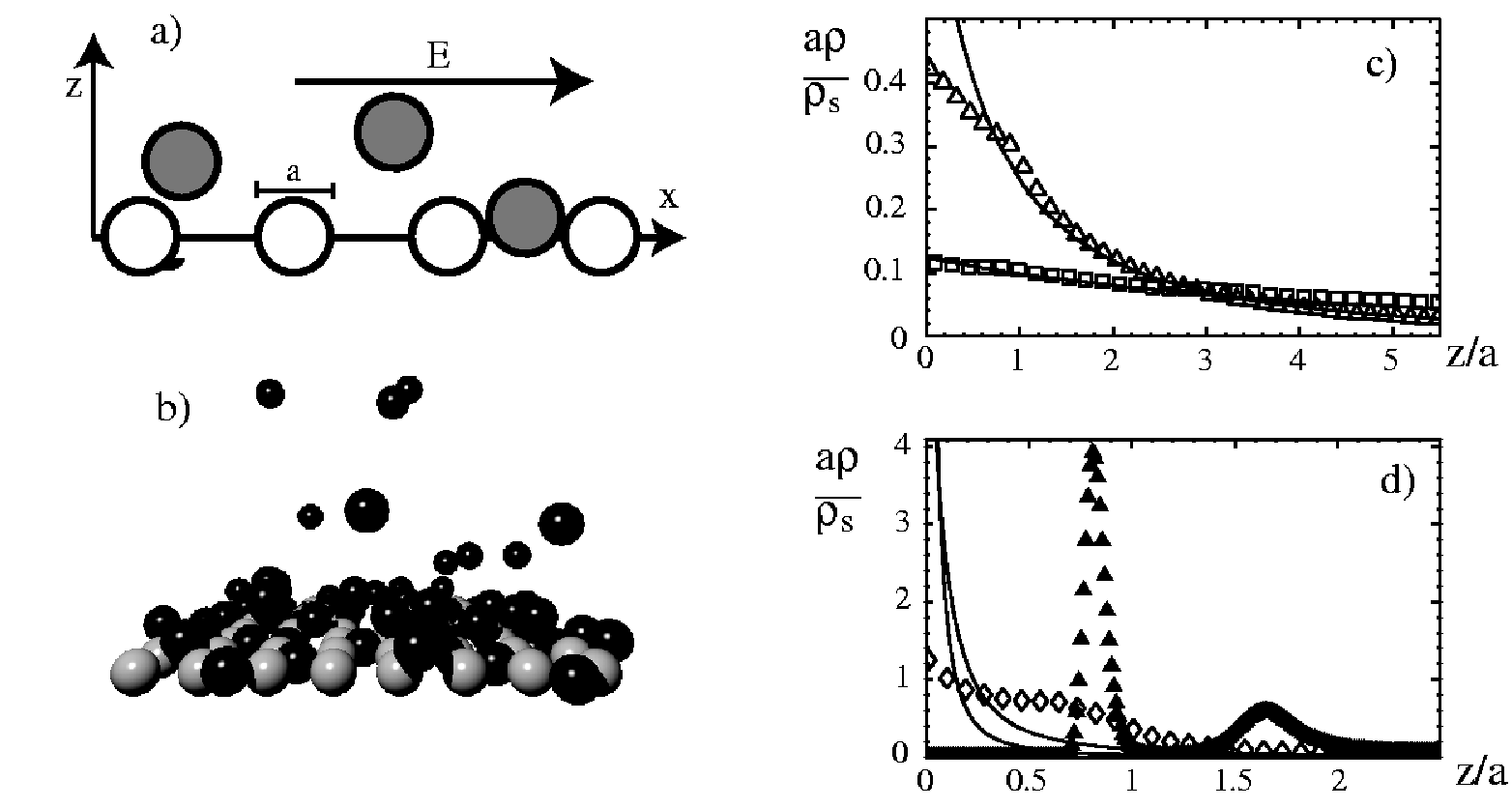}}
\end{center}
\vspace*{-.5cm}
\caption{
a) In the simulations a two-dimensional
layer of fixed surface ions
is in contact with oppositely charged
counterions of the same diameter $a$ and valency $q$.
b)  Snapshot of an ion configuration for a surface
density $\rho_s a^2 =0.5$ and Coulomb parameter  $\zeta =q^2 \ell_b / a =  2.5$.
c) Laterally  averaged counterion-density profiles
for Coulomb coupling $\zeta = 2.5$
as a function of the rescaled distance
from the surface (surface ions are fixed on a square lattice). 
Shown are results for 
surface ion densities $\rho_s a^2 = 0.0079$ (open squares), 
$\rho_s a^2 = 0.05$ (open triangles), d) $\rho_s a^2 = 0.5$ (open diamonds) 
and $\rho_s a^2 = 2$ (filled triangles) together with the
mean-field predictions for the laterally homogeneous case (solid lines).}
\vspace*{-.5cm}
\label{ciocco1} 
\end{figure}

In this section we 
consider a two-dimensional layer of $N$
charged spheres of valency $q$ and diameter $a$ (at $z=0$), 
together
with $N$ oppositely charged counterions of the same valency and diameter,
which are confined to the upper half space ($z>0$)
in a cubic simulation 
box of length $D$, see Fig.\ref{ciocco1}a. The number density of surface ions is
$\rho_s = N / D^2$.
The other important parameter is
$\zeta = q^2 \ell_B /a$ which measures the ratio
of the Coulomb interaction and the thermal energy at the minimal
inter-ionic  distance $a$.
Collapse of counterions and 
surface ions is prevented by a truncated Lennard-Jones term
acting between all particles.
The model we consider includes the combined 
effects of discrete surface charges, 
surface corrugations, and counterion excluded volume\cite{rolPRL}, which 
are all neglected in the classical mean-field approaches
but have been considered quite 
recently\cite{andreepl,vanmegen,bo,attsph1,itamar5,messina,Dima,Dima2,Foret1,Foret2,fleck3,Dias}.
We employ Brownian-dynamics simulations
where the velocity of all particles follows from
the position Langevin equation. The proper canonical
distribution functions are obtained by adding a suitably 
chosen Gaussian noise force acting on all particles and
expectation values  are obtained by averaging particle trajectories over time.
In Fig.\ref{ciocco1}b we show a snap shot of the counterion-configuration obtained
during a simulation. In Fig.\ref{ciocco1}c and d we show
laterally averaged 
counterion density
profiles for fixed Coulomb strength $\zeta = 2.5$  and various 
surface ion densities. This Coulomb strength corresponds to a distance
of closest approach between ions of 3 \AA~ which is a quite realistic value
for normal ions.
We also show the mean-field (MF) prediction for the
laterally homogeneous case, Eq.(\ref{PBone}), 
which reads in normalized form 
$a  \rho(z) / \rho_s = \mu^{-1} /(1+z/\mu)^2$. 
As before, the  Gouy-Chapman length $\mu= a/(2 \pi \zeta a^2 \rho_s)$
is a measure 
 of the decay length of the profiles. 
For small surface-ion densities,  Fig.\ref{ciocco1}c, 
the measured profiles agree quite well with the
 MF predictions, as expected,
since the Gouy-Chapman length is larger than the lateral surface-ion
separation and the charge modulation and hard-core repulsion 
matter little. However, even 
for the smallest density considered (open squares) there are some deviations
in the distance range $z/a < 1$ which we attribute to the hard-core
repulsion between surface ions and counterions. 
For the larger surface
densities in Fig.\ref{ciocco1}d the deviations become more pronounced
(simply shifting the MF profiles does not lead to
satisfactory agreement). For 
$\rho_s a^2 = 0.5$ (open diamonds) some counterions still reach the
surface at $z=0$, but the profile is considerably shifted to larger
values of $z$ due to the impenetrability of surface ions and counterions.
Finally, for $\rho_s a^2 = 2$ (filled triangles) the surface ions
form an impenetrable but highly corrugated layer, and the counterion
profile is shifted almost by an ion diameter outwards (and a second
layer of counterions forms).
 These results remind us that in experimental
systems a number of effects are present which make comparison
with theories based
on laterally homogeneous charge distributions difficult. 
As a 
side remark, the coupling constant $\Xi = 2 \pi \rho_s a^2 \zeta^2$ 
(which measures deviations
from MF theory due to fluctuations and correlations, see previous section) 
 is for the data in Fig.\ref{ciocco1}d in a range where
deviations from MF theory are becoming noticeable for the smeared-out
case\cite{moreiraepl}; for $\rho_s a^2 = 2$ one finds $\Xi \approx 75$ which 
means that Poisson-Boltzmann theory is invalid for almost all relevant
surface distances. But it is important to note that the deviations from
Poisson-Boltzmann we talked about in the previous section, as illustrated
in Fig. \ref{fig2}b  for smooth substrates,
are totally overwhelmed by the more drastic effects illustrated in Fig.\ref{ciocco1}. 

\begin{figure}[t]
\begin{center}
\resizebox{12cm}{!}{\includegraphics{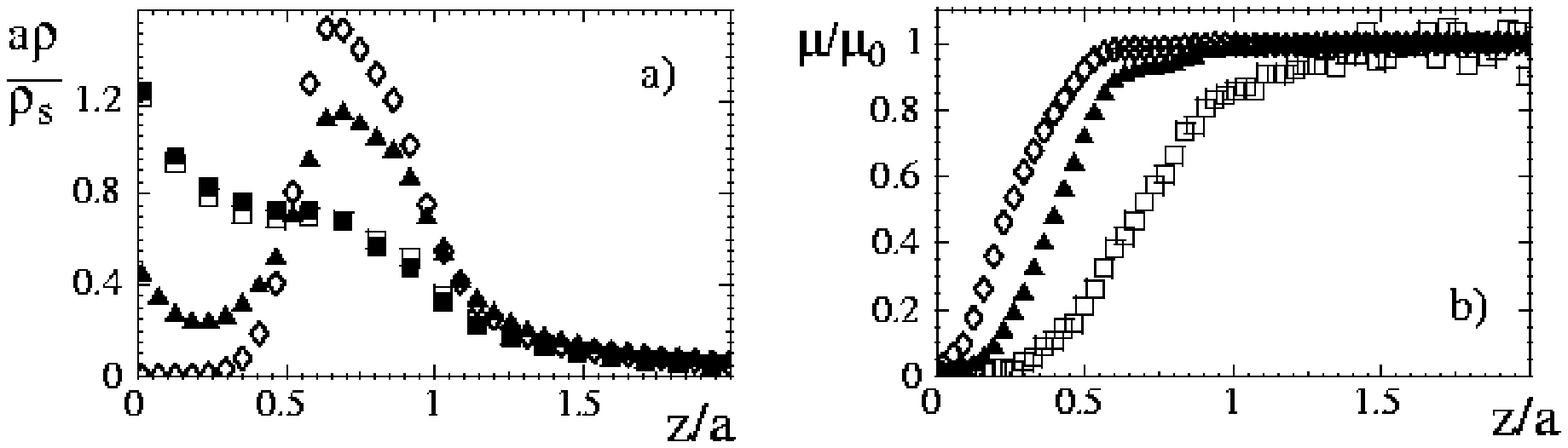}}
\end{center}
\caption{
a) Counterion density profiles for  Coulomb parameter 
$\zeta = 2.5$ and surface ion density $\rho_s a^2= 0.5$ 
for different electric field strengths 
$\tilde{E}= qae E/k_BT =0$ (filled squares), 
$\tilde{E}=1$ (open squares), 
$\tilde{E}=10$ (filled triangles), 
$\tilde{E}=40$ (open diamonds) 
for a square lattice of  fixed surface ions. 
b) Shown are the rescaled counterion mobility profiles for 
three different electric field strengths, where
$\mu_0$ is the bulk ion mobility; same symbols are used as in a).} 
\label{field}
\end{figure} 

The main advantage of the Brownian-dynamics technique is that dynamic quantities
can be calculated in the presence of externally applied fields even far
from equilibrium. As an illustration, we shown in Fig. \ref{field}a counterion density profiles 
for various values of a tangentially applied  electric field 
$\tilde{E}= qae E/k_BT $. The field acts on the mobile counterions and sets them in 
motion. This is the fundamental setup of electroosmotic and electrophoretic
experiments for large colloidal particles.
Fig. \ref{field}a shows that  the density
profiles shift to larger distances in the $z$-direction
for increasing
field strength. By doing this, the counterions avoid being trapped 
within the surface ion layer, and the conduction is maximized 
(though hydrodynamic interactions play a role 
at such elevated field strengths, as has been confirmed recently\cite{Woon}). 
In Fig. \ref{field}b the corresponding counterion mobility profile is shown.
For the smallest field considered,
$\tilde{E} = 1$ (open diamonds), which belongs to the linear 
quasi-static  regime,
the mobility is highly reduced for distances below roughly
$z/a =1$, which is plausible 
since in this distance range surface 
ions and counterions 
experience strong excluded-volume interactions and thus friction.
The maximal mobility of $\mu / \mu_0 = 1$ 
is reached quickly for
larger separations from the surface. For larger fields the crossover
in the mobility profiles moves closer to the surface, and
since the density at the wall decreases,
the total fraction of immobile counterions goes drastically down.
Since the decrease of the mean electrophoretic mobility
is caused by a fairly localized layer of immobilized counterions,
the integrated   relative mobility  can be interpreted as 
the fraction of mobile ions, or, in other words, the fraction of
counterions that are not located within the stagnant
Stern layer. 
This gives a  dynamical definition of the Stern layer which is unambiguous
and connects to the experimentally relevant Zeta potential\cite{rolPRL}. 
The effects seen at elevated field strengths are not relevant experimentally
for mono-valent ions, since they correspond to unrealistically
high electric field strengths  where in fact water is fully oriented; for 
highly charged objects, however, similar non-equilibrium effects in electric fields
do occur. A drastic example of a far-from-equilibrium phase transition
is the structural bifurcation that is observed in a two-dimensional 
electrolyte solution at large electric fields\cite{Netzconduct}.
Here the ions spontaneously form interpenetrating 'traffic lanes' at large field 
strengths, which tend to maximize the possible current that is supported
by the system. Whether such flux-maximizing states are always
realized when one moves far away from equilibrium is presently not clear.

The main message of this example is that the counterion mobility with respect
to  a tangential field is highly reduced by the presence of surface corrugation\cite{rolPRL},
which is plausible  since counterions are dynamically trapped within the
surface-charge layer. 
The resultant modified boundary condition
is relevant for  a whole collection of experimental results
on the electrophoretic mobility of charged colloids.
Simulations that include hydrodynamic interaction essentially confirm the
present results and allow to directly connect to experimentally 
measurable quantities\cite{Woon}

\section{Polyelectrolytes at charged planes: overcharging and charge reversal }

\begin{figure}[t]
\begin{center}
\resizebox{12cm}{!}{\includegraphics{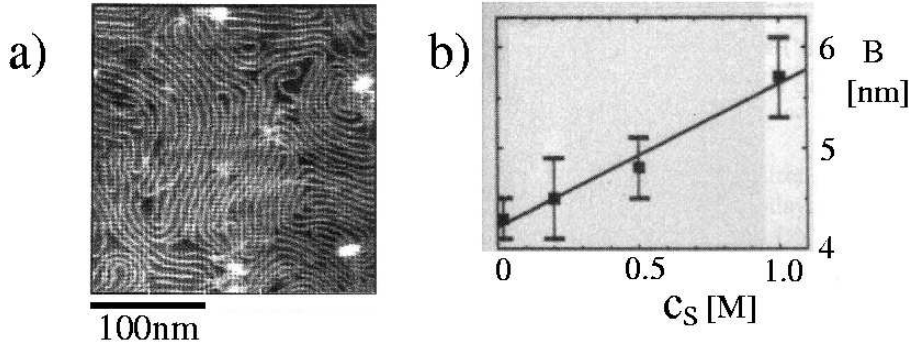}}
\end{center}
\caption{
a) AFM picture of a DNA  layer adsorbed at an oppositely charged substrate 
at a salt concentration $c_s = 1 M$. The average lateral distance between
DNA strands is $B \simeq 6 nm$.  
b) Mean lateral distance between DNA molecules, $B$,  as a function of the 
salt concentration (reproduced after \cite{fang}).}
\label{fig7}
\end{figure}

For many applications, it is important to adsorb highly charged polymers
in a controlled
way on planar substrates, for example for the production of DNA 
chips\cite{dnachip} or the fabrication of charge-oscillating 
multilayers\cite{Decher,Donath,Caruso}. 
Various experiments have been performed with DNA\cite{fang,Maier}
and synthetic polymers with comparable charge 
density\cite{Kerstin,Berlepsch}.
Important issues are the structure of the adsorbed
layer or the amount of adsorbed material at a given set of parameter
values (such as salt concentration of the ambient solution, 
polymer concentration, etc.). Fig. \ref{fig7}a shows  
atomic-force-microscope pictures of an adsorbed DNA layer
on a positively charged substrate, obtained at relatively high
salt concentration of $1 M$\cite{fang}. The analysis of the AFM pictures
shows that the adsorbed layer is extremely thin, which is in contrast
to the rather diffuse layers that are obtained with neutral polymers.
The individual DNA strands have a rather well-defined mutual distance
of $B \approx 6 nm$ at a salt concentration $c_s = 1 M$, which is 
larger than the DNA diameter of $D \approx 2 nm$\footnote{It is important
to note that the DNA layer shown in Fig.\ref{fig7}a has been prepared at
a salt concentration of $c_s = 1 M$ but imaged
at a much smaller salt concentration (presumably without changing its 
structure), since at high 
salt the layer becomes extremely fuzzy and is impossible to image
with an AFM.}. At length scales above $100 nm$ the DNA strands 
change their orientation, the structure resembles a finger print. 
The lateral distance between DNA strands grows with increasing 
salt concentration, see Fig.\ref{fig7}b\cite{fang}. 
All these findings can be theoretically explained
by considering the competition between electrostratic attraction to
the substrate and electrostatic and entropic
repulsion between neighboring DNA strands\cite{joanny}, 
as will be shown in the following.

A polyelectrolyte (PE) characterized by a
linear charge density  $\tau$, is subject to an electrostatic
potential created by $\sigma_s$, the homogeneous surface charge
density (per unit area) on the substrate. 
Because this potential is attractive for
an oppositely charged substrate (which is the situation that
was considered in all above-mentioned experiments), 
it is the driving
force for the adsorption and we neglect complications due to 
additional interactions between surface and PE which have been 
considered recently\cite{joanny2,Rubin}.
One example for additional effects are  interactions
due to the dielectric discontinuity at the
substrate surface\footnote{An ion in solution has a repulsive interaction from the
surface when the solution dielectric constant is higher than that
of the substrate. This  effect can lead to desorption for highly
charged PE chains. On the contrary, when the substrate is a metal
there is a possibility to induce PE adsorption on non-charged
substrates or on substrates bearing charges of the same sign as
the PE. See Ref.~\cite{Netz1} for more details.}
and to the impenetrability of the substrate for
salt ions\cite{Netz1}.
Within the linearized Debye-H\"uckel (DH) theory, the electrostatic attractive
force acting on a PE section at a distance $\delta$ from the
homogeneously charged plane is in units of $k_BT$ and per PE unit length 
\begin{equation}
f_{\rm att} (\delta) =  -4\pi \ell_B \sigma_s \tau 
{\rm e}^{-\kappa \delta}.
\end{equation}
The screening length $\kappa^{-1}$ depends on the salt concentration
$c_s$ and ion valency $q$ and is defined via $\kappa^2 = 8 \pi q^2 \ell_B c_s$.
Assuming that the polymer is adsorbed over a layer of width
$\delta $ smaller than the screening length $\kappa^{-1}$, the
electrostatic attraction force per PE  unit length becomes constant and can be
written as
\begin{equation}
\label{fatt} f_{\rm att}  \simeq -4 \pi \ell_B \sigma_s
\tau ~.
\end{equation}
For simplicity, we neglect non-linear effects due to counter-ion condensation
on the PE (as obtained by the Manning counterion-condensation argument\cite{Mann})
and on the surface (as obtained within the Gouy-Chapman theory). 
Although these effects
are important for highly charged systems\cite{fleck},
most of the important features of single PE adsorption already appear on 
the linearized Debye-H\"uckel level.

\begin{figure}[t]
\begin{center}
\resizebox{12cm}{!}{
\includegraphics{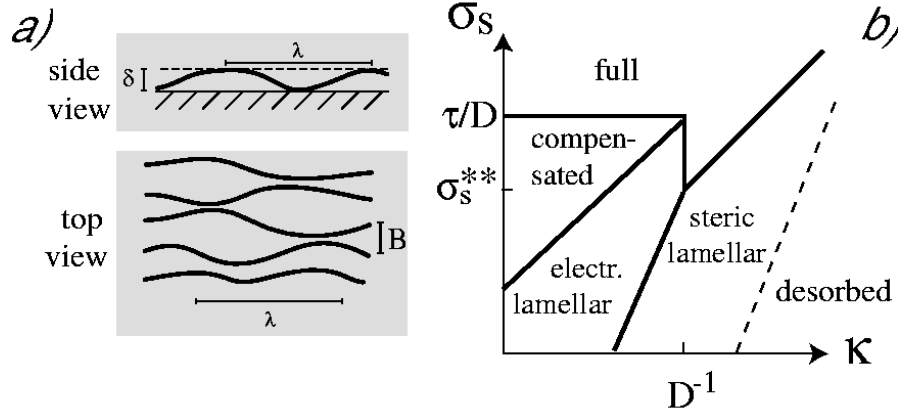}
}
\end{center}
\caption{
a) Schematic structure of the adsorbed DNA layer in a side view
and in a top view, exhibiting the deflection length $\lambda$,
the adsorbed layer height $\delta$, and the lateral distance between 
adsorbed DNA strands, $B$. 
b) Scaling diagram of the adsorption behavior
of a highly charged polymer using logarithmic axes
as a function of the surface charge density $\sigma_s$ and inverse
screening length $\kappa= \sqrt{ 8 \pi \ell_B c_s}$. 
The electrostatically stabilized lamellar phase is strongly charge-reversed.}
\label{fig8}
\end{figure}

Because of the confinement in the adsorbed layer, the polymer
feels an entropic repulsion. If the layer thickness $\delta$ is much
smaller than the effective persistence length of the polymer,
$\ell_{\rm eff}$, as depicted in the side view of
Fig.~\ref{fig8}a, a new length
scale, the so-called deflection length $\lambda$, enters the
description of the polymer statistics. 
We call the persistence length an effective one, because it in principle
contains effects due to the electrostatic repulsion between monomers.
The deflection length
$\lambda$ measures the average distance between two contact
points of the polymer chain with the substrate. As shown by
Odijk, the deflection length scales as $\lambda \sim \delta^{2/3}
\ell_{\rm eff}^{1/3}$ and is larger than the layer thickness $\delta $
but smaller than the persistence length $ \ell_{\rm
eff}$~\cite{Odijk}. The entropic repulsion follows in a simple
manner from the deflection length by assuming that the polymer
loses roughly an entropy of one $k_B T$ per deflection length.

On the other hand, if $\delta > \ell_{\rm eff}$, 
 the polymer forms a random coil with many loops
within the adsorbed layer. The chain can be viewed as an assembly
of decorrelated blobs, each containing a subchain of length
$L \sim \delta^2/\ell_{\rm eff}$, within which 
the polymer obeys Gaussian statistics.
The decorrelation into blobs has an entropic cost of roughly one $k_B T $
per blob. The entropic repulsion force per polymer unit length is
thus ~\cite{Odijk}
\begin{equation} \label{Odijkrep}
\label{frep} f_{\rm rep}  \sim
     \left\{ \begin{array}{llll}
     & \delta^{-5/3} \ell_{\rm eff}^{-1/3}
         & {\rm for} &  \delta \ll \ell_{\rm eff}  \\
     & \ell_{\rm eff} \delta^{-3}
         & {\rm for }     & \delta \gg \ell_{\rm eff}~,   \\
                \end{array} \right.
\end{equation}
where we neglected a logarithmic correction factor which is not
important for our scaling arguments.

The equilibrium layer thickness follows from equating the
attractive and repulsive forces, Eqs.~(\ref{fatt}) and
(\ref{frep}). For rather stiff polymers and   small layer
thickness, $\delta < \kappa^{-1} < \ell_{\rm eff}$, we obtain
\begin{equation}
\label{delta}
\delta \sim \left(
\ell_B \sigma_s \tau \ell_{\rm eff}^{1/3} \right)^{-3/5} ~.
\end{equation}
For a layer thickness corresponding to the screening length, $\delta
\approx \kappa^{-1}$, scaling arguments predict a rather abrupt
desorption transition~\cite{joanny}. Setting $\delta \sim
\kappa^{-1}$ in Eq.~(\ref{delta}), we obtain an expression for the
adsorption threshold (valid for $\kappa \ell_{\rm eff} > 1$)
\begin{equation}
\label{strong} \sigma^*_s \sim \frac{\kappa^{5/3} } {\tau \ell_B
\ell_{\rm eff}^{1/3}  } ~.
\end{equation}
For $\sigma_s   > \sigma_s^*$ the polymer is adsorbed and localized
over a layer with a width smaller than or comparable to
the screening length (and
with the condition $\ell_{\rm eff}> \kappa^{-1}$ also satisfying
$\delta < \ell_{\rm eff}$, indicative of a flat layer).
As $\sigma_s$ is decreased, the polymer
abruptly desorbs at the threshold $\sigma_s \simeq \sigma_s^*$. 

From Eq.~(\ref{delta}) we see that the layer thickness
$\delta$ is of the same order as $\ell_{\rm eff}$ for $ \ell_B
\sigma_s \tau \ell_{\rm eff}^2 \sim 1$, at which point the
condition $\delta \ll \ell_{\rm eff} $ used in deriving
Eq.~(\ref{delta}) breaks down. 
Let us now consider the opposite limit of a rather flexible chain, 
$\ell_{\rm eff} <\kappa^{-1}$.
If the layer thickness is larger than the persistence length but
smaller than the screening length, $\ell_{\rm eff} < \delta <
\kappa^{-1} $, the prediction for $\delta$ obtained from
balancing Eqs.~(\ref{fatt}) and (\ref{frep}) becomes
\begin{equation}
\label{delta2}
\delta \sim \left( \frac{\ell_{\rm eff}}
{\ell_B \sigma_s \tau} \right)^{1/3}~.
\end{equation}
From the expression  Eq.(\ref{delta2}) we see that $\delta$ has the same size as
 the screening length $\kappa^{-1}$ for
\begin{equation}
 \label{weak}
\sigma_s^*  \sim \frac{\ell_{\rm eff} \kappa^3} {\tau \ell_B  } ~.
\end{equation}
This in fact denotes the location of a continuous adsorption
transition at which the layer grows to infinity \cite{joanny}. The
scaling results for the adsorption behavior of a flexible polymer,
Eqs.~(\ref{delta2})-(\ref{weak}), are in agreement  with previous
results~\cite{Wiegel}.

Now we
generalize the discussion of the single PE chain
adsorption and consider the effect of interactions
between different adsorbed polymers on a simple scaling level. In
order to do so, we assume that the adsorption energy is  strong enough
such that the polymers essentially lie flat on the substrate, which is
the relevant case for describing the experiments shown in Fig.\ref{fig7}a.
Chain crossings are disfavored, and
lateral chain correlations are large enough to induce the
polymers to form some type of locally ordered lattice. The formation
of this two-dimensionally correlated adsorbed layer is accompanied by 
a loss  of energy and entropy, which will turn out to be important to understand
the density of adsorbed chains on the substrate.
We follow here the
original ideas of Ref.~\cite{joanny}, which were in parallel developed by
 Nguyen et al.\cite{Nguyen0,Nguyen}. To understand the idea,
consider the top view in Fig.~\ref{fig8}a, where a
{\em lamellar} phase is shown  where different polymer strands are parallel
locally, characterized by an average lamellar spacing $B$.
The lamellar phase is stabilized
either by steric or by electrostatic repulsions between
neighboring polymers; we will in fact encounter both stabilization
mechanisms for different values of the parameters.
We calculate the free energy and other characteristics of
the adsorbed lamellar phase, 
assuming that we are inside the adsorbed regime of a single
polymer. We will also assume, later on,
that the desorption transition obtained for the single-chain
case also applies to the case of  many-chain adsorption.
 As was shown in Ref.~\cite{joanny}, to obtain the complete
phase diagram it is sufficient to consider the lamellar phase
depicted in Fig.~\ref{fig8}a, since other possible phase
morphologies are metastable or degenerate. We assume that the
distance between neighboring polymer strands, $B$,  is much
smaller than the effective persistence length, $B < \ell_{\rm
eff}$ (which can be checked self-consistently).
Since the possible conformations of the adsorbed polymers are
severely restricted in the lateral directions, we have to include,
in addition to the electrostatic interactions, a repulsive free
energy contribution coming from steric interactions between stiff
polymers~\cite{Odijk}. This is the same type of entropic
repulsion that was used before  to estimate the pressure
inducing desorption from a substrate, Eq.(\ref{Odijkrep}), but
now including the previously neglected logarithmic cofactor. The total free energy
density per unit area and per $k_BT$ in the lamellar phase is given by
\begin{equation}
\label{freelam}
F_{\rm lam}  \simeq   -\frac{2 \pi \ell_B \sigma_s \tau}{B \kappa}
+ \frac{1}{\ell_{\rm eff}^{1/3} B^{5/3}} \ln
\left(\frac{\ell_{\rm eff}}{B} \right) + F_{\rm rep}~,
\end{equation}
where the first term comes from the electrostatic
attraction to the oppositely charged surface (which in accord with
the potential used for the repulsion between polymers later on,
is taken to be penetrable to ions), the second term is the
 Odijk entropic repulsion between polyelectrolyte chains \cite{Odijk} and
$F_{\rm rep}$ is the electrostatic repulsion within the  lamellar array.

To obtain the electrostatic repulsive energy, we first note that the
reduced potential created by
a charged line with line charge density $\tau$ is at
a distance $B$ within the Debye-H\"uckel approximation
given by
\begin{equation}
\label{Vline}
V_{\rm line}(B)  = \tau
\int_{-\infty}^{\infty} {\rm d}s\;
v_{\rm DH}(\sqrt{B^2+s^2})  =
2 \ell_B \tau K_0[\kappa B]~,
\end{equation}
with the Debye-H\"uckel potential $v_{\rm DH}(r) = \ell_B {\rm e}^{-\kappa r}/r$
and where $K_0$ denotes the modified Bessel function.
The repulsive electrostatic free energy density of an array of parallel
lines  with a nearest-neighbor distance of $B$ and line charge density
$\tau$ can thus be written as
\begin{equation}
\label{selfsum}
F_{\rm rep} = \frac{2 \ell_B \tau^2}{B} \sum_{j=1}^{\infty}
K_0[j B \kappa]~.
\end{equation}
This expression is also accurate for rods of finite radius $D$ as long as
$D \ll B$ holds.
In the limit $B \kappa \ll 1$, when the distance between strands is
much smaller than the screening length, the sum can be transformed
into an integral and we obtain
\begin{equation}
\label{selflam1}
F_{\rm rep} \simeq \frac{2 \ell_B \tau^2}{B}
\int_0^{\infty} {\rm d}s\; K_0[s B \kappa] =
\frac{ \pi \ell_B \tau^2}{B^2 \kappa} ~.
\end{equation}
This expression neglects effects due to the presence of a solid substrate.
For example, and as discussed in Ref. \cite{Netz1},
for a low-dielectric substrate the electrostatic
interactions are enhanced by a factor of two close to the substrate
surface, a rather small effect which will be neglected
in the following.
Corrections to the approximation in Eq.(\ref{selflam1}) have
been treated in \cite{Nguyen0,Nguyen}.
Since the average adsorbed surface charge density is given
by $\sigma_{\rm ads} = \tau/B$, it follows that the self energy
Eq. (\ref{selflam1}) in the limit  $B \kappa \ll 1$  is
given by $F_{\rm rep}  \simeq  \pi \ell_B \sigma_{\rm ads}^2 \kappa^{-1}$ and
thus is identical to the self energy of a totally
smeared-out charge distribution~\cite{joanny}.
In this case, lateral correlations therefore do not matter.

In the opposite limit, $B \kappa \gg 1$, when the polymers are
much farther apart than the screening length, the
sum in Eq. (\ref{selfsum}) is dominated by the first term and (using the
asymptotic expansion of the Bessel function) the free energy density (in units of $k_B T$)
becomes
\begin{equation}
\label{selflam2}
F_{\rm rep} \simeq \frac{ \sqrt{2 \pi} \ell_B \tau^2
{\rm e}^{-B \kappa}}{B^{3/2} \kappa^{1/2}} ~.
\end{equation}
In this limit, it is important to note that the smeared-out repulsive
energy Eq. (\ref{selflam1}) is much larger and thus
 considerably overestimates the
actual electrostatic repulsion between polymer strands.
Conversely, this reduction of the electrostatic repulsion between
polymers results in an enormous overcharging of the substrate,
as we will see shortly.

In order to determine the equilibrium distance between
the polymer strands,
we balance the electrostatic attraction term,
the first term in Eq.(\ref{freelam}),
with the appropriate repulsion term. There are three choices
to do this.
For $D< \kappa^{-1}  < B^* <  B$ (with some crossover length $B^*$
to be determined later on),
the electrostatic repulsion between the polymers is irrelevant
(i.e. the last term in Eq.(\ref{freelam}) can be neglected),
and the lamellar phase is {\em sterically} stabilized in this case.
The equilibrium  lamellar spacing  is given by
\begin{equation}
\label{stericB}
B \sim \left[ \frac{\kappa }
{\tau \sigma_s \ell_B \ell_{\rm eff}^{1/3}}
\ln \left( \frac{\tau \sigma_s \ell_B \ell_{\rm eff}}{\kappa}
\right) \right]^{3/2}~.
\end{equation}
In all what follows, we neglect the logarithmic cofactor.

For $ D< \kappa^{-1} < B < B^*  $,
the steric repulsion between the polymers is irrelevant
(i.e. the second term in Eq.(\ref{freelam}) can be neglected).
The free energy is minimized by balancing the electrostatic
adsorption term , the first term in Eq.(\ref{freelam}), with the
electrostatic repulsion term appropriate for the case
$B \kappa > 1$, Eq. (\ref{selflam2}),
which leads to the {\em electrostatically }
stabilized lamellar spacing
\begin{equation}
\label{electB}
B \sim \kappa^{-1} \ln
\left[ \frac{\tau \kappa}{\sigma_s} \right] .
\end{equation}
The adsorbed charge density then follows from
$\sigma_{\rm ads}  \sim \tau /B$  as
\begin{equation} \label{chargereverse}
\sigma_{\rm ads}  \sim \sigma_s \frac{ \tau \kappa \sigma_s^{-1}  }{
\ln(\tau \kappa \sigma_s^{-1})} ~.
\end{equation}
Therefore, the electrostatically stabilized lamellar phase shows
always strong charge reversal, since the polymer spacing  $B$ is 
larger than the screening length and thus $ \tau \kappa \sigma_s^{-1} >1$.
This can be seen from comparing the two equations (\ref{electB})
and (\ref{chargereverse}).
The crossover between the sterically stabilized lamellar phase,
described by Eq.(\ref{stericB}), and
the lamellar phase which  is stabilized by electrostatic repulsion,
Eq. (\ref{electB}), occurs when the predictions for $B$ become
simultaneously equal to the crossover spacing 
$B^*$, leading to a crossover for a surface charge density of
(without logarithmic cofactors)
\begin{equation}
\label{st/el}
\sigma_s^* \sim
\frac{\kappa^{5/3} }{\tau \ell_{\rm eff}^{1/3}\ell_B } ~.
\end{equation}
For $\sigma_s $ larger than the crossover value in Eq.(\ref{st/el})
the distance
between neighboring polymer strands is smaller than $B^*$ and
the electrostatic stabilization mechanism is at work,
for  $\sigma_s$ smaller than the crossover value in Eq.(\ref{st/el}) the
lamellar spacing $B$ is larger than the characteristic
crossover length $B^*$ and the Odijk entropic repulsion dominates.
We obtain the interesting result that in the sterically stabilized 
adsorbed phase the strand separation increases with increasing salt concentration,
see Eq.(\ref{stericB}), while in the electrostatically stabilized 
phase the strand separation decreases with increasing salt, see Eq. (\ref{electB}).
The intuitive reason for this is clear: in the sterically stabilized phase
adding salt diminishes the electrostatic attraction to the substrate, 
while in the electrostatically stabilized phase the predominant effect
of salt is to weaken the repulsion between neighboring PE strands.

The electrostatically stabilized lamellar phase crosses
over to the {\em charge-compensated} phase when $B$
as given by Eq. (\ref{electB}) becomes of the order of the screening
length $\kappa^{-1}$.
In the charge-compensated phase, the lamellar spacing is obtained by balancing the
electrostatic adsorption energy with the repulsion in the smeared-out limit
Eq.(\ref{selflam1}) and  is given by
\begin{equation}
\label{compB}
B \simeq \frac{\tau}{\sigma_s} ~.
\end{equation}
In this case  the adsorbed surface charge density $\sigma_{\rm ads} = \tau/B$
exactly neutralizes the substrate charge density,
\begin{equation}
\sigma_{\rm ads} \sim \sigma_s ~.
\end{equation}
The crossover between the  charged-reversed phase and charge-compensated phase
is obtained by matching Eqs. (\ref{electB}) and (\ref{compB}), leading to
a threshold surface charge density of
\begin{equation}
\label{comp/el}
\sigma_s^* \sim
\tau \kappa ~.
\end{equation}

Finally, taking into account that the polymers have some width $D$,
there is an upper limit for
the amount of polymer that can be adsorbed in a single layer. Clearly, the
lateral distance between polymers in the {\em full} phase is given by
\begin{equation}
\label{fullB}
B \simeq D
\end{equation}
and thus the adsorbed surface charge density in the full phase reads
\begin{equation}
\sigma_{\rm ads} = \frac{\tau}{D}.
\end{equation}
The crossover between the full phase and the compensated phase is obtained by
comparing Eqs. (\ref{compB}) and (\ref{fullB}), leading to
\begin{equation}
\label{comp/full}
\sigma_s^* \sim \tau /D .
\end{equation}

In Fig.~\ref{fig8}b we show the adsorption diagram for
strongly charged polymers (defined by $\tau \sqrt{\ell_B \ell_{eff}}
> 1$) as a function of the substrate charge density $\sigma_s $ and
the inverse screening length $\kappa $. The electrostatically
stabilized lamellar phase shows strong charge
reversal as described by Eq. (\ref{chargereverse}). At slightly
larger surface charge densities we predict a charge-compensated phase
 which is not full (i.e. $B < D$) for a  range of surface 
 charge densities  as determined
by Eqs. (\ref{comp/el}) and (\ref{comp/full}). At even larger
substrate charge density, the adsorbed polymer phase becomes
close packed, i.e. $B \approx  D$. We note that since the full phase is
not charge reversed, the full phase can consist of a second
adsorbed layer (or even more layers, as discussed in \cite{joanny,Nguyen3}). 
It should however be clear
that close to charge compensation the effective substrate charge
density an additional layer feels is so small that the condition
for adsorption is not necessarily met. At low substrate charge densities the
distance between adsorbed polymer strands becomes so large that
the entropic repulsion between polymers dominates the
electrostatic  repulsion, and finally, at even
lower charge densities, the polymers desorb. One
notes that the transition between the electrostatically and
sterically stabilized adsorbed phases, Eq.(\ref{st/el}), has the
same scaling form (disregarding logarithmic factors) as the
desorption transition of semi-flexible polymers,
Eq.(\ref{strong}). We have shifted the desorption transition to
the right, though, because typically there are attractive
non-electrostatic interactions as well, which tend to stabilize
adsorbed phases. This is also motivated by the fact that
 the sterically stabilized phase
has been seen in experiments on DNA adsorption, as will be discussed below.
The critical charge density $\sigma_s$ where the full phase, the
electrostatically and the sterically stabilized phases meet at one point,
is given by $\sigma^{**}_s \sim 1/(D^{5/3} \ell_{\rm eff}^{1/3} \tau \ell_B)$.
In the phase diagram we have assumed that the charge density threshold for
 the full phase Eq.(\ref{comp/full}), $\sigma_s^* \sim \tau/D $, 
 satisfies the inequality $\tau/D > \sigma_s^{**}$, which for a fully
charged PE at the Manning threshold, $\tau = 1/\ell_B$,
amounts to the condition $\ell_{\rm eff} > \ell_B^3 / D^2$,
which is true for a large class of PE's (especially stiff biopolymers such as DNA).
As one increases at fixed substrate charge density $\sigma_s$
the salt concentration, one moves through the compensated, 
electrostatically and the sterically stabilized adsorbed phases, before one
finally induces desorption. The polymer separation is predicted to first
stay constant, then decrease and finally increase, before desorption takes place.

One important result of our discussion
is that in the electrostatically stabilized
phase the substrate charge is strongly reversed by the adsorbed polymer layer.
This can give rise to a charge-oscillating multilayer formation if the adsorption
of oppositely charged polymer is done in a second step\cite{Decher,Donath,Caruso}.
The general trend that emerges is that charge reversal is more
likely to occur for intermediate salt concentrations and rather
low substrate charge density. For too high salt concentration
and too low substrate charge density, on the other hand, the
polymer does not adsorb at all.
In essence, the salt concentration and the substrate charge
density have to be tuned to intermediate values in order
to create charge multilayers.

In experiments on DNA  adsorbed on oppositely charged substrates
one typically observes a lamellar phase \cite{fang,Maier}. In
one experiment, the spacing between DNA strands was found to
increase with increasing salt concentration~\cite{fang}. One
theoretical explanation invokes an effective interaction between
neighboring DNA strands mediated by elastic deformations of the
membrane, which forms the substrate in these experiments
\cite{Dan}. In the sterically stabilized regime, the distance
between adsorbed polymers increases  as $B \sim
\kappa^{3/2}$ with the salt concentration, see
Eq.(\ref{stericB}), which offers an alternative explanation for
the experimental findings. It would be interesting to redo DNA
adsorption experiments on rigid substrates, where the elastic
coupling to the membrane is absent. For high enough substrate
charge densities and by varying the salt concentration one should
be able to see the crossover from the electrostatically
stabilized phase, Eq.(\ref{electB}), where the DNA spacing
decreases with added salt, to the sterically stabilized phase,
Eq.(\ref{stericB}), where the DNA spacing increases with added
salt.

\subsection{Overcharging by  spherical polyions}

\begin{figure}[t]
\begin{center}
\resizebox{12cm}{!}{\includegraphics{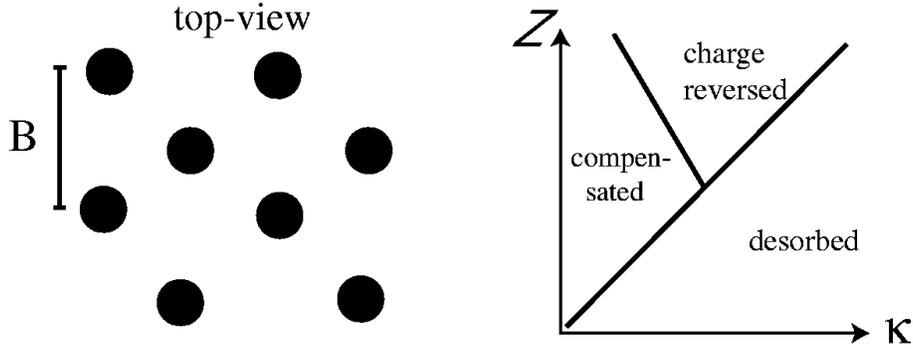}}
\end{center}
\caption{
Schematic structure of a layer of adsorbed polyions is shown 
to the left, characterized by the lateral distance $B$ between
adsorbed ions.
The right shows the resulting scaling diagram using logarithmic axes as a
function of polyion charge $Z$ and inverse screening length $\kappa$, 
featuring  the desorbed phase, a charge reversed phase where more ions adsorb
than needed to neutralize the surface charge, and a compensated phase
where the adsorbed ions exactly neutralize the surface. }
\label{chargerevers}
\end{figure} 

The arguments from the last section for the overcharging of a charged plane 
by charged polymers can be straightforwardly adapted to the adsorption
of spherical polyions on surfaces\cite{Nguyen,Tanaka,Perez}. 
Using the DH approximation, the adsorption energy
of a polyion of charge $Z$ on a surface of charge density
$\sigma_s$ in units of $k_BT$ is 
$W_{att}  \simeq - 4 \pi Z \ell_B \sigma_s / \kappa$  where we assume that the 
particle radius is smaller than the screening 
length so that the full particle charge contributes to the attraction.
Neglecting logarithmic factors depending on the bulk particle concentration,
the desorption threshold is reached when the adsorption energy equals
thermal energy, i.e. $W_{att} \simeq 1$. The condition for adsorption is therefore
$Z > \kappa / (\ell_B \sigma_s)$. Assuming the particles to form a correlated
liquid arrangement on the surface (as depicted in Fig. \ref{chargerevers})
with a distance between particles
larger than the screening length, the repulsion between two nearest neighbors
is $W_{rep} \simeq Z^2 \ell_B e^{-B \kappa}/B$. The equilibrium distance
of particles is obtained by minimizing the free energy per area,
\begin{equation}
F  =(W_{att}+W_{rep})/B^2 = 
 \frac{Z^2 \ell_B e^{-B \kappa}}{B^3 } - \frac{  Z \ell_B \sigma_s }{ \kappa B^2}
\end{equation}
which assumes that particles are obtained from some reservoir at vanishing chemical potential.
The resulting equilibrium separation is obtained as
\begin{equation}
B \sim \kappa^{-1} \ln (Z \kappa^2 / \sigma_s)
\end{equation}
and the adsorbed charge density as
\begin{equation}
\sigma_{ads}  \simeq \frac{Z}{B^2}  \sim  \sigma_s
\frac{Z \kappa^2 / \sigma_s}{\ln^2 (Z \kappa^2 / \sigma_s)}.
\end{equation}
It can be easily seen that for $Z \kappa^2 / \sigma_s > 1$ the surface
is overcharged and the distance between adsorbed particles is larger
than the screening length (in agreement with our assumption).
Thus the condition for overcharging is $Z > \sigma_s \kappa^{-2}$. 
Conversely, for $Z <  \sigma_s \kappa^{-2}$, the separation between
particles becomes smaller than the screening length. A calculation similar to the 
one in the preceding section shows that the separation in this case
is $B \sim \sqrt{Z/\sigma_s}$ and the surface is exactly
neutralized by the adsorbed layer. The three regimes are shown 
schematically in Fig. \ref{chargerevers} which demonstrates that 
overcharging is obtained with multivalent ions only above a certain threshold
and only at intermediate salt concentrations. Overcharging
of charged particles by multivalent counterions is important
for a multitude of applications and can change the sign of the
electrophoretic mobility\cite{Tanaka,Perez} and induce macrophase
separation \cite{attsph9, attsph10}.
Note that the present argument
is analogous to the derivation in \cite{Nguyen}.  Effects such as 
non-linear electrostatics (including counterion release) have
been included in the literature\cite{Shklovrev}.

\section{Polyelectrolytes at  charged spheres}

\begin{figure}[t]
\begin{center}
\resizebox{12cm}{!}{\includegraphics{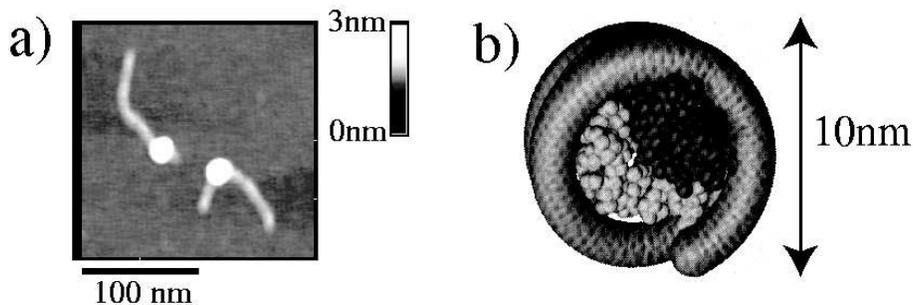}}
\end{center}
\caption{
a) AFM pictures of two complexes consisting of a positively
charged histone protein and a partially wrapped DNA strand of length
130 $nm$. Adapted after \cite{yoshikawa}.
b) View of the three-dimensional structure of the histone protein;
the approximate path of the DNA consists of two full turns
(approximately 146 base pairs) and is indicated by a tube
(adapted after \cite{Alberts}).}
\label{fig9}
\end{figure} 

Although DNA  is a quite stiff biopolymer and thus resists bending,
and although it is highly negatively charged and thus prefers 
coiled and open structures,  the 2m 
of human DNA is packaged into the cell nucleus which only has
a diameter of a few micrometers. 
To make things even worse, the DNA is not just sitting in the nucleus,
but it is all of the time being read, repaired and reshuffled.
The mechanism for the folding is ingenious: The DNA is wrapped around a large number
of small, highly positively charged almost spherical proteins
(called histones), it is thereby partially neutralized and
greatly compactified. Fig.\ref{fig9}a shows an atomic-force-microscopy
picture of two reconstituted complexes consisting of
histone proteins and DNA strands of length 130 $nm$\cite{yoshikawa}. 
The structures
were obtained at a salt concentration of $c_s = 50 mM$. The precise
path of the DNA on the proteins can not be resolved, but from 
the length of the unwrapped DNA portion it can be deduced that
roughly one whole turn of DNA is wrapped around the proteins. 
Fig.\ref{fig9}b shows the structure of the complex as obtained
from X-ray diffraction on crystallized DNA-histone complexes.
Indeed, in the crystalline state (which is not necessarily exactly 
equal to the solubilized state at room temperatures)
 the DNA is wrapped twice around the proteins. A huge body
 of experimental evidence\cite{Yager,Gordon}-\cite{Khrapunov} suggests
that the complex is only stable for intermediate
salt concentrations between $20 mM$ and $500 mM$ (the so-called
physiological salt concentration in the body is roughly $150 mM$
which corresponds to the salt concentration in the sea).
For salt concentrations outside this finite range the complex falls 
apart. As was explained in detail in Section 3.3,
salt modulates the electrostatic interactions;
it follows that an electrostatic mechanism is the cause for this interesting behavior. 

Designing a simple model that can explain the salt and charge dependent behavior of 
the nucleosomal core particle requires a number of approximations concerning
the structure of the DNA, the  histone octamer and their interactions. 
Our strategy was to formulate the simplest possible model which captures
the characteristic features responsible for the salt-dependent nucleosomal
core structures\cite{Netz2,kunzeprl,kunzepre,Hoda1,Hoda2}.
In the following we report calculations of the ground state of a single 
semiflexible polyelectrolyte of finite length which is in contact with 
a single oppositely charged sphere in the presence of 
salt.
Extensive literature exists on similar model calculations
for stiff chains\cite{Marky1,Marky2}, flexible chains\cite{Goeler,Guro,Gole,Mat},
for multiple spheres compexing with one polymer\cite{Schie,Schie2,Ngusph},
for interacting complexes\cite{Pod1,Pod2,Pod3,Dzu} and
on simulation studies for complex formation\cite{Skep,Stoll,Lin1,Lin2,Lin3}.
Polymer form fluctuations away from the ground state
will be considered at the end of this section, where it will
be shown that the ground state
approximation for rather stiff or highly charged polymers
is justified. We specifically
consider parameters appropriate for the DNA-histone system, approximating
the histone as a uniformly charged, impenetrable and solid sphere and the 
DNA as a uniformly charged semiflexible polymer of length 
$L$\cite{Grosbergbook,Hagerman1988}. We 
deal with the so-called nucleosomal core particles consisting of DNA with
146 base pairs (bp) of length $0.34\,nm$ each, leading to a total DNA 
length of $L = 49.6 nm$. The stiffness of the DNA contains charge independent and charge 
dependent contributions. The former are due to the energy associated with 
the deformation of hydrogen bonds which stabilize the double helical geometry of 
DNA and are incorporated by using a semiflexible polymer model with a mechanical
bending stiffness.  The latter stem from the
fact that the negatively charged monomers of the DNA tend to maximize their
mutual distance and thus prefer an extended configuration. This 
electrostatic contribution decreases with increasing salt concentration of the 
solution and vanishes at the hypothetical limit of infinite salt 
concentration\cite{Odijk1,Skolnick}.
The salt independent mechanical persistence length of DNA is therefore
the infinite salt limit of the total persistence length, which has 
experimentally been determined
as $\ell_0 \approx 30\,nm$~\cite{Borochov,Manning81,Sobel1991}. 
A discussion of different methods to 
determine the persistence length of DNA is found in~\cite{Hagerman1988}. 
The electrostatic contribution to the persistence length we take into
account by explicitly including the electrostatic self energy of DNA 
conformations, we therefore accurately include the scale-dependence
of the electrostatic contribution to the persistence length\cite{Barrat}
which is particularly important in our case since the scale of
bending (the histone diameter) becomes of the order or even smaller
than the screening length.

The histone octamer is approximated as a rigid sphere of radius 
$R_{\rm hist}=4\,nm$.
This is of course only a very rough approximation of the 
real structure, which is not a perfect sphere and also possesses a
corrugated surface with specific binding sites for the DNA. Also, any
conformational changes of the histone octamer, which do occur for
extremely low or high salt concentration, are neglected. 
The DNA is modeled as a polymer with radius of $1 nm$. In the actual
calculation we fix the minimal distance between the sphere center and the 
DNA monomer centers to be $R=5 nm$, which is the sum of histone and
DNA radii. 
The electrostatic interactions between charges on the DNA with 
each other and the sphere are described by Debye-H\"uckel (DH) potentials
that neglect non-linear effects (such as counterion condensation 
or counterion release, for which one would need to use the full non-linear 
Poisson-Boltzmann theory\cite{Mann,Bret,Fixman}). 
The main reason for this approximation
is that the calculation of the optimal DNA configuration within the
PB approach is at present numerically not feasible\footnote{Even 
for the relatively simple problem of the adsorption
of a single DNA, modelled by a rigid charged cylinder,
on an oppositely charged plane, the accurate numerical
solution of the PB equation is nontrivial, see
\cite{fleck}. A possible way out could be the recently introduced
field-theoretic non-linear charge-renormalization theory, which takes
non-linear effects at charged polymers
 into account and expresses them in terms
of a DH theory with renormalized polymer charge density\cite{Netzrenorm}.}.
For large salt concentrations, the DH approximation becomes
valid, as has been shown by calculating the electrostatic contribution
to the bending rigidity of a charged cylinder\cite{Bret,Fixman}.

The energy functional for  a given DNA configuration of contour length $L$, 
parameterized by the space curve ${\bf r}(s)$ and in units
of $k_BT$, reads
\begin{eqnarray} \label{freeenergy}
H &  = &   \frac{\ell_0}{2}
\int_0^L {\rm d} s \; \ddot{\bf r}^2(s)
- \frac{\ell_B Z \tau}{1+ \kappa R}
  \int_0^L {\rm d} s \;
\frac{{\rm e}^{- \kappa  (|{\bf r}(s)|-R) } }
{{|\bf r}(s)|} \nonumber \\
& &{}+ \ell_B \tau^2 \int_0^L {\rm d} s \; \int_s^L {\rm d} s' \;
\frac{{\rm e}^{- \kappa  |{\bf r}(s) - {\bf r}(s')|} }
{{|\bf r}(s)- {\bf r}(s')|}
\end{eqnarray} 
where we implicitly assume that the DNA molecule is inextensible,
i.e., $|\dot {\bf r}(s)| =1$.
The first term describes the mechanical bending energy, 
proportional to the bare persistence length  $\ell_0$, where the curvature
$\ddot{\bf r}(s)$ is given by the second derivative of ${\bf r}(s)$ with 
respect to the internal coordinate s.  The second term 
describes the electrostatic attraction between the sphere and the DNA\cite{Verwey}. 
The charge of the sphere in units of the elementary charge e is 
denoted by $Z$ and  the linear charge density of
the DNA (in units of e) is denoted by $\tau$. 
 A key ingredient of the
Debye-H\"uckel-theory is the screening of electrostatic  interactions, which
is quantified by the Debye-H\"uckel screening length $\kappa^{-1}$. 
It measures the distance beyond which the interaction between 
two charges is exponentially damped.
For monovalent salt   one finds
$\kappa^2=8\pi \ell_B c_s$, where $c_s$ is the salt concentration.  
At 0.1 molar concentration in 
water, i.e. at physiological conditions, one has
$\kappa^{-1}\approx 1\,nm$. 
The third term describes the electrostatic repulsion 
 between charges on the DNA. 
We therefore have two terms that tend to straighten the
DNA, namely the mechanic bending energy and
the electrostatic repulsion between DNA monomers. 
The former is salt independent, whereas the latter
looses importance with increasing salt concentration. 
These repulsions 
are balanced by the sphere-DNA electrostatic attraction, which  
favors bending of the DNA in order to wrap it around the sphere, but 
also becomes weaker for increasing salt concentration.   
It transpires that salt will determine the DNA structure in a rather
subtle way, as will be demonstrated by our numerical results.

\begin{figure}[t]
\begin{center}
\resizebox{12cm}{!}{\includegraphics{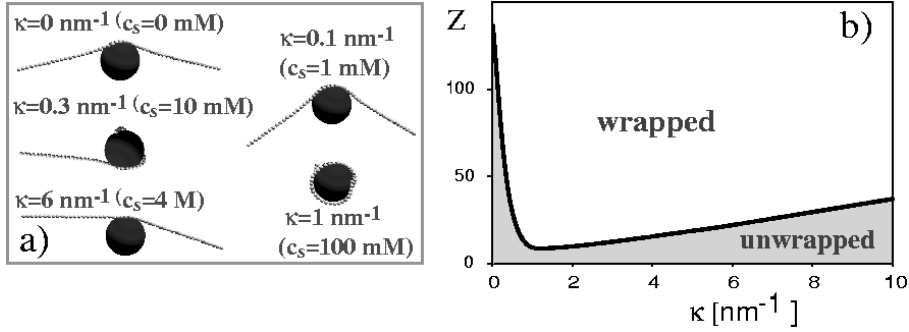}}
\end{center}
\caption{
a) DNA configurations as obtained by numerical minimization
of the free-energy expression Eq.(\ref{freeenergy}) for fixed
sphere charge $Z=25 $ and various salt concentrations. As can be seen,
the DNA is only wrapped for intermediate values of the salt concentration.
b) Global phase digram, featuring the wrapping transition as a function of the
inverse screening length (proportional to the square root of the 
salt concentration, $\kappa \sim c_s^{1/2}$), and sphere charge $Z$. 
}
\label{fig10}
\end{figure} 

In analyzing the model defined by Eq.(\ref{freeenergy}) 
we take advantage of the short length of
the DNA strand, $L= 50\,nm$, 
compared to the effective persistence length, which is at least 
of the order of the bare persistence length, $\ell_0 = 30\,nm$. As a
starting point, we 
therefore neglect fluctuations and undulation forces~\cite{Netz2} between DNA and sphere
and apply a ground state analysis to the model. This consists of finding the 
configuration of the DNA that minimizes Eq. (\ref{freeenergy}) with the constraints
$|\dot{\bf r}(s)|=1$ (no stretching) and $|{\bf r}(s)|\ge R$ (impenetrability of sphere).
We also use  Debye-H\"uckel-potentials with the full
DNA line charge $\tau=2/0.34\, nm$, 
corresponding to the maximal degree of dissociation of DNA (and neglecting
counterion-condensation effects\cite{Mann} which have been investigated
in \cite{kunzepre}). 

In Fig. \ref{fig10} we show a series of DNA configurations obtained
for a fixed sphere charge $Z = 25$ and for inverse screening
lengths ranging from 
$\kappa = 0\, nm^{-1}$ to $6\,nm^{-1}$. 
In the pure Coulomb case with no added salt, the sphere is located 
in the middle of the DNA, and the two arms are slightly bent towards the sphere.
As the salt concentration is increased from $\kappa=0$ to  $\kappa=0.1\,nm^{-1}$,
the deflection of the DNA arms increases continuously. 
Interestingly, the binding of the DNA onto the sphere becomes stronger as
one weakens the electrostatic interactions.
Upon further increase of salt concentration, the two-fold rotational and
the mirror symmetries are broken, see 
the configuration at $\kappa=0.30\,nm^{-1}$. One of the DNA arms is totally wrapped
around the sphere.
As $\kappa$ is increased further, the extended arm is more and more pulled 
onto the sphere until at 
$\kappa=1\,nm^{-1}$ the DNA is fully  adsorbed onto the sphere and the
 two-fold rotational symmetry is restored.
At $\kappa = 6 \, nm^{-1}$ a strongly discontinuous 
transition occurs in which the DNA completely dewraps from the sphere.
The dewrapped state at high salt concentration is markedly different from 
the state at zero salt. There is only one short region of 
nonzero bending of the DNA connecting two basically straight arms.
This sequence of complexation structures demonstrates one of our main results,
namely that the wrapped DNA conformation is only stable for intermediate
salt concentrations, explaining a large set of experimental results for
nucleosomal core particles\cite{Yager}.

We summarize our results for a DNA length of $L=50 nm$ 
in the phase diagram presented in Fig.\ref{fig10}b.
In the absence of salt, $\kappa =0$, the wrapping  transition occurs
at $Z=133$ (in agreement with previous theoretical predictions\cite{Mat}). 
In agreement with experiments~\cite{Yager}, complexation is
most pronounced at intermediate salt concentrations.
For low salt concentration, the strong DNA-DNA repulsion
prevents complexation, for high salt screening weakens 
the DNA-sphere attraction sufficiently so that the mechanical
bending resistance induces dewrapping.
The minimal sphere charge to wrap the DNA,
$Z \approx 10$, is obtained for
$\kappa^{-1} \approx 1\,nm$ ($c_s \approx 0.1\, M$ for monovalent salt), 
corresponding to physiological 
conditions. 
Since the total charge on the DNA is about $300$, the complex is strongly
overcharged for all $Z<300$, i.e., in the whole wrapped region
shown in the phase diagram. The  high-salt
prediction for the wrapping transition can be
obtained analytically by locally balancing the various terms in the energy functional,
Eq. (\ref{freeenergy}), namely the bending energy per unit length,
$H_{\rm bend} \simeq \ell_0/2R^2$,  and the electrostatic attraction
per unit length in the limit $\kappa R > 1$, 
$H_{\rm att} \simeq \ell_B Z \tau/\kappa R^2$, leading to 
$Z_{\rm wrap} \simeq \ell_0 \kappa /2 \ell_B \tau$, in agreement with the
numerical results\cite{Netz2}.

These results highlight a peculiarity of electrostatic complexation phenomena,
and is mirrored by an at first sight perplexing approximation used in our
model calculation: We do use the Debye-H\"uckel approximation for the interaction
between charges on the sphere and on the DNA, which amounts to taking into account
positional fluctuations of the salt ions within a Gaussian approximation\cite{Netz-1}.
However, we do not take into account positional fluctuations of the DNA itself,
but concentrate on the ground state instead. The reason for the different
treatment of salt ions and DNA monomers is that the total amount of charge
per statistically independent unit is $q=1$ for monovalent ions but 
roughly $q\simeq 180$ for one persistent segment of DNA of length $L \simeq \ell_0$.
Therefore fluctuations are rather unimportant for the DNA structure
(except for very large salt concentrations where a desorption transition does
occur which can be treated using similar methods as used for the desorption
of polyelectrolytes on planar substrates in the preceding section\cite{Netz2})
but are of extreme importance for the counterion clouds. 
To make these statements quantitative, we estimate now the
conformational entropy which 
can be obtained by a normal-mode analysis  of the
chain fluctuations around the ground state configuration. 

\begin{figure}[t]
\begin{center}
\resizebox{12cm}{!}{\includegraphics{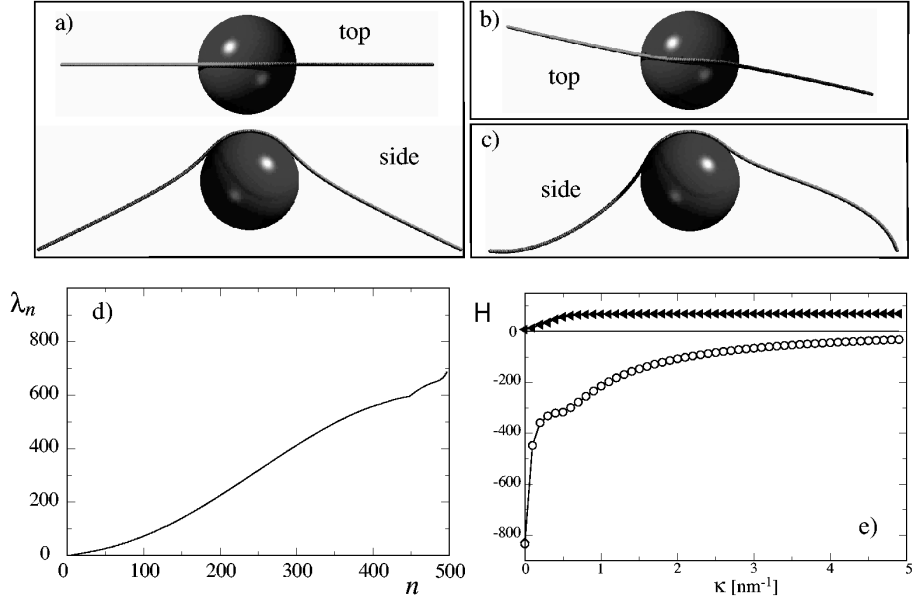}}
\end{center}
\caption{a) Ground-state configuration of the complex for a sphere charge $Z=40$
and zero salt concentration $\kappa =0$. b) Second and c) sixth excited state.
The second eigen mode  breaks the mirror symmetry but leaves the two-fold rotational
symmetry intact, while the sixth eigen mode breaks the two-fold rotational symmetry.
d) Eigenvalue spectrum for  $Z=20$ and $\kappa =0$. e) Ground state
energy (open symbols) and Gaussian fluctuation contribution (filled symbols)
for charge $Z=20$ as a function of the inverse screening length $\kappa$. 
The fluctuation contribution is obtained by integrating out the 
complete eigenmode spectrum shown in d). }
\label{excited}
\end{figure} 

Let us assume that the ground state configuration of the chain is given by the
space curve ${\bf r}_0(s)$ which minimizes the energy functional
$H$ in Eq.(\ref{freeenergy})
according to $\delta H /  \delta {\bf r}(s) = 0$. 
In the actual calculation the space is discretized
and the actual degrees of freedom are bond angles,
 but this is not important for the present presentation. 
For small fluctuations around the ground state, the effective Hamiltonian
of the complex  may be expanded around the minimum up to second order as 
\begin{equation}
H = H[{\bf r}_0] + \frac{1}{2} \int ds ds' [{\bf r}(s) - {\bf r}_0(s)]
H^{(2)}(s,s') [{\bf r}(s') - {\bf r}_0(s')]
\end{equation}
where 
the Hessian matrix associated with the effective Hamiltonian, is defined as 
 \begin{equation}
   H^{(2)}(s,s') = \bigg(\frac{ \delta ^2 H} {\delta {\bf r}(s)  \delta {\bf r}(s')  }\bigg)_{{\bf r}={\bf r}_0}
 \end{equation} 
It characterizes the spectrum of chain excitations within the harmonic (or Gaussian) approximation. 
 The normal modes of a complex are obtained by diagonalizing the Hessian,
 using the discrete notation 
  ${H}^{(2)}_{mn} A_{np} = \Lambda_{mn} A_{np}  $, which is solved numerically for
the matrix of eigen-modes, $A_{np}$, and  
the diagonal matrix of eigen-values, $ \Lambda_{mn} = \lambda_n \delta_{mn}$.

In Figure \ref{excited}, we show in a) the ground state and in b) and c)
the second and the sixth  excited states of the complex, respectively\cite{Hodatobe}.
The parameter values chosen are  $Z = 40$ and $\kappa = 0$. 
As seen, the ground state's mirror symmetry is broken in the second excited state
and  the two-fold rotational 
symmetry is broken in the sixth excited state.
The latter eigen-mode can be viewed as a translational motion of the
 bound sphere along the PE chain. 
In Fig. \ref{excited}d we show the mode spectrum obtained from the Hessian
via diagonalization. We discretize the chain by 250 beads, we therefore
have a total of 500 eigenmodes. From the fluctuation spectrum the 
configurational entropy can be obtained since the configurational 
integral can be performed on the Gaussian level exactly
in tems of the normal modes\cite{Hodatobe}. 
In Fig. \ref{excited}e we show the ground state energy of the 
complex (open symbols) and the configurational entropy contribution
(filled symbols) as a function of $\kappa$. 
For both contributions the reference state of a free polymer has been subtracted.
The entropy is positive, as it costs configurational freedom to bind
a polymer onto a sphere, the ground-state energy is negative as sphere
and polymer do attract. When the two contribution have equal magnitude
(roughly at $\kappa \approx 3 nm^{-1}$)  the total free energy gain upon complexation
is zero.  This signals an entropy-driven
unbinding transition which of course depends on the solution concentration of
DNA strands and histone spheres via the law-of-mass action\cite{Hodatobe}.
 It follows that in some regions of the phase diagram shown
in Fig.\ref{fig10} the complex  is dissolved into its constituting
parts, depending on the bulk concentrations.

\section{Polyelectrolytes at  charged cylinders}

When semiflexible charged polymers are mixed with much stiffer oppositely
charged polymers, a complex forms where the more flexible polyelectrolyte
(PE) wraps around the stiff 
polymer\cite{Odijk2,Park,kunzeepl,Nguyen,Messinacyl,Winklercyl,UllnerLinse}.
Experimentally, such complexes itself fold up into toroidal or stem-like
structures under dilute conditions\cite{Laemmli,Golan};
in more concentrated solutions, bundles and networks are observed\cite{Jason}. 
In this section we try to understand 
the morphology of the underlying molecular complex, 
namely the conformation of the wrapping polymer: does it form a helix
or does it adsorb in (one or more)  parallel straight strands onto
the cylinder?
Using linear (Debye-H\"uckel) theory, supplemented by nonlinear (counterion 
release) arguments\cite{Park}, we find transitions between
both morphologies as the salt concentration (or other parameters) are varied.

Most of the current interest in such complexes comes from their potential 
applications in gene therapy:
The main problem here is to introduce genetic material into patients' 
cell nuclei, a process called DNA transfection.
The classical viral strategies are highly effective in transfecting DNA but 
may provoke immune reactions of the body, switch back
to their lethal origin or lead to a stable transformation 
of target cells (advantageous in some cases though in general 
undesirable)\cite{Felgner,Kabanov}.        
Nonviral transfection strategies avoid these difficulties at the price 
of much reduced effectiveness\cite{Kabanov}. Still, they
hold promising potential for further development and refinement. 
The polyfection scheme
consists of complexing DNA with physiologically tolerated polycations,
such as  polypeptides or synthetic polycations~\cite{Kabanov,Pouton}
and shows relatively high efficiency,  especially 
with confluent (non-dividing) cells\cite{Zauner}.
As a major advantage, the  
properties of these self-assembling {\em polyplexes}  can be controlled rather 
reliably by for example varying the mass or charge density of the 
polycations\cite{Erbacher}, by using block-copolymers with a cationic 
block and an uncharged block which forms some type of protection layer 
against coagulation or degradation\cite{Dash}, and finally by linking 
target-specific ligands to the polymer chains\cite{Wagner}.

The microscopic structure of polyplexes is not very well understood.
Electron micrographs of DNA-polylysine complexes exhibit
highly condensed torus or stemlike 
structures\cite{Laemmli}, very similar to what is seen with DNA condensed 
by multivalent counterions\cite{Bloomfield}.
More recent AFM studies demonstrated that condensation involves five to sixfold 
overcharging of the DNA by peptide charges at elevated salt 
concentration\cite{Golan}. 
The underlying 
molecular structure of polycation-DNA complexes (toroidal or stemlike), 
which  involve multiply packed DNA loops,   is not resolved in these 
experiments. X-ray diffraction measurements, on the other hand,
 showed that polylysine wraps helically around the DNA 
molecule (and at low salt concentrations neutralizes the DNA charges),
while polyarginine, a cationic polypeptide with a different backbone 
flexibility, shows a different wrapping mode\cite{Suwalsky}. 
Similar complexation is obtained by mixing DNA with a rather bulky 
cationic dendrimeric polymer: AFM pictures demonstrate that in this case
the DNA wraps around the dendro-polymer\cite{goessl}. This essay, with
potential gene-therapeutical applications, holds the advantage that the 
physical properties of the complex and the effects of various parameters
can be studied in great detail and with comparative ease.
From all the above listed experiments, it is clear that the
salt concentration of the surrounding medium, the charge of the complexing
polycations, and their flexibility  can induce different morphologies 
of the polyplex. 
The possibility and mechanism of
DNA-overcharging by adsorbed  polycations
is interesting from a fundamental point of 
view\cite{Park,Mat,Nguyen,Netz2}, although 
it has been shown that optimal transfection yield is obtained with neutral 
complexes\cite{Erbacher,Wagner} (despite the naive expectation
that cationic complexes would interact more favorably with the
typically negatively charged cell and endosomal walls).

\begin{figure}[t]
\begin{center}
\resizebox{12cm}{!}{\includegraphics{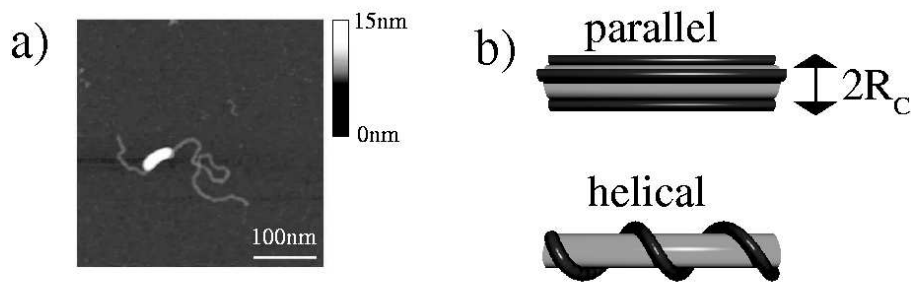}}
\end{center}
\caption{
a) AFM pictures of a DNA strand of lengh $L_P=850 nm$ which is partially wrapped
around a $L_C=50nm$ long cationic dendropolymer (reproduced from Ref.\cite{goessl}).
b) The two posibble wrapping morphologies that are theoretically studied
and compared.
}
\label{fig11}
\end{figure} 

In Fig.\ref{fig11}a a DNA-dendropolymer complex is shown which consists of a DNA strand of length 
$L_P\approx 850 nm$  wrapped around a cylindrical dendropolymer of length 
$L_C \approx 50 nm$ and radius $R_C \approx 3 nm$. The DNA is complexed as  a single
strand with the dendropolymer, from the wrapped DNA length (which can be determined
by measuring the length of the non-wrapped DNA sections) it follows that the DNA
almost fully covers the dendropolymer surface. Since no loops are seen that emerge
from the complex, it is suggested that the DNA wraps helically around the core.
In fact, more DNA wraps than is needed to neutralize the dendropolymer core, 
the precise amount of wrapped DNA turns out to be salt dependent\cite{goessl2}. 

In this section we analyze the complexation between 
a charged rigid cylinder and an oppositely charged semiflexible polyelectrolyte. 
The parameters in our model are the linear charge densities of the
uniformly charged 
cylinder and the PE, $\tau_{C}$ and $\tau_P$, the cylinder radius $R_C$
and the bare PE persistence length $\ell_0$. 
On the linear level the interaction between all charges
is given by the bulk Debye-H\"uckel (DH) potential
$v_{DH}(r)=\ell_B \exp(-\kappa r)/r$
where $r$ is the distance between charges,
   $\ell_B$ is the Bjerrum length, $\ell_B=e^2/\!4 \pi \varepsilon 
\varepsilon_0\,k_BT$,  and $\kappa^{-1}$ is the screening length. 
The DH approximation is valid  for weakly charged PEs, elevated
salt concentrations and, as is explained in Ref.\cite{kunzeepl},
for complexes close to electroneutrality. Effects due to 
dielectric boundaries, additional non-electrostatic interactions,
inhomogeneous charge distributions on the cylinder and on the PE,
polymer confinement (which are all neglected)
and counterion release are of only secondary
importance for the resulting phase diagrams, since we always compare
different morphologies of roughly the same amount of adsorbed PE. 
It is the free-energy difference between different morphologies
that we are most interested in, not their absolute values\cite{kunzeepl}. 
The thermodynamic ensemble we consider is the one where PE is present
in excess, i.e., we minimize the free energy per cylinder unit length,
treating the non-adsorbed PE as a reference state the electrostatic
self energy of which therefore has to be subtracted.
We also neglect end effects which will only be important if the screening
length becomes larger than the cylinder length.
This is the ensemble that is indeed relevant to describe the experimental
situation in Fig.\ref{fig11}a.
The thermodynamic  ensemble considered in Ref.\cite{Winklercyl} is different
since the cylinder self-energy was not subtracted; it  consequently gives results
very different from our scheme.
We compare two morphologies, namely a helical arrangement of the PE, 
where a single helix,
characterized by the length ratio of the wrapped polymer section
and the cylinder length, wraps around the 
cylinder (see Fig.\ref{fig11}b), and the straight morphology, where $n$ 
parallel strands of PE adsorb on the cylinder. We minimize both
configurational energies with respect to the relative amount of
wrapped polyelectrolyte
(neglecting configurational fluctuations around the ground state
which are unimportant for rather stiff and highly charged PEs)
and compare the two resulting free energies to determine the stable phase.
This comparison, which is done numerically in the general case, shows
that both morphologies compete closely with each other. 

\begin{figure}[t]
\begin{center}
\resizebox{12cm}{!}{
\includegraphics{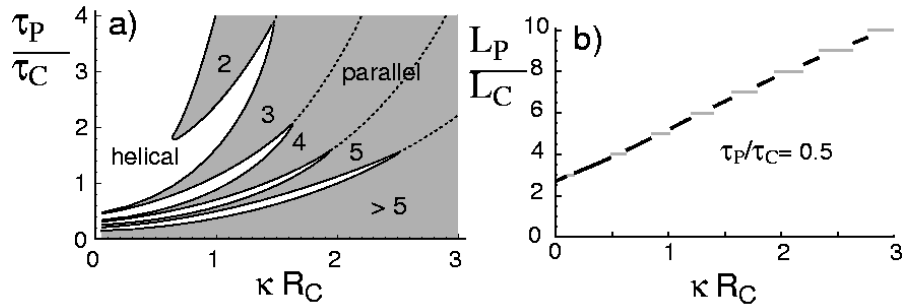}
}
\end{center}
\caption{
a) Phase diagram of the polyelectrolyte-cylinder complex 
as a function of the linear-charge density ratio $\tau_P/\tau_C$ 
and the inverse rescaled screening length $\kappa R_C$ for vanishing
 persistence length of the wrapping polymer, $\ell_0=0$.
The helical phase (white) dominates at low salt concentration
(small $\kappa R_C$) while the straight configuration (shaded; number 
of adsorbed PE strands is indicated and direct
transitions between states with different numbers
of adsorbed strands are denoted by broken lines) is realized at large 
$\kappa R_C$.
b) Relative amount of wrapped PE, $L_P/L_C$,
for zero persistence length $\ell_0=0$ for a line
charge density ratio of $\tau_P/\tau_C=0.5$, 
corresponding to a horizontal cut
through the phase diagram Fig.\ref{fig12}a. In the helical phase the 
wrapping parameter (and thus the overcharging of the cylinder)
continuously increases as $\kappa R_C$ grows, while the
straight configuration is characterized by integer values.
}
\label{fig12}
\end{figure}

As can be seen in the phase diagram Fig.\ref{fig12}a, which is obtained in the limit
when the wrapping polymer is totally flexible and has no bending stiffness,
$\ell_0 =0$, the helical phase is favored at low salt concentrations (to the left)
and highly charged wrapping polymers ($\tau_P/\tau_C \gg 1$), while the
parallel morphology is favored at high salt concentrations. 
As the charge density of the wrapping polymer increases, as one moves up in the
phase diagram, the number of adsorbed strands in the parallel phase goes down. 
Fig.\ref{fig12}b shows for the specific charge density ratio $\tau_P/\tau_C=0.5$ that
the amount of wrapped polymer, characterized by the ratio of the contour
length of the wrapped polymer and cylinder length, $L_P/L_C$, 
grows with increasing salt concentration (the desorption transition, which is 
expected to occur at high salt concentrations in the absence of additional
non-electrostatic attractive forces is not shown but follows
the same rules as outlined in  Section 5). For line charge 
ratio $\tau_P/\tau_C=0.5$ the complex would be neutral for a wrapping 
ratio  $L_P/L_C=2$. As a matter of fact, more polymer wraps around
the core cylinder than is needed to actually neutralize the complex
(in agreement with the experimental results\cite{goessl}). As the salt concentration
increases the overcharging is even further enhanced, also in agreement
with experiments\cite{goessl2}.

The overcharging in the low-salt limit is easily understood analytically.
For simplicity we consider the parallel morphology with $n$ adsorbed
polymers with line charge density $\tau_P$ at a cylinder of line
charge density $\tau_C$. In the limit of low salt, $\kappa R_C \rightarrow 0$,
the electrostatic potential of a charged cylinder, 
which is derived  in Eq.(\ref{Vline}) on the Debye-H\"uckel level, 
shows a logarithmic behavior. The potential at the
cylinder surface is  given by  $-2 \ell_B \tau_C \ln(\kappa R_C)$. The total attractive
free energy of $n$ adsorbed polyelectrolytes is thus in the low-salt limit
per unit length given by
\begin{equation}
F_{\rm att} \simeq 2 n \ell_B \tau_C \tau_P \ln(\kappa R_C).
\end{equation}
Between the $n$ adsorbed polymers there are $n(n-1)/2$ repulsive pair interactions,
all of the same logarithmic type. Clearly, the distances between the various
pairs are all different, but since the repulsion is logarithmic these differences
give negligible additive contributions to the resulting total repulsive free energy, 
which can be written as
\begin{equation}
F_{\rm rep} \simeq - n(n-1) \ell_B \tau_P^2 \ln(\kappa R_C).
\end{equation}
The number of adsorbed polymers results from minimization of the sum 
of the repulsive and attractive contribution, 
$\partial (F_{\rm att}+F_{\rm rep})/\partial n =0$, and is given by
\begin{equation} \label{pred}
n = L_P/L_C = 1/2 + \tau_C/\tau_P.
\end{equation}
It equals the ratio of the wrapped polymer length and the cylinder length,
$L_P/L_C$, which is the quantitiy that
is plotted in Fig.\ref{fig12}b. The number of adsorbed strands $n$ is an integer
quantity. However, the analogous calculation for the helical phase
in the low-salt limit gives the same result as Eq.(\ref{pred}).
Both phases turn out to be degenerate for integer values of the
wrapping ratio $L_P/L_C$. The result Eq.(\ref{pred})
is in agreement with the numerical data displayed
in Fig.\ref{fig12}b, and predicts for the case $\tau_P/\tau_C=0.5$
the wrapping length ratio $L_P/L_C=5/2$  in the 
zero salt limit $\kappa R_C \rightarrow 0$ (note that the asymptotic
approach of this limit is logarithmically slow, see \cite{kunzeepl}).
The effective charge of the complex is in the same low-salt limit 
from Eq.(\ref{pred}) predicted to be
\begin{equation} \label{pred2}
\tau_{\rm eff} = n \tau_P - \tau_C = \tau_P/2
\end{equation}
and was also obtained using an alternative approach\cite{Nguyenfrac}. 
This result shows that in the low-salt limit the complex will have 
the same charge sign as the wrapping polymer, the usual wording for this
is that the complex is overcharged. The effective charge density of the complex
amounts to half the one of the wrapping polymer.
Therefore, if the negatively charged DNA wraps around a cationic 
dendropolymer, the complex will have a net negative charge, 
if however a flexible cationic polypeptide wraps around the negatively
charged DNA, the resulting complex will be positively charged. 
These qualitative trends are in agreement with experiments, and they
show how to tune the charge of a polyelectrolyte complex by changing
the ratio of the bending rigidities of the cationic and anionic polymers
involved in forming the complex.

\section{Polyelectrolytes in electric fields}

The behavior of flexible polyelectrolytes (PE) exhibits a number of remarkable
features which are due to the electrostatic coupling between 
polymeric and counterion degrees of freedom. 
Noteworthy is the sequence of PE conformations which is observed in
simulations as the electrostatic coupling between the charges on the 
PE and the counterions is increased\cite{Stevens,Winkler,Khan3,Liu}. 
Experimentally, the coupling can be tuned by changing temperature,
dielectric constant of the solvent, counterion valency/size and
charge density of the PE. For very small coupling the PE resembles
a neutral polymer since the electrostatic repulsion between monomers is
very small. As the coupling increases, the monomer-monomer repulsion
leads to a more swollen configuration (the standard PE effect).
However, as the coupling further increases, counterions condense
on the PE, decrease the repulsion between monomers and the PE 
starts to shrink. Finally, at very large electrostatic coupling,
the PE is collapsed to a close-packed, almost charge-neutral 
condensate which contains most of its counterions. A similar sequence is
experimentally seen with synthetic PEs\cite{Cruz} and
DNA\cite{Bloomfield2,Raspaud,Yamasaki}.
\begin{figure}[t]
\begin{center}
\resizebox{14cm}{!}{
\includegraphics{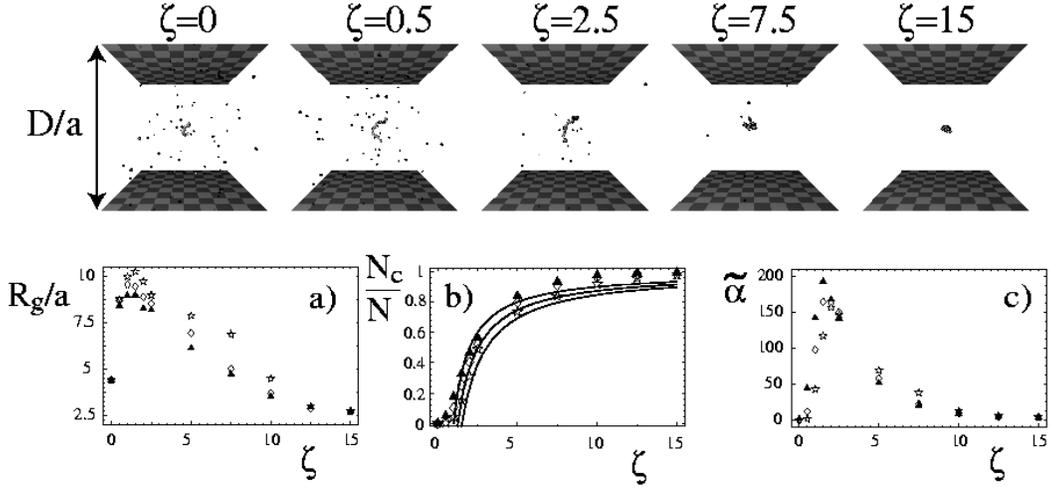}
}
\end{center}
\caption{ Simulation snapshots of a PE with $N=50$ monomers 
in a cubic box of diameter $D/a=100$ 
for various values of the coupling parameter $\zeta$.
a) Radius of gyration $R_g/a$, 
b) average rescaled number of condensed counterions $N_c/N$, 
and c) polarizability $\tilde{\alpha}=k_BT \alpha /(q e a)^2$ 
for a PE of monomer number  $N=50$ and box sizes
$D/a=200$ (stars), $D/a=100$ (diamonds), 
$D/a=50$ (filled triangles).   }
\label{fig13}
\end{figure} 
This well-known behavior is visualized in
Fig.\ref{fig13} where we show in the top panel a few snapshots
of a Brownian dynamics simulation for
a PE chain with $N=50$ monomers
for different values of the Coulomb parameter
\begin{equation}
\zeta = q^2 \ell_B /a
\end{equation}
where $a$ is the diameter of monomers and counterions (which are both
modelled as hard spheres) and
$q$ is the valence of monomers and counterions (which are the same).
The PE behavior is best quantified by
the  radius of gyration  $R_g$, defined as 
\begin{equation} \label{Rg}
\tilde{R}_g^2 = \frac{1}{2N^2} \sum_{i,j,=1}^N 
\langle ({r}_i-{r}_j)^2 \rangle
=\frac{1}{N} \sum_{i=1}^N 
\langle ({r}_i-{R}_{com})^2 \rangle
\end{equation}
where the sum includes PE monomers only and
${R}_{com}$ denotes the center of mass in rescaled 
coordinates defined as
\begin{equation}
{R}_{com} 
=\frac{1}{N} \sum_{i=1}^N 
{r}_i .
\end{equation}
The rescaled radius of gyration
in Fig.\ref{fig13}a
shows for the various box sizes used a maximum at $\zeta \approx 3/2$.
In the simulations, the PE and all counterions are confined in a cubic box of width $D$.
There is a systematic trend in the data showing that the radius of gyration
is larger for larger box sizes. This can be understood by studying the 
degree of counterion condensation.
The number of condensed counterions (rescaled by the total
number of counterions), $N_c/N$, is shown in Fig.\ref{fig13}b 
and depends weakly on the box size $D$: The 
bigger the box the smaller the number of condensed counterions.
We rather arbitrarily define a counterion as condensed when 
its center is closer than $2a$ to any monomer center, i.e. when 
there is at least  one monomer
closer than two times the diameter (we checked that our results depend
only very weakly on the precise distance chosen to discriminate
between condensed and uncondensed counterions). Since the 
number of condensed counterions goes down with increasing box size,
it is fairly easy to understand that the effective PE repulsion goes up
and thus the radius of gyration increases. The solid lines in Fig.\ref{fig13}b
denote the standard Manning prediction for the number of condensed ions\cite{Mann}
corrected by the finite length of the PEs\cite{Netztobe}.

In Fig.\ref{fig13}c the polarizability according to the fluctuation-dissipation 
theorem, 
\begin{equation}
\tilde{\alpha} = \frac{ k_BT \alpha }{(q e a)^2}= 
\frac{ \langle P^2 \rangle }{3(q e a)^2}
\end{equation}
is shown. 
The dipole moment of the complex is denoted by $P$  and is measured
with respect to the PE center of mass.
The resultant polarizability shows a strong $\zeta$ dependence
and qualitatively follows the trend of the radius of gyration shown 
in Fig.\ref{fig13}a. 
The classical result for the polarizability 
of a sphere with radius $R$
and uniformly distributed charge $Q$ around an 
opposite point charge $Q'$ is for $Q=Q'$ given by $\alpha = 4 \pi 
\epsilon R^3$\cite{Boettcher} or, in rescaled units with $\tilde{R} = R/a$, 
$\tilde{\alpha}  = \tilde{R}^3 /\zeta$. 
Identical results are obtained
from the Clausius-Mossotti equation\cite{Mossotti} or using different, more complicated
charge distributions. In fact, by comparing the 
radius of gyration and the polarizability,
one can show that these simple relations also hold for the collapsed
PE chain.

\begin{figure}[t]
\begin{center}
\resizebox{14.cm}{!}{\includegraphics{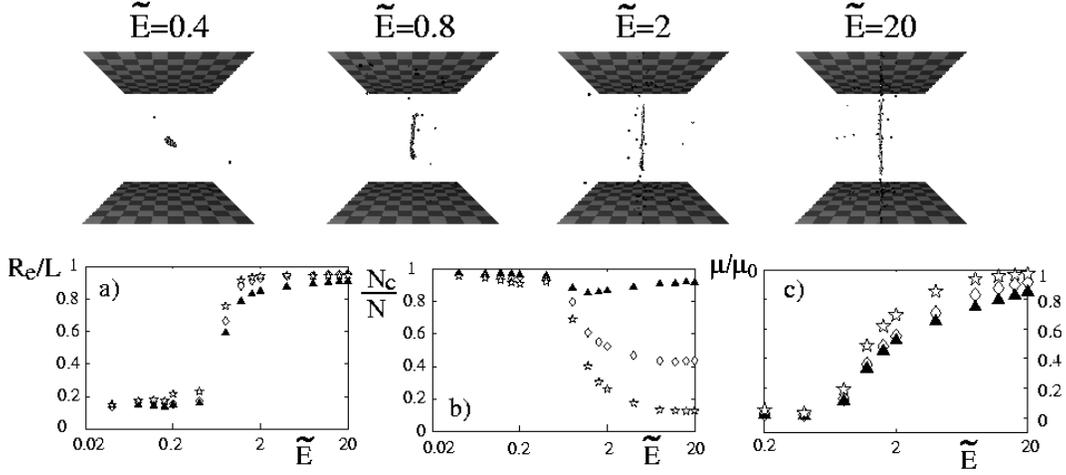}}
\end{center}
\vspace*{-.5cm}
\caption{ PE snapshots for fixed Coulomb parameter
$\zeta=10$ and box size $D/a=100$ and various rescaled
field strengths, exhibiting an unfolding transition at 
$\tilde{E}  = qea E/k_BT \approx 0.4$. a) Rescaled end-to-end radius $R_e/L$,
where $L = a N$,
b) number of condensed counterions $N_c/N$, and
c) relative mobility of PE monomers
for box sizes
$D/a=200$ (stars), $D/a=100$ (diamonds), 
$D/a=50$ (filled triangles) for $\zeta=10$.
 } 
\label{fig14}
\end{figure}

All these phenomena concern the static, equilibrium
behavior of PEs. In electrophoretic experiments, PEs are subject
to external electric fields and the resulting mobility is 
measured\cite{Stellwagen,Viovy}. 
Such techniques are widely used to separate DNA and charged proteins
according to their molecular weight. In these situations, the electric
field induces motion of ions and PEs, thus dissipation of energy,
and one is facing a non-equilibrium problem. 
In the following, we briefly discuss the effects of electric
fields on PE condensates. 
In contrast to previous theories, where the counterions are not taken 
into account explicitly or their coupling
to the PE is rather weak\cite{Sem,Mann2,Long}, we start from a strongly
coupled (collapsed) PE-counterion system
and investigate the resultant effects for large electric fields 
(i.e. far from equilibrium).
We choose a Coulomb coupling of $\zeta=10$.
Fig.\ref{fig14} shows a few snapshots for increasing field strength,
exhibiting an unfolding transition of the PE condensate at a critical
field strength. 
The non-equilibrium unfolding transition manifests itself as a
rather abrupt increase of the  rescaled end-to-end 
radius $R_e /L$, which
in Fig.\ref{fig14}a is shown  for $\zeta=10$ and
various box sizes as a function of the rescaled applied field
$\tilde{ E} = qea { E} /k_BT$. The contour length
of the PE is denoted by $L$.
The number of condensed counterions 
in the high-field extended configuration exhibits a dramatic dependence
on the box size, see Fig. \ref{fig14}b. It approximately
equals the ratio of polymer length and box size, $N_c/N \approx
L/D$,
since the counterions in the large-electric-field limit
are distributed almost evenly along the electric-field direction.
In the absence of interactions between PE monomers and counterions,
or in the limit of infinite dilution, the electrophoretic mobility
$\mu$ (which is equivalent to the conductivity)
equals the bare mobility $\mu_0$
for all charged particles. In Fig. \ref{fig14}c we show the PE monomer mobility
for different box sizes as a function of the external field. For small
fields the mobility is almost zero, i.e., the condensed counterions
slow down the PE considerably. As the field strength increases,
the rescaled mobility $\mu/\mu_0$
slowly approaches unity. This is an extreme example of the Wien effect,
which was originally observed for simple electrolyte solutions. 
It transpires that  the  non-equilibrium
effect of strong external driving fields can, together with strong
electrostatic interactions, lead to qualitatively new features such as 
field-driven conformational transitions and complex dissolution.
At a critical field strength, a collapsed PE unfolds and 
orients in the direction of the field. 
Since the PE mobility is expected to change drastically at the
unfolding transition, this transition should be detectable by 
mobility measurements and in turn could be used for efficient separation
of PEs of different length\cite{Netztobe}.

\begin{figure}[t]
\begin{center}
\resizebox{6.cm}{!}{\includegraphics{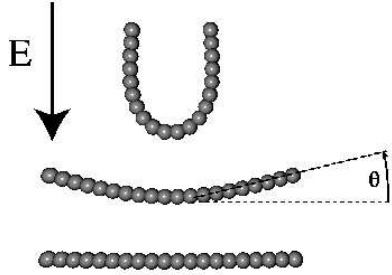}}
\end{center}
\vspace*{-.5cm}
\caption{ Snapshots for an elastic, semiflexible rod,
characterized by a bare persistence length $\ell_0$,
that is driven down in a quiescent, viscous liquid
by a total force $F$ that is acting equally on 
all monomers.
The stationary shapes are obtained for 
rescaled driving forces of 
$L^2 F /( \ell_0 k_BT )$ = 760, 76,7.6,
from top to bottom. 
Hydrodynamic effects lead to a deformation and,
consequently, to an orientation of the rod.} 
\label{rods}
\end{figure}

So far we neglected hydrodynamic effects, that means that the 
flow can freely penetrate the polymer and no hydrodynamic interactions
occur when objects move past each other. Clearly, 
one would expect a whole 
number of interesting kinetic effects to turn up when polymers
move in strong fields in viscous fluids. 
Probably the most basic effect concerns a hydrodynamic
mechanism for the orientation of polymers as they move through
quiescent liquids. The arguments presented above show that
charged polymers align along their long axis along an electric field.
What happens when hydrodynamics are taken into account?

For orthotropic bodies, that is for bodies with three mutually orthogonal planes of symmetry
such as cylinders, the hydrodynamic coupling term between translational
and rotational motion vanishes\cite{Happel}.
This result is valid only in the hydrodynamic
limit of small Reynolds numbers (i.e. small bodies and low velocities), which is the 
relevant regime for almost all viscous solvent effects of nanoscopic materials.
It implies that a {\em rigid } cylindrical particle, such as a rod-like synthetic or biological
polymer, that is  driven by a homogenous external force (be it 
gravitational or electric in the case when the particle is charged)
and as a result is moving through a quiescent fluid, is hydrodynamically 
not oriented in any particular
way: the orientational distribution function will be uniform, not favoring 
parallel or perpendicular orientation with respect to the direction of motion.
In contrast, for {\em elastic} rods
hydrodynamic effects lead to a 
bending  and therefore reduction in  symmetry\cite{Elving,Mazur,xaver}.
The hydrodynamic translational-rotational coupling becomes finite and 
in consequence sedimenting rods orient perpendicularly to 
the direction of motion. 
This is graphically illustrated in Fig.\ref{rods} where we show 
stationary shapes of a semiflexible polymer that
is moving downwards under the action of a force acting
uniformly on the monomers. The structures are obtained with a simulation code
that takes hydrodynamics into account on the Stokes or creeping-flow level,
including the effects of elasticity and thermal fluctuations\cite{xaver}.
The relevant parameters are the mechanical bending
rigidity, expressed in terms of the bare persistence length $\ell_0$, the total
force acting on the rod, $F$, 
related to an electric field $E$ via $F = q e E N$,
and the length $L$. The structures depicted
in Fig.\ref{rods} are obtained for 
rescaled driving forces of 
$L^2 F /( \ell_0 k_BT )$ = 760, 76,7.6,
from top to bottom. 
It is seen that the rod is deformed, but also that it is oriented with respect
to the direction of motion.
The mechanism is quite simple to understand:
The external  force drives all monomers
in the same way; due to hydrodynamic interactions, the effective force pushing
the rod is larger
in the  middle of the rod than at the two ends, because the middle receives
hydrodynamic thrust from both sides of neighboring segments. 
This imbalance in 
driving thrust is balanced by an elastic {\em deformation}, the rod bends.
The rod bending reduces the symmetry and
hydrodynamically couples translational and rotational degrees of
freedom. The shift between the center of force (i.e. mass) and 
hydrodynamic stress produces an orienting torque, 
and as a result the bent rod is oriented perpendicularly to the direction of motion
with the opening pointing backwards. 
Since all materials have  finite elastic moduli,
this hydrodynamic orientation mechanism is universal and 
should be directly
observable for sedimenting rods  in an ultracentrifuge\cite{Buiten}.
In an ultracentrifuge, one can infer the change in average orientation
since the mobility of a perpendicular long rod is smaller by a factor
of two as compared with the mobility of a parallel rod. 
A variation in orientation can be induced by either changing the rod length
or the driving field strength, the complete scaling functions (including
numerically determined prefactors) are reported in \cite{xaver}.

The average orientation of a variety of charged rod-like particles
(such as Tobacco-Mosaic\cite{Konski} and FD viruses\cite{Kramer}, 
or different synthetic polyelectrolytes\cite{Opper}) in electric
fields has been determined in birefringence experiments.
The anisotropic electric polarizability  favors
an orientation with  the direction of the maximal  polarizability
parallel to the electric field, as seen in Fig.\ref{fig14}. The largest contribution
to the polarizability is furnished by the easily deformable counterion
cloud accompanying each charged particles, which is maximal 
along the long axis of the particle\cite{Yoshida,Netztobe}. 
As a result, charged rod-like
particles are (at not too low fields) oriented in parallel with the electric
field. This is called the {\em normal birefringence} of charged rods. 
{\em Anomalous birefringence}, meaning perpendicular orientation
of rods, is typically obtained for long particles, 
low salt concentration or particle
concentrations beyond mutual overlap\cite{Opper,Kramer,Hoffmann}
and at present only partially understood\cite{Cates}.
It seems likely  that  the anomalous electric
birefringence of charged polymers is caused by  the above-mentioned 
hydrodynamic orientation in cases  when the typically
much stronger electric polarizability orientation is weakened 
due to the overlap or evaporation of counterion clouds. 

\section{Charge regulation}

So far we treated  surface charges as fixed and invariable  and only 
considered Coulomb interactions (and possibly excluded-volume interactions)
between charged groups and counterions. 
In an aqueous environment, all chemical groups are interacting
chemically with each other, and in specific, there is a certain binding
energy that is released when e.g. a proton is binding to an acidic rest
and making it charge neutral {\em which goes beyond the Coulomb potential}.
We will deal with the microscopics of these binding forces in Section 11
where we will in fact determine binding energies using quantum-mechanical methods. 
For the present consideration we shall assume that an equilibrium reaction constant
exists which controls the reaction between the dissociated (charged) state and
the associated (uncharged) state of a chemical surface group. 
The chemical equilibrium between the charged and uncharged versions
of the surface groups can be tuned by the $pH$ of the solution, which is a measure
of the bulk concentration of protons.
Strong acids are typically fully charged
whereas weak acids are only partially charged at normal conditions 
($pH \approx 7$). Electrostatic repulsion between neighboring charged groups
tends to decrease the effective charge of an object, since the charge repulsion
acts like a chemical potential favoring association. Another
way of looking at this is to realize that the counterion concentration 
in the vicinity of a surface group increases when there are other charged surface groups
close by; this concentration increase of  counterions (among them protons)
drives the dissociation reaction backwards.
This repulsion effect is stronger at low salt concentrations (i.e. for
long-ranged electrostatic interactions)\cite{Katchalsky}. 
The situation is more complicated at dielectric boundaries\cite{netzregulate} 
or when  charged macroions interact with each other\cite{Bohmer,yoram},
since here the charge on each group is interacting with its immediate neighbors
but also with  image charges and charged groups on macroions in the vicinity.
In this Section we explicitly consider a charged polymer, where charged groups
are arrayed on a line, and a charged surface which
consists of an ordered two-dimensional array of dissociable surface groups.
In order to treat the effects of added salt on a manageable level, we use
screened Debye-H\"uckel (DH) interactions between all charges,
$v_{\rm DH}( r) = \ell_B {\rm e}^{- \kappa r}/ r$.

A surface group, which in all that follows is assumed to be an acid,
can be either charged (dissociated)
or neutral (associated), which is described by a chemical reaction
\begin{equation} \label{dissreac}
 AH + H_2O \rightleftharpoons A^- + H_3O^+ 
 \end{equation}
where $AH$ denotes the associated (neutral) acidic group and
$A^-$ denotes the dissociated (charged) group. 
At infinite dilution, the law of mass action relates the concentrations
to the equilibrium constant
\begin{equation}
K = \frac{ [A^-][H_3O^+]}{[AH][H_2O]}.
\end{equation}
Since the water concentration is for most purposes a constant, 
one defines an acid-equilibrium constant as
\begin{equation} \label{dissreac2}
 K_a =  K  [H_2O]  = \frac{ [A^-][H_3O^+]}{[AH]}
 \end{equation}
which now has units of concentration. Defining the negative common
logarithm of the $H_3O^+$ concentration and the acid constant as
$pH=-\log_{10}[H_3O^+]$ and $pK_a= -\log_{10} K_a$, the law of mass action
can be rewritten as
$ [A^-]/[AH] = 10^{pH-pK_a}$.
The degree of dissociation $\alpha$, defined as
$\alpha = [A^-]/([AH]+[A^-])$, follows as 
\begin{equation} \label{diss}
\alpha = \frac{1}{1+10^{pK_a-pH}}.
\end{equation}

\begin{figure}[t]
\begin{center}
\resizebox{14.cm}{!}{\includegraphics{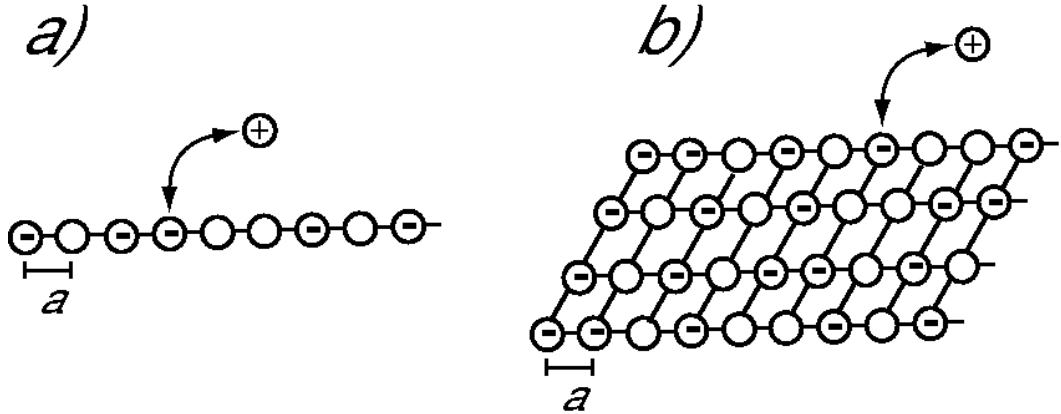}}
\end{center}
\caption{
Schematic representation of the geometry used in
the charge-regulation mean-field theory.
a) A straight polyelectrolyte chain
consists of $N$ dissociable monomers 
which can be either  charged or neutral.
b) A two-dimensional surface contains dissociable surface groups
that are positioned on some regular lattice, here chosen to be
a square lattice. The nearest-neighbor distance in both cases
is $a$.}
\vspace*{-.5cm}
\label{schemregul}
\end{figure}

Let us first consider the case of a single charged polymer or polyelectrolyte (PE).
In the present simplified model,
 we neglect conformational degrees of freedom 
of the PE and assume a straight polymer consisting of $N$
monomers with a bond length (i.e. distance between dissociable groups)
$a$, as is depicted in Fig. \ref{schemregul}a. 
This model is applicable to stiff 
PEs and for strongly adsorbed 
PEs, since they are indeed flat. The statistical mechanics of this problem
is quite involved as it involves summation over all possible charge distributions
on the line; 
the partition function  reads
\begin{equation} \label{part}
Z=\sum_{\{s_i\}=0,1} {\rm e}^{-{\cal H}}
\end{equation}
where the Hamiltonian (in units of $k_BT$)  is for monovalent monomers defined as 
\begin{equation}
{\cal H}  =  \mu \sum_i s_i  + 
\sum_{i>j}s_i s_j v_{\rm DH}(a|i-j|)
\end{equation}
and $s_i$ is a spin variable which is 1 (0) if the i-th monomer
is charged (uncharged). 
The chemical potential for a charge on a monomer is given by
\begin{equation}
\mu=-2.303(pH-pK_a)-\ell_B \kappa.
\end{equation}
The first term is the chemical free energy gained by dissociation,
which contains the chemical binding energy including Coulomb attraction
between acidic rest and proton and also the concentration of
protons as explained above;
the second term  is the sum of the  self-energies
of the released proton and the dissociated
acid.  The self-energy of a single charge  follows from the expression
\begin{equation}
\lim_{r \rightarrow 0} \frac{ v_{\rm DH}(r) - \ell_B/r}{2}= -  \ell_B \kappa /2 ,
\end{equation}
where the bare Coulomb self energy (which is a divergent constant) 
has been subtracted; it measures the free-energy gain associated
with the build-up of the counterion cloud. It  can be derived by
integrating the electrostatic field energy over the entire space,
or, which is simpler, by a thermodynamic charging procedure.

All different charge distributions are 
explicitly summed in Eq.(\ref{part}), which together with the 
long-ranged interaction $ v_{\rm DH}(r) $
between charged monomers makes the problem difficult. 
From the partition function, Eq.(\ref{part}), one derives 
the rescaled free energy per monomer, 
$f =-\ln (Z)/N$, from which  the fraction of charged monomers
is obtained  as $\alpha= \partial f/ \partial \mu $. 
Previously, similar problems have been solved using continuous mean-field 
theory\cite{Katchalsky,Itamar,Castelnovo}, restriction 
to only nearest-neighbor repulsions\cite{Harris,Marcus,Lifson,Borkovec} 
and computer simulations\cite{Ullner,Zito}.
In the presence of long-ranged charge  repulsions, however,
nearest-neighbor approximations break down, while the continuous mean-field theories
neglect the short-range structure of the charged species (they do not know about the
minimal distance between dissociable sites). 
Simulations provide accurate and specific answers, but for practical purposes
a closed-form solution in terms of a formula is desirable.
Lattice mean-field theory provides a simple 
close-form solution to the problem; the accuracy of this approach has been
demonstrated by extensive comparison with exact enumerations \cite{yoram} . 
To proceed, one defines a variational Hamiltonian which is chosen to be so simple that
closed-form solution of the partition function is possible. Standard mean-field theory 
employs a single-site Hamiltonian of the form 
\begin{equation}
{\cal H}_0  =  h \sum_i s_i  
\end{equation}
with $h$ being an as yet undetermined variational parameter.
The variational free energy is defined as 
\begin{equation}
f_{var} = f_0 +\langle {\cal H} - {\cal H}_0 \rangle_0 
\end{equation}
where $f_0= - \ln (1-e^{-h})$ is the free energy per site of the variational Hamiltonian.
All expectation values appearing in the variational free energy can be explicitly calculated.
Due to its construction, the variational Hamiltonian is a strict upper bound 
to the true free energy, i.e. $f_{var} \geq f$, and therefore the best possible 
estimate of the true free energy is reached by minimizing $f_{var}(h)$ with respect 
to  the variational parameter $h$. The resultant free energy 
$f_{MF} = \min_h f_{var}(h) $ is
the mean-field approximation, from which the mean-field dissociation degree 
$\alpha$ can be calculated via the already presented formula
$\alpha= \partial f_{MF}/ \partial \mu $. The resultant expression has been 
extensively compared with exact enumeration studies and found to 
be very accurate, especially at low salt concentration\cite{yoram}.
The result can be given as an implicit expression,
 \begin{equation} \label{result}
-\mu = 2.303 (pH-pK_a) + \ell_B \kappa  = \ln \frac{\alpha}{1-\alpha} +  \Delta \alpha 
\end{equation}
which can numerically or graphically be inverted.
Here,
\begin{equation}
\Delta= 2 \sum_{n=1}^{\infty} v_{DH}(na)= 
\frac{2 \ell_B}{a}  \sum_{n=1}^{\infty}  \frac{e^{-a \kappa n}}{n} = 
-2 (\ell_B/a) \ln (1-e^{-\kappa a}) 
\end{equation}
is the charge regulation parameter which takes charge-repulsion
between neighboring monomers into account. 
It has the limiting behavior $\Delta \simeq - 2 (\ell_B/a) \ln (\kappa a)$
for $\kappa a \rightarrow 0$ and 
$\Delta \simeq 2 (\ell_B/a) e^{-\kappa a}$ 
for  $\kappa a \gg 1$.
For $\Delta=0$
(obtained for large salt concentration $\kappa a \gg 1$)
the usual 'law-of-mass-action' dissociation behavior  is obtained,
for $\Delta > 0$ the dissociation is much reduced.
It is interesting to compare our result Eq.(\ref{result}) with previous 
heuristic formulas which  include the effects
of charge repulsion on the dissociation by phenomenological 
fitting parameters\cite{Kat2,Mandel}. Our expression is different,
but in fact not more complicated.

\begin{figure}[t]
\begin{center}
\resizebox{14.cm}{!}{\includegraphics{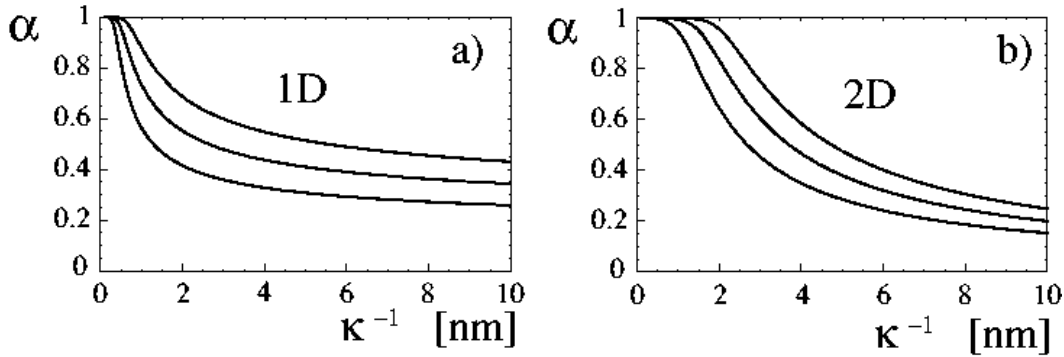}}
\end{center}
\vspace*{-.5cm}
\caption{ a) Mean-field results for the dissociation fraction $\alpha$
of a poly-anion in the bulk
with monomer separation $a=0.25 nm$ for fixed 
$pH-pK_a = 2,3,4$ (from bottom to top) as a function
of the screening length $\kappa^{-1}$.
b) Dissociation fraction of charge groups on  a surface with 
lateral group separation of $a = 1nm$ for fixed 
$pH-pK_a = 2,3,4$ (from bottom to top) as a function
of the screening length $\kappa^{-1}$.} 
\label{regulate}
\end{figure}

In Fig.\ref{regulate}a the fraction of charged monomers is presented
for a polyacid
with monomer-separation $a=0.25 nm$, as applicable
to vinyl-based polymers,
for fixed $pH-pK_a = 2,3,4$ (from bottom to top)
as function of the screening length in the bulk.
As is well known, the dissociation for all but
very high salt concentrations (small $\kappa^{-1}$) is incomplete
and further decreases with increasing $\kappa^{-1}$,
a phenomenon called {\em charge regulation}.
 As a main result, even rather strong PEs are only partially
 charged at low salt concentrations\cite{Katchalsky} 
 (where we have not taken additional
 complications due to chemical binding of metal ions 
 into account\cite{Lifson,Helm1}). 

On a two-dimensional surface, the mean-field formalism works just as well. 
The only modification concerns the charge regulation parameter $\Delta$, which 
now takes interactions of one charge with all neighbors in two dimensions into
account. Assuming the dissociable sites to be located on a 2D square lattice with
lattice constant $a$, as schematically depicted
in Fig.\ref{schemregul}b,  it is given by 
\begin{equation}
\Delta= \sum_{n=-\infty}^{\infty}  \sum_{m=-\infty}^{\infty} v_{DH}(\sqrt{m^2 +n^2} a)= 
\frac{\ell_B}{a}  \sum_{n=-\infty}^{\infty}  \sum_{m=-\infty}^{\infty}
 \frac{e^{-a \kappa \sqrt{m^2 +n^2}  }}{\sqrt{m^2 +n^2}}.
\end{equation}
In contrast to the one-dimensional case, the double sum (where the origin is excluded)
cannot be performed exactly. Since we are mostly interested in the behavior
when the screening length is large, we can use the isotropic form of the summand and
make the simplification 
$  \sum_{n=-\infty}^{\infty}  \sum_{m=-\infty}^{\infty} \approx  
2 \pi   \sum_{n=1}^{\infty}  $
and obtain the result 
\begin{equation}
\Delta = \frac{2\pi \ell_B}{a} \frac{1}{e^{\kappa a}-1}
\end{equation}
with the limits $\Delta \simeq 2 \pi \ell_B /(\kappa a^2)$ for 
low salt $\kappa a \rightarrow 0$ and 
$\Delta \simeq 2 \pi \ell_B e^{-\kappa a }/a$  for high salt
$\kappa a \gg 1$. The main difference to the 1D case is the behavior for 
small salt concentrations or large screening lengths, 
where the coupling is much larger and thus the 
dissociation is even further  reduced as compared to the 1D case. 
This is demonstrated in Fig.\ref{regulate}b, where 
we show the 
dissociation fraction of charge groups, $\alpha$,  on  a surface 
as a function
of the screening length $\kappa^{-1}$ for fixed 
$pH-pK_a = 2,3,4$ (from bottom to top) .
We chose a  lateral surface group separation of $a = 1nm$ 
which is much larger than the charge distance
for the polymer in Fig.\ref{regulate}a, which reflects that
surface charges are typically more sparse than charges 
on polyelectrolytes. Still, the surface dissociation is highly 
reduced, especially for large screening lengths where the
interaction between groups that are distant comes into play.
Comparing the 1D and 2D case one indeed sees that on 
the surface the dissociation degree drops faster as the screening
lengths increases, which has to do with the  functional dependence
of the charge regulation parameter $\Delta$ on $\kappa a$
(inverse power law in 2D as opposed to logarithmic in 1D).

In summary, charge regulation happens and reduces the charge
of all objects with dissociable surface groups, especially
at low salt concentrations. It should be taken into account
especially when interactions between charged objects are considered.
In MD simulations of weakly acidic groups, 
force fields have to be used which take 
the chemical binding effects into account. 
Quantum-chemistry methods can be of help, as they allow
to determine those chemical binding forces, as will be explained 
in Section 11.

\section{Water at hydrophobic substrates}

In the simplified models used in the previous sections the presence
of water was accounted for only by the presence of a uniform relative dielectric constant
with the value $\varepsilon = 78$. 
As is routinely done in most theoretical considerations,
no other water effects were included, which works for many cases, but especially
at surfaces  is a highly questionable concept, as will be discussed now.
For very large non-polar objects or in the limiting case of a planar 
hydrophobic substrate in contact with water, it is known since a long time
that the water density
is reduced at the hydrophobic surface and the structure
of the interfacial water is very different from the bulk\cite{Stillinger, Lee,LinseBenzene}. 
Clearly, this has important consequences for all material constants 
characterizing the solution
(dielectric constant, viscosity, screening length, $pK_a$) close to the surface.
It seems fair to say that without a proper characterization of the 
behavior of water at hydrophobic surfaces, no true understanding of 
the  properties of such surfaces and their interactions will be possible. Quite possibly,
many of the features interpreted as being inherent to surfaces themselves, might
in fact reflect properties of the interfacial water layer instead (e.g. 
protein adsorption resistivity\cite{grunze}, zeta potentials\cite{werner},
surface potentials\cite{kreuzer}, polymer-adsorption energies\cite{Hugel,gaub}, 
just to mention a few). We will now consider the water density profile close
to a planar surface using Molecular Dynamics (MD)  simulations. 
In MD, one basically integrates Newton's equation of motion for an assembly of
molecules or atoms, using heuristically chosen force parameterizations. 
The constant pressure ensemble is realized by adjusting the system volume.
For simplicity, we only consider neutral walls, and in order to bring out the
consequences of the presence of a wall most clearly, we deal with the
special case of a very hydrophobic wall.

The work described here was motivated
by recent scattering experiments where the water density depletion at planar
non-polar substrates was determined. As the main result of those
experiments, it was shown that the effective depletion thickness
(defined as the thickness of a step-like depletion layer consisting of vacuum 
with the same integrated depleted amount as the rounded and smeared-out 
depletion profiles found in experiments)
is roughly 2.5 Angstroms on hydrophobic poly-styrene substrates\cite{Steitz}
and  5 Angstroms on hydrophobic self-assembled monolayers\cite{Schwendel} using
neutron reflectivity methods, and about 1 Angstrom on paraffin substrates
using X-ray reflectivity measurements\cite{Jensen}.
The reason for the discrepancies among different experiments
is not well understood, but since
the strength of  water depletion is reduced with decreasing
radius of curvature of the hydrophobic solutes\cite{shavkat}, it is
clear that surface roughness is one important factor (among many others, as e.g.
small traces of attractive interactions between wall and water molecules) and will, 
if present, reduce the depleted amount.

\begin{figure}
\begin{center}
\resizebox{12.7cm}{!}{
\includegraphics{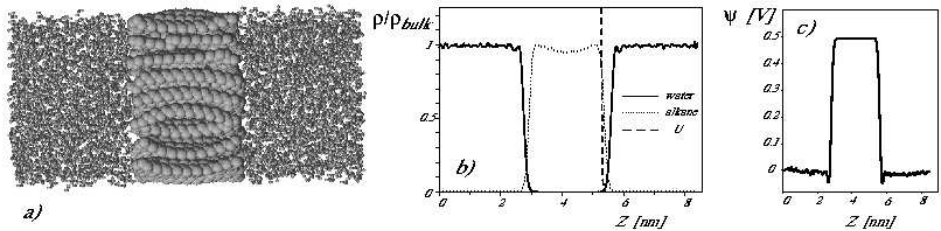}}
\caption{\label{wall}
a) Snapshot of the MD simulation of a planar hydrophobic slab
(made up of 64 alkane molecules)
in contact with a water slab, consisting of 2781 SPC/E water molecules.
b) Normalized density profile of the hydrophobic alkane slab (dotted line, in the middle)
and the water layers (solid line, to the left and to the right) at 
constant pressure of 1 bar and temperature T=300K.
The broken line denotes $U(z)$,
the laterally averaged Lenard-Jones potential
felt by the water molecules in units of $k_BT$.  
 The system was thermalised
for 100 ps and averaged for 2 ns.
c) Local electrostatic potential across the water-alkane interface,  exhibiting a potential drop
of about 0.5 Volts.}
\end{center}
\end{figure}

In the simulations, we
built up the hydrophobic substrate by self assembled alkane chains
which seems to be an acceptable representation of the substrate structure
used in recent experiments\cite{Steitz,Schwendel,Jensen}. 
In Figure \ref{wall}a)  a snapshot of the MD simulation is shown, 
which serves to illustrate the geometry of the system\cite{shavkat}. 
The alkane molecules form a compact slab in the middle
of the simulation box. They are only allowed to fluctuate in the 
z-direction and thus allow for fast pressure equilibration. The water slab
has a thickness of about 4 Nanometers, which should be large enough
such that bulk water properties are reproduced. We therefore
interpret our results as being caused by the single hydrophobic substrate --
water interface and neglect interactions between the two interfaces through the
finite water slab.

In Figure \ref{wall}b) 
we show the normalized densities of the alkane slab (dotted line)
and the water layer (solid line) at atmospheric pressure and at a temperature T=300K.
It is clearly seen that between the alkane slab and the water layer a region 
of reduced density appears. The density profiles are calculated
using point-like atomic form factors and denote the nuclear density;
they therefore correspond to what would be seen in a neutron scattering experiment.
The broken line denotes the laterally averaged interaction potential due to the alkane
slab, in other words, this is the potential energy felt by the water molecules.
 At room temperature
and normal pressure, we obtain for the
depletion thickness of a planar hydrophobic substrate the value
$d_2 = 2.565$ \AA, which is of the order of  the length
obtained in recent neutron scattering experiments\cite{Steitz,Schwendel}
and twice the length obtained with X-rays at hydrophobic substrates\cite{Jensen}.
In Fig. \ref{wall}c the electrostatic potential across the interface is 
shown to change by about 0.5 V, which is due to the almost
complete orientation of the topmost water layer. 

What do these results imply  for charged surfaces? Many of the charged surfaces used
in experiments are in fact, if one forgets about the charged groups for a moment,
of hydrophobic nature. Let us assume for the sake of argument that the hydrophobic
effects discussed above persist, even when charges are present.
A layer of reduced water density at such surfaces means that the effective dielectric 
constant is reduced; this suggests that i) the association-dissociation equilibrium of 
surface charges is perturbed and ii) that electrostatic interactions at hydrophobic 
surfaces in general might be stronger than anticipated based on simpler models. 
The viscosity at the surface will most likely be reduced which might be important 
in connection with electrokinetic effects. Finally, the polarity of the top water 
layer is such that in principle (i.e. neglecting ion-water interactions) negative ions 
should be preferentially adsorbed on the surface. This is indeed borne out by recent 
MD simulations\cite{Tobias}, which however 
demonstrated an additional  subtle dependence on the presence 
of ionic polarizabilities.
For interactions between charged surfaces it is conceivable
 that the hydrophobic attraction
due to the overlap of the depletion layers might very well dominate the resulting
behavior.

\section{Ion-specific effects}

In the preceding sections, ions were either treated as point-like or as hard spheres.
However, as has been recently reviewed\cite{Hofintro,Ninham97},
a large number of phenomena in colloid, polymer, and interface science
that involve electrolytes show pronounced ion-specificity,
as categorized in the famous Hofmeister series\cite{Hof1,Hof2,Hof3,Hof4,Hof5}. 
A striking experimental example of
counterion specificity is obtained for the cationic surfactant discussed in Section 3.
Exchanging the Bromine ion in DDAB by a Chlorine ion, the phase diagram changes dramatically
and the phase coexistence disappears 
completely\cite{Dubois}.\footnote{This is difficult to understand
based on dispersion or polarization effects, since 
their combined contribution to the effective inter-ionic potential in water is quite small.
The reason for the big difference in the surfactant phase diagram
is probably a steric coordination effect, allowing it for one of the ions to 
come into close contact with the cationic surfactant head region.}
A subset of these unresolved issues is typically associated with the so-called 
hydrophobic force, a rather long-ranged attraction between hydrophobic
surfaces, which is much stronger than predicted from standard van-der-Waals
calculations and is also strongly ion-type-dependent\cite{Spalla}. 
Previous theoretical explanations invoked 
solvent-structure effects\cite{stepan,Rudi,Kusalik,Lyubartsev,Buri}, and
surface-specific ion interactions\cite{stern,Podgornik1}
or charge-regulation phenomena\cite{Belloni}.
The presence  of excess ionic polarizabilities was proposed to 
lead to corrections to the usual van-der-Waals interaction energy, which 
could be one of the factors determining ion-specific 
interactions\cite{Ninhamnew,Netz3,kunz,netzopinion}.
But it was also shown that even with pure Coulomb interactions, one obtains
strong deviations from the standard mean-field approaches if one takes into account
that the charge distribution on all charged surfaces is laterally 
modulated\cite{andreepl,vanmegen,bo,messina,Dima}.
Specifically, the counterion density right at a charged surface
(which for a homogeneously charged surface and in the absence of additional
interactions is exactly given by Eq.(\ref{condit2}) because
of the contact-value theorem) is for a modulated surface charge distribution
increased. Surface-charge inhomogeneities in that sense act like additional
attractive interactions between surfaces and ions, but it should be clear
that all the above-mentioned effects are present simultaneously.
It is therefore not easy to disentangle these various factors, especially since
experimentally one typically measures macroscopic quantities such as 
surface tension, ionic activities, ionic osmotic coefficients, etc. and not
ionic distribution functions from which  effective interactions between ions
and surfaces could be deduced. 
As an additional complication, computer simulations, which would include
all above mentioned effects, are still difficult to perform, 
even using coarse-grained models where one replaces explicit solvent by some
suitably chosen  dielectric constant plus solvent-induced effective interactions
between solute molecules. One therefore has to rely on various approximations,
and it is often not easy to tell to what degree the approximation or the model parameters
are responsible for the outcome.

\begin{figure}[t]
\begin{center}
\resizebox{14.cm}{!}{\includegraphics{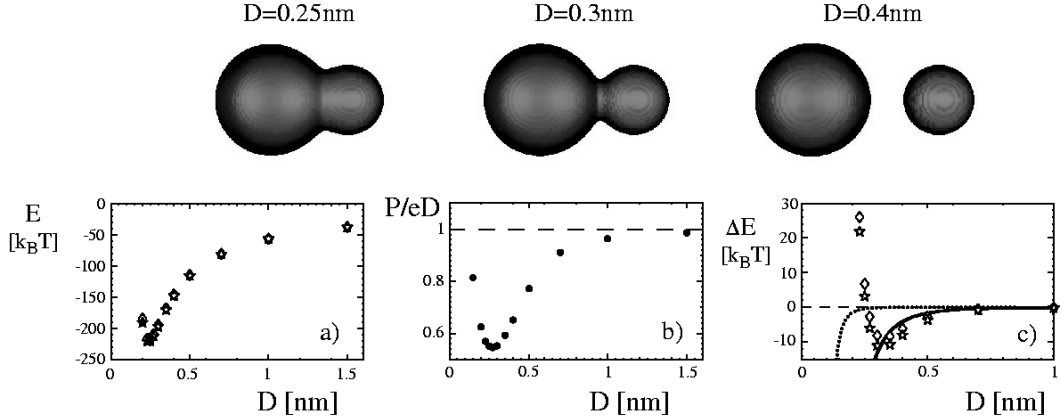}}
\end{center}
\vspace*{-.5cm}
\caption{ Top panel: Surfaces of constant electron density (roughly
corresponding to the density at the Pauling surface) for three
different values of the distance $D$ between a sodium (to the right)
and a chloride ion (to the left) in vacuum.
a) Interaction energy obtained using Hartree-Fock with a TZV basis set
(open diamonds) and 
including electron-correlations on the MP2 level (open stars).
b) Dipole moment divided by distance 
 as a function of the ion distance, indicating that the two ions
are fully charged for large separations.
c) Non-coulombic interaction obtained by subtracting the Coulomb energy from 
the data in a). Note that a deep minimum is present which can be explained by
the charge-induced dipole interaction (solid line) for large distances.
The dispersion interaction (dotted line) is irrelevant for all distances.
All data are obtained in vacuum.} 
\label{fig15}
\end{figure}

How do we expect ion-specific effects to come into play on a microscopic level?
At any charged surface, one has chemical groups which carry most of the
surface charge, i.e., at which location the charge density is locally 
increased. 
The counterions (or any other oppositely charged
molecule) will, due to Coulomb interactions, be on average quite close
to these surface groups (in fact much closer than would be predicted according
to Poisson-Boltzmann theory assuming a laterally homogeneous surface charge, compare
Section 4), and it seems natural to surmise that it is the interaction
between oppositely charged chemical groups that will be most susceptible to
chemical specificity. This is not to say that solvent effects (i.e. water structuring
which of course is different for different ion types) is unimportant, on the contrary,
but it does not make sense to separate solvent-induced effects from the bare
interactions between oppositely charged groups.
One experimental example where the specificity is exhibited very clearly 
is with AFM experiments for polyelectrolytes adsorbed on a variety of different 
substrates and in different electrolyte solutions\cite{Hugel,gaub}, 
where the plateau desorption force can be directly
converted into the binding energy per unit length.

\subsection{Interactions between ions}

Fortunately, progress in the available quantum-computational methodology 
 allows to calculate effective interactions between charged species in an essentially
 ab-initio manner\cite{Gamess}, including solvent effects\cite{tomasi}. 
As a prototype for the effects studied, we show in Fig.\ref{fig15} ab-initio
results for the interaction between a sodium and a chloride ion in vacuum.
In Fig.\ref{fig15}a we show the resulting interaction energy between the two ions
as a function of their mutual distance in units of $k_BT$. The open diamonds are obtained
using Hartree-Fock (HF) methods, i.e. each electron 
sees a mean-charge-distribution due to the
other electrons, otherwise the Schr\"odinger equation is explicitly
solved for all 28 electrons involved (treating the nuclei as being fixed), using
an expansion in TZV basis functions with added polarization and diffuse wave functions. 
The open stars denote results where electron
correlations have been taken into account on the one-loop level (HF-MP2), the energy is slightly
lower. One would expect that most of the long-ranged attraction seen in Fig.\ref{fig15}a is due to
the Coulomb attraction between the separated charges on the two ions. 
To make that notion quantitative, we first have to find out how much charge is transferred
between the two ions.  In Fig.\ref{fig15}b 
we plot the ratio of the dipole moment of the whole charge distribution (including both ions and in
units of the elementary charge) and the ion-ion separation as a function of the separation. 
This ratio can be interpreted as the effective charge transfered between the two ions.
It is seen  that indeed charge transfer
is almost perfect at large distances in which limit both ions are fully charged.
For smaller distances polarization effects lead to a decrease of the transfered charge.
In Fig.\ref{fig15}c we show the same data as 
in Fig.\ref{fig15}a but with the Coulomb attraction $v(r) = -\ell_B/r$ subtracted 
(note that the Bjerrum length in vacuum measures $\ell_B = 55.73 nm$). Quite 
surprisingly, the resulting energy shows a pronounced attractive minimum of 
depth $\simeq -10 k_BT$ in a distance range $ 0.25 nm < D< 0.4 nm$. Both HF and MP2 
calculations give roughly the same result, the difference between the two
can be viewed as an estimate for the systematic error in the calculation 
(an additional source of error is introduced due to
the necessarily incomplete basis sets used). What is the reason for this quite strong 
attraction between the ions? To the mind come two contributions, namely the 
polarization attraction due to the charge-induced dipole interaction and the 
van-der-Waals interaction. Let us discuss both in some detail.

The static polarizability of the isolated ions can be calculated using the same
ab-initio methods by applying a small electric field and measuring the induced
dipole moment (or by measuring the polarization energy).
We obtain $\alpha_{Cl^-}/(4 \pi \varepsilon_0) = 3.404$ \AA$^3$
within HF and  $\alpha_{Cl^-}/(4 \pi \varepsilon_0) = 3.666 $ \AA$^3$ within HF-MP2 for
the $Cl^-$ ion and $\alpha_{Na^+}/(4 \pi \varepsilon_0) = 0.132$ \AA$^3$
within HF  and $\alpha_{Na^+}/(4 \pi \varepsilon_0) = 0.143$ \AA$^3$ within HF-MP2
for the $Na^+$ ion. Other data are collected in Table 1.
The charge-induced dipole interaction between the $Cl^-$ and $Na^+$ ions is
in units of $k_BT$ \cite{Israel}
\begin{equation}
w_{\rm ind} (r) = -\frac{ \ell_B (\alpha_{Cl^-}+\alpha_{Na^+})}{8 \pi \varepsilon_0 r^4}
\end{equation}
and is plotted in Fig.\ref{fig15}c as a solid line. It describes the ion-ion interaction quite well
except for very small distances where the electron clouds overlap strongly. So it seems
fair to say that corrections to the bare Coulomb interaction between ions 
in vacuum (which has been subtracted off in Fig.\ref{fig15}c) 
 are at large distances mostly dominated by polarization effects.

The dispersion interaction between two atoms has been calculated using 
quantum-mechanical perturbation theory\cite{London,Mahan} and is given by
(in units of $k_BT$)
\begin{equation} \label{disp}
w_{\rm disp} (r) = -\frac{ 3 \alpha_{Cl^-} \alpha_{Na^+} I_{Cl^-} I_{Na^+}}
{2 (4 \pi \varepsilon_0)^2 (I_{Cl^-}+I_{Na^+})  r^6}
\end{equation}
where the ionization energies $I$ of the two ions are measured in units of
$k_BT $ also. For the ionization potentials we obtain using the same
level of HF-MP2 the results
$I_{Cl^-} = 138.57 k_BT $ and $I_{Na^+} = 1820.2  k_BT$,
other data can be found in Table 1.\footnote{
It is instructive to compare our ab-initio calculations with experimental values.
For ions only few  data are available in the literature. 
Our ab-initio results for the polarizability and ionization potential of a 
neutral $Na$ atom, which are well tabulated, are  
$\alpha_{Na}/(4 \pi \varepsilon_0) = 25.04 \AA^3$ and
$I_{Na} = 5.086 eV $, which have to be compared with the 
experimental values $\alpha_{Na}/(4 \pi \varepsilon_0) = 23.6 \AA^3$
and $I_{Na} = 5.139 eV $\cite{Israel}. The agreement is sufficient for the present
purpose. In the table one notes big differences between the data for ions and 
the corresponding neutral atoms. Approximating ionic properties by the neutral-atom data
is therefore a bad idea.} 
The resulting dispersion interaction is plotted in Fig.\ref{fig15}c as 
a dotted line. It is basically negligibe for the whole relevant range of distances. 
This tells us that the attraction that appears in the quantum-mechanical calculation
is at large distances mostly due to polarization effects.
It has to be considered as an 
important factor in the interaction between charged surfaces, since such additional interactions
between ions and surfaces lead to charge regulation and thus to varying effective 
surface charges\cite{kreuzer}.
Likewise, the interaction between ions in the bulk modifies the
osmotic coefficients, the screening length,  ionic activities
and therefore gives an additional shift of the surface-group dissociation equilibrium.
Clearly, this interaction
is highly specific and different for different ion types,
especially at small distances. A hand-waving
explanation why the very short-ranged properties of this
 interaction will be important is that oppositely charged ions are 
 squeezed  together such that  the 
electron orbitals overlap
to a degree where quantum-mechanical effects come into play.
This might be intuitively understood by looking at the electron-density contours
shown in the upper panel in Fig.\ref{fig15}. In all three pictures, the electron densities
on the contour surfaces are the same and roughly correspond to the density on
the Pauling surfaces (the Pauling ionic radius as deduced from crystal structures
for the $Cl^-$ ion is 
$R_{Cl^-} \simeq 0.181 nm$ and for the $Na^+$ ion 
it is $R_{Na^+} \simeq 0.095nm$). It is seen that for the range
of distances where the attraction is strongest the electron distributions overlap.
A similar short-ranged interaction between ions
had been introduced in an ad-hoc fashion in order to accurately
fit activity coefficients of alkali-halide solutions\cite{Rasaiah,Ursenbach},
but we argue here that it is a general feature of oppositely charged groups 
and not restricted to simple ions  but also applies to the interaction
between macroscopic charged bodies. 
A similar interaction should also be present for the case of similarly charged ions,
though here we would in general expect the effects to be small since the Coulomb
repulsion in this case will make close contacts between ions unlikely in the general case. 
In previous theories which concentrated on 
water-structure effects for electrolyte  behavior, 
the bare interionic potential has been typically 
regarded as rather structureless\cite{Lyubartsev}. It might be 
interesting to  reconsider such calculations by adding quantum-mechanical
potentials as we have calculated.

\begin{figure}[t]
\begin{center}
\resizebox{14.cm}{!}{\includegraphics{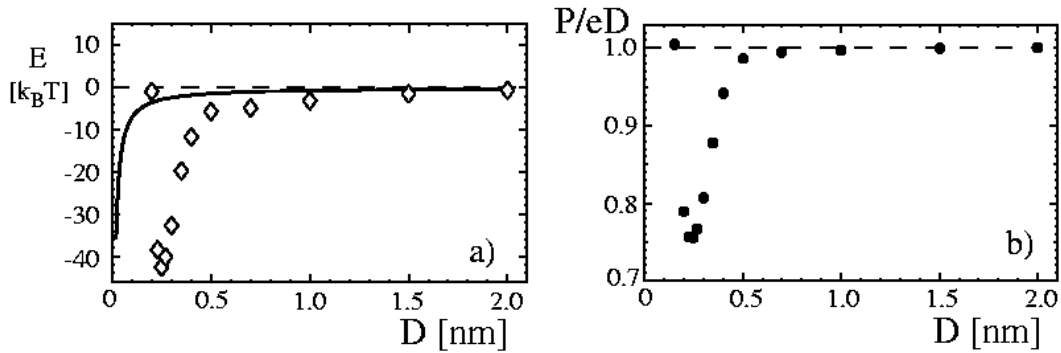}}
\end{center}
\vspace*{-.5cm}
\caption{ a) Interaction energy between a sodium and a chloride ion
using Hartree-Fock methods and
a polarizable continuum model with parameters representing liquid water.
Open diamonds represent the full interaction energy in units of $k_BT$. 
The solid line corresponds to the Coulomb interaction between two unit charges 
in a homogenous material of 
relative dielectric constant of $\varepsilon = 78.39$, which underestimates the true 
 attraction considerably.
b) Dipole moment divided by distance 
 as a function of the ion distance, indicating that the two ions
are fully charged already for intermediate separations.} 
\label{fig16}
\end{figure}

\begin{table}
\caption{\label{tab1} Ab-initio results for ionization energies $E_{ion}$  and
polarizabilities $\alpha$ for various atoms and ions in vacuum. 
The cavity radii $R_{cav}$ are by a factor 1.2
larger than the Pauling  radii. The effective ionic dielectric constant  $\varepsilon$
and the excess polarizabilities $\alpha_{exc}^{ \infty}$ at large frequencies
and at zero frequency   $\alpha_{exc}^{ 0}$
are calculated from the Clausius-Mossotti/Lorenz-Lorentz equation, 
see text.  The shift of the ionization energy in water, $ \Delta E_{ion} $, is
calculated using a simple Born self-energy model.
For conversion, note that $ 1eV=96.4516 kJ/mol$ and
$1eV=38.610 k_BT$. Numbers in parantheses are experimental values.
 }
\begin{tabular}{cccccccc}
& $ E_{ion} $ [eV] & $\frac{\alpha}{ 4 \pi \varepsilon_0} $ [\AA$^3$]&
$R_{cav}$  [\AA] & $\varepsilon$ & 
$\frac{ \alpha_{exc}^{\infty}}{  4 \pi \varepsilon_0} $ [\AA$^3$] &
$\frac{\alpha_{exc}^{0}}{  4 \pi \varepsilon_0} $ [\AA$^3$] &
$ \Delta E_{ion} $ [eV]   \\
\hline
$Li$   & 5.37 (5.39) & 24.9 (24.3) & && &&\\
$Li^+$ & 75.0 & 0.022 & 0.72 & 1.186 & -0.040 &-0.18 & - 29.7 \\ 
$Na^+$ & 47.1 & 0.14  & 1.14 & 1.32 & -0.11 & -0.72 & -18.8\\
$K^+$ &  31.5 & 0.81 & 1.596 & 1.75 & 0.062 & -1.97 & -13.4\\
$Fl^-$ & 3.54 & 1.01 & 1.632 & 1.90 & 0.195 & -2.10 & 4.37\\
$Cl^-$ & 3.59 & 3.67 & 2.172 & 2.62 & 1.63 & -4.87 & 3.28\\
$Br$ & 13.6 (11.81) & 2.91 (3.05)& && &&\\
$Br^-$ & 3.39 & 6.61 & 2.34 & 4.19 & 4.29 & -5.91 & 3.04 \\
$NO_3^- $& 3.69 & 4.69 &2.45 &  2.40 & 1.88 & -7.04 & 2.91\\
$SCN^- $& 2.42 & 7.48 & 2.62 & 3.14 & 4.08 & -8.45 & 2.72\\
$H_2 O$ & & 1.31 (1.45) & 1.93 & 1.669 (1.78) & &&\\
\end{tabular}\tabularnewline
\end{table}

Conversely, it is important within our approach to critically check how the 
interaction we obtain will be modified in the presence of water. To do so
we performed Hartree-Fock calculations using  the so-called polarizable-continuum-model
where the ions are embedded in spherical shells outside of which a dielectric
medium with relative dielectric constant $\varepsilon = 78.39$ is 
assumed. The choice of the radii of these cavities is critical, we chose the cavity
radii to be bigger than the Pauling  radii of the ions by a factor 1.2.
This means that the two dielectric cavities for the case of $Na^+$ and
$Cl^-$ start touching at a distance of 
$D=1.2 (0.095nm + 0.181 nm) = 0.331 nm$.
The Schr\"odinger equation is then solved taking into account the effects
of polarization charges\cite{tomasi}. The open diamonds in Fig.\ref{fig16}a
represent the full interaction obtained between the two ions. One notes that the
long-ranged attractive tail has almost disappeared. 
The solid line corresponds to the Coulomb interaction between two unit charges 
with a relative dielectric constant of $\varepsilon = 78.39$.
This Coulomb potential is quite small,
 which is understandable since the Bjerrum length now takes the 
value $\ell_B \simeq 0.71 nm$. 
It is important to note that the true electrostatic interaction is quite
involved because of the complicated geometry:
for distances larger than $D=0.331nm$ one has two separate spherical
dielectric cavities which are immersed
in a high-dielectric background mimicking water. For smaller distances, the spherical
cavities overlap. The  Coulomb interaction between two charges inside the cavities
shows a complicated crossover from a weak interaction at large distances, characterized
by a Bjerrum length $\ell_B \simeq 0.71 nm$, to a strong interaction
at distances smaller than the cavity-overlap distance, where the Bjerrum length
becomes closer to the vacuum value $\ell_B = 55.73 nm$.
This is more or less what one sees in the data in Fig.\ref{fig16}a.
As a consequence, the short-ranged 
attraction is even stronger and now has a depth of $40 k_BT$. 
It should be noted that this short-ranged attraction is mostly due to 
the modification of the Coulomb potential
in the presence of dielectric boundaries (for similarly charged ions, 
the effective interaction will be predominantly repulsive). 

What would we expect
for the charge-induced-dipole and the dispersion interaction in this case? 
To make progress we first need to evaluate the effective dielectric constant 
of the ion-containing cavity, which follows from the Clausius-Mossotti equation\cite{Mossotti}
(or the frequency-dependent analogue, the Lorenz-Lorentz equation\cite{Boettcher})
by 
\begin{equation} 
\varepsilon = \frac{2 \alpha/(4 \pi \varepsilon_0 R^3)+1}
                 {1-\alpha/(4 \pi \varepsilon_0 R^3)}.
\end{equation}
For the chloride ion with $R_{Cl^-} \simeq 0.2172 nm$ 
and  $\alpha_{Cl^-}/(4 \pi \varepsilon_0) = 3.666 \AA^3$ one obtains
$\varepsilon_{Cl^-} \simeq 2.62$ and for
the sodium ion with $R_{Na^+} \simeq 0.114 nm$ 
and $\alpha_{Na^+}/(4 \pi \varepsilon_0) = 0.143 \AA^3$
one obtains $\varepsilon_{Na^+} \simeq 1.32$, where these
numbers are equally valid in the static and dynamic case. 
Other results for different atoms and ions are given in Table 1.
For the charge-induced dipole
interaction what counts is the static excess polarizability of the ions in water,
which again can be calculated from the inverted  Clausius-Mosotti/Lorenz-Lorentz 
equation,
\begin{equation} \label{LL}
\alpha_{exc}/(4 \pi \varepsilon_0 )= R^3 
\frac{\varepsilon_{ion} - \varepsilon_{water}}
{\varepsilon_{ion} +2  \varepsilon_{water}}
\end{equation}
and is denoted by $\alpha^0_{exc}$ and given in Table 1.
Since water has a much higher static dielectric constant than the ions,
$\varepsilon_{water}^0 =78.39$,  it is clear that
the excess static polarizability is negative and thus the charge-induced dipole
contribution to the interaction energy is repulsive. 
It is therefore ruled out as a possible explanation for the observed
attraction between the ions seen in the data in Fig.\ref{fig16}a.
For the dispersion interaction we have a static contribution,
which is attractive but rather weak (since it is at most of the order of 
$3 k_BT $ at contact\cite{Israel})  and a dynamic contribution. For the dynamic
dispersion interaction what counts is the frequency-dependent dielectric constant
of the ions, given above, and of water, which follows from the refractive index 
$n \simeq 1.33$ as $\varepsilon_{water}^{\infty} = n^2 \simeq 1.78$.
According to the Lorenz-Lorentz equation (\ref{LL}), the excess polarizability
is reduced, such that the 
dynamic dispersion interaction will be even smaller than the one in vacuum
(which is  shown  in Fig.\ref{fig15}c as a dotted line). 
To get explicit numbers for the dynamic excess polarizabilities of ions in water,
we have calculated the high-frequency dielectric constant of water within 
our ab-initio technique using the same method as for the ions. The result
is $\varepsilon^{\infty}_{water} = 1.669$ and thus smaller than the experimental value
by 10 \% (see Table 1). For consistency reasons, we have estimated the finite-frequency
ion excess polarizabilities with the calculated value of the water dielectric constant.
The results are given in Table 1. The resulting excess polarizabilities are
always smaller than the ones in vacuum.
We also estimate the ionization energies in the water environment using
a simple Born self-energy argument.
For the anion, the ionization energy is increased by the term
\begin{equation}
\Delta E_{ion} = (\ell_B^{vac} - \ell_B^{water})/(2 R_{cav})
\end{equation}
which measures the electrostatic self-energy difference of a charged sphere in vacuum
and in water.
The vacuum Bjerrum length is given by $\ell_B^{vac} = 55.73 nm$ 
and the Bjerrum length in water is $\ell_B^{water}= 0.71 nm$.
For the cations, the ionization energy is reduced by the term
\begin{equation}
\Delta E_{ion} = -(4-1) (\ell_B^{vac} - \ell_B^{water})/(2 R_{cav})
\end{equation}
which is the self-energy difference of a divalent and a monovalent charged sphere in vacuum
and in water. The resulting numerical values  are given in Table 1.
The effect of the ionization energy change on the dispersion interaction is
roughly to increase the dispersion strength by a factor of two (this follows
from the fact that the sum of ionization energies in the denominator of 
equation  (\ref{disp})  is
dominated by the larger cationic energy which therefore cancels the 
cationic energy in the numerator).
The reduction of the polarizability in water however is larger than the increase
of the ionization term, so that in essence the dispersion interaction in water
is even weaker than in vacuum. 
Similarly to the situation in vacuum, therefore,
the dispersion interaction is only a negligible contribution to the full interaction
obtained within the ab-initio calculation. As a main result, we find that,
owing to 
the shape and size dependent crossover of the effective Coulomb interaction,
the effective interaction between ions in a polarizable continuum medium 
is thus quite specific and depends sensitively on the shape and size of the ions.
 It remains
to be checked how these results will be modified if discrete water molecules are included
in the calculation, but it seems likely  that specific short-ranged interactions
between oppositely charged chemical groups play an important role in the 
physics of strongly charged systems.

\subsection{Dissociation constants}

\begin{figure}[t]
\begin{center}
\resizebox{10.cm}{!}{\includegraphics{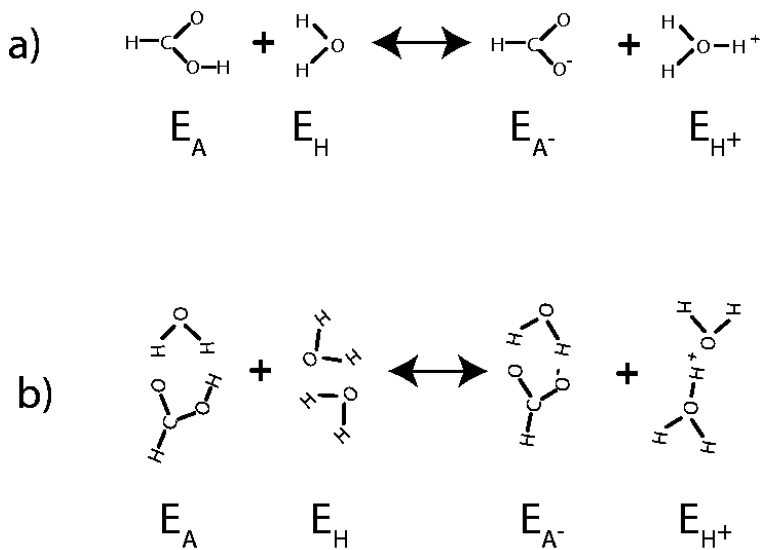}}
\end{center}
\vspace*{-.5cm}
\caption{  Dissociation reaction of a carboxylic acid, in
a) without bound water and in b) with one associated water molecule.
In the PCM calculations a cavity is formed consisting of  spheres
centered at the heavy atoms. Outside the cavity the dielectric constant 
is that of water, inside it takes the vacuum value. } 
\label{reaction}
\end{figure}

The dissociation of an acid is a special case of the interaction between
two oppositely charged ions, namely the acidic rest group and the proton.
We will specifically consider the dissociation of the carboxylic acid, 
which has been quite extensively studied in the literature and serves
as a good model to compare different approaches with each 
other\cite{Schuur,Silva,Klamt}.
The basic chemical reaction,
according to the formula Eq.(\ref{dissreac}),
 is depicted in Fig.\ref{reaction}a.
In the quantum chemical calculation we optimize the conformation of
each 'molecule' and calculate the energy in the electronic and conformational
ground state. The results for all energies for a vacuum calculation of
reaction a) are given in
the first row in Table 2. All energies are expressed  in units of $k_BT$. The actual
numbers are quite large, since all core electrons contribute. It is clear that in 
order to extract the binding energy of the proton, high precision  is needed
since we are interested in the small difference between large numbers. We
define the energy differences $\Delta E_A = E_{A-} - E_{A}$ and 
$\Delta E_H = E_{H+} - E_{H}$, from which the binding energy 
is obtained as $\Delta E = \Delta E_A  + \Delta E_H$. According to
Eq.(\ref{dissreac2}) the acidic dissociation constant is 
\begin{equation}
K_a = e^{-\Delta E } [H_2 O],
\end{equation}
or, after taking the negative common logarithm and using that the water
concentration is roughly $[H_2 O] = 55 mol/l$,
\begin{equation}
pK_a = \frac{ \Delta E }{2.303} -1.74.
\end{equation}
The $pK_a$ that comes out from the vacuum calculation
is $pK_a = 122$ and disagrees wildly with  the experimental value
$pK_a = 3.77$ for carboxylic acid\cite{Schuur,Silva,Klamt}. 
The deviation is caused by the neglect
of the surrounding water, which tends to support the dissociation reaction 
(and thus lowers the $pK_a$ value). In the second row we show the reaction
b) pictured in Fig.\ref{reaction}, which involves one coordinated water molecule
but otherwise occurs in vacuum. The $pK_a$ is lowered down to $pK_a = 92.7$
but is still much too high.  Building larger and larger water clusters is possible, but not
entirely satisfactory because the orientational freedom of liquid water is not 
preserved in a zero-temperature quantum-chemistry 
calculation.\footnote{In principle
a Car-Parrinello calculation, where the force fields in a MD simulation
are determined quantum-chemically, would cure this problem\cite{Car}.
However, the present-time accuracy of such calculations is not sufficient to 
predict absolute $pK_a$ values. In principle, the proton should also 
be treated quantum-mechanically, as is indeed possible with ab-initio
path-integral simulations\cite{Marx}.} As already described in  Section 11.1,
the dieletric properties of water can be approximately taken into account
by enclosing all molecules in a cavity outside of which a dielectric 
medium  is present.\footnote{Other methods for evaluating
polarizabilities based on ab-initio calculations are introduced 
in\cite{PolJung,PolJensen,PolMorita,PolMikkelson}.}
 The free parameter here is the radius of the cavity, which
consists of spheres that are centered around all heavy atoms. In the next
four  rows in Table 2 we show a set of results where the ratio between
the Pauling radii of the ions and the cavity radius 
is changed from 1.2 (the standard value) down to 0.9. 
It is seen that a smaller cavity radius brings down the $pK_a$ value,
until finally for a ratio of 0.9 a result close to the experimental value 
$pK_a \approx 3.77 $ is found.
In a previous calculation a similar problem of obtaining agreement 
 between experimental
$pK_a$ values and calculated ones was detected and discussed at length\cite{Klamt}.
In a sense, we use the cavity radius factor  as an adjustable parameter
to reproduce experimental results. 
It is important to note that we do not attach much physical significance to this
adjustment and  leave the whole problem of predicting $pK_a$ values 
to future investigations.
Our heuristic  viewpoint is utilized  in the
last row of Table 2, where we show a calculation where instead of the proton 
a sodium atom is allowed to bind to the carboxylic acid. As expected,
the resultant value $pK_a=-7.5$ shows that sodium binding can totally
be neglected. But for different acids and other ions the binding constants
might well be such that chemical binding must be taken into
account as a possible alternative to protonation events.
One example includes the case of Calcium ion binding to polyacrylic acids\cite{Fantinel}.

\begin{table}
\caption{\label{tab2} Ab-initio results for
the ground-state energies involved in the dissociation  of 
a carboxylic acid. The first two rows are for the reactions a) and b) in vacuum,
as shown in Fig.\ref{reaction}. The next four rows are for reaction a) but enclosed
in a dielectric cavity, mimicking an aqueous environment. 
Here the cavity radius scaling factor is changed from 1.2 to 0.9.
The last row contains results for the binding of a sodium ion. All energies are given 
in units of $kB_T$. }
\begin{tabular}{ccccccccc}
& $E_A$  		&   $E_{A-}$ 	& $ \Delta E_A$ 	& 	
$E_H$ 		& $E_{H+}$ 	&  $\Delta E_H $&
$ \Delta E $ 	& $pK_a $  \\
\hline
reaction a)    & -198677 & -198086 &  591.4 & -80025.0 &-80331.4  & -306.4& 285.0 & 122.0 \\
\hline
reaction b)  &&&  578.9 &&& -361.4 & 217.5 & 92.7 \\ 
\hline
a) cavity 1.2 &&& 482.2  &&& -430.8 & 51.4 &20.6 \\ 
a) cavity 1.1 &&& 477.1  &&& -437.6 & 39.5 &15.4 \\ 
a) cavity 1.0 &&& 473.3  &&& -445.4 & 28.0 &10.4 \\ 
a) cavity 0.9 &&& 471.9  &&& -454.9 & 17.0 & 5.6 \\ 
\hline
Na cavity 0.9 &&&  &&& & -13.1 & -7.5 \\ 
\end{tabular}\tabularnewline
\end{table}

\section{Summary and perspectives}

A number of different situations have been reviewed which have in common
that electrostatic effects play a dominant role. This is achieved for highly
charged surfaces and for  highly charged polymers. Among the most salient
results we find simple explanations for the puzzling phenomena of
attraction between similarly highly charged surfaces and overcompensation
of charged surfaces by adsorbing polyelectrolytes. In general terms, it is
the long-range nature of the Coulomb interaction which lies at the heart
of these effects. We also briefly talk about the effect of electric fields
on strongly coupled charged systems. For the specific case of a collapsed
charged polymer, an electric field induces motion of ions and charged monomers
and for high enough fields disolves the complex. This is an intrinsic 
non-equilibrium phenomenon. Finally, the interaction between oppositely
charged chemical groups has been investigated using quantum-mechanical ab-initio
methods. Since in highly charged systems one often has intimate contacts between
such groups, the short-ranged bonding we find is quite relevant for the
understanding of experiments where ion-specific effects are present.
One has only started to bridge the gap between the quantum-mechanical world
at small distances and the mesoscopic world of primitive models 
(where ions are replaced by hard spheres, and the solvent by a dielectric 
constant plus possibly effective interactions between the ions). What needs
to be fully elucidated is the coupling between water structure close to ions and
at charged surfaces and the effective interaction between such charged groups,
which probably involves effective many-body interactions. 
Experimentally evidenced ion-specific effects
will turn out to be a stringent test for such theories. 
Non-equilibrium phenomena are receiving more and more attention by theorists 
over the last 
years\cite{Loewenrev}. However,
the whole field of electrophoresis and electroosmosis still contains many open questions.
This is even more true for non-stationary non-equilibrium situations. 
Here simulation techniques are currently the method of choice, although 
field-theoretic and other analytical tools will prove useful as well.

\section*{Acknowledgements}

I thank Y. Burak, M. Dubois,
I. G\"ossl, M. Grunze, C. Holm, J.F. Joanny, A. Klamt, H.J. Kreuzer, 
W. Kunz, K.K. Kunze,  Y. Levin, P. Linse, S. Mamatkulov, A. Moreira,
H. Orland, P. Pincus, R. Podgornik, J.P. Rabe, B. Shklovskii, K. Takeyasu, 
K. Yoshikawa, T. Zemb
for discussions and for sending data and pictures.
Financial support of the German Science Foundation (DFG)
through SFB 486, SFB 563, and Schwerpunkt program
"Nano and Microfluidics"
 is acknowledged.

\end{document}